\documentclass{aa}

\usepackage[varg]{txfonts}

\defcitealias{Zoutendijk-2020-A&A-635-A107}{Paper~I}
\defcitealias{Zoutendijk-2021-A&A-651-A80}{Paper~II}

\begin{document}
    \title{The MUSE-Faint survey}
    \subtitle{III.~No large dark-matter cores and no significant tidal stripping in ultra-faint dwarf galaxies\thanks{Based on observations made with ESO Telescopes at the La Silla Paranal Observatory under programme IDs 0100.D-0807, 0101.D-0300, 0102.D-0372, 0103.D-0705, and 0104.D-0199.}}
    \titlerunning{The MUSE-Faint survey. III.}
    \author{%
        Sebastiaan~L. Zoutendijk\inst{1} \and
        Mariana~P. J\'ulio\inst{2,3} \and
        Jarle Brinchmann\inst{2,1} \and
        Justin~I. Read\inst{4} \and
        Daniel Vaz\inst{2,3} \and
        Leindert~A. Boogaard\inst{5} \and
        Nicolas~F. Bouch\'e\inst{6} \and
        Davor Krajnovi\'c\inst{7} \and
        Konrad Kuijken\inst{1} \and
        Joop Schaye\inst{1} \and
        Matthias Steinmetz\inst{7}%
    }
    \authorrunning{S.~L.~Zoutendijk et al.}
    \institute{%
        Leiden Observatory, Leiden University, P.O.~Box~9513, 2300~RA~Leiden, The Netherlands\\\email{zoutendijk@strw.leidenuniv.nl} \and
        Instituto de Astrof\'{\i}sica e Ci\^encias do Espa\c{c}o, Universidade do Porto, CAUP, Rua das Estrelas, PT4150-762~Porto, Portugal \and
        Departamento de F\'{\i}sica e Astronomia, Faculdade de Ci\^encias, Universidade do Porto, Rua do Campo Alegre 687, PT4169-007~Porto, Portugal \and
        University of Surrey, Physics Department, Guildford, GU2 7XH, UK \and
        Max Planck Institute for Astronomy, K\"{o}nigstuhl 17, D-69117 Heidelberg, Germany \and
        Univ. Lyon, Univ. Lyon1, ENS de Lyon, CNRS, Centre de Recherche Astrophysique de Lyon UMR5574, 69230, Saint-Genis-Laval, France \and
        Leibniz-Institut f\"ur Astrophysik Potsdam (AIP), An der Sternwarte 16, D-14482 Potsdam, Germany%
    }
    \date{Received date / Accepted date}
    \abstract{}{%
        The lowest-mass galaxies, ultra-faint dwarf galaxies, promise unparalleled constraints on how feedback regulates galaxy formation, and on the small-scale matter power spectrum.
        Their inner dark-matter densities can also be used to constrain dark-matter models and to determine the most promising targets for potential signals from dark-matter annihilation or decay.
        However, these goals are limited by the current large uncertainties on the properties of the dark-matter haloes that these galaxies inhabit.
        In this paper, we present 201 new stellar line-of-sight velocities from the MUSE-Faint survey for the faint and ultra-faint dwarf galaxies Antlia~B, Leo~T, Hydra~II, and Grus~1.
        Combining these with literature data, we obtain the tightest constraints to date on their dark-matter halo masses and inner dark-matter densities.%
    }{%
        We use the Jeans equations implemented in CJAM to model the density profiles and constrain the presence of dark-matter cores and solitons (a prediction of fuzzy dark-matter models).
        Further modelling is done with GravSphere to test the influence of the choice of modelling tool.
        We calculate masses, concentrations, and circular velocities from the profiles, include results for Eridanus~2 from our previous work, and compare these properties to theoretical scaling relations, deriving constraints on tidal stripping in the process.%
    }{%
        We find that dark-matter cores as large as those of more massive dwarf galaxies are ruled out for our galaxies (core radius $r_\mathrm{c} < 66$--$95\,\mathrm{pc}$ at the 68\% confidence level).
        We constrain the soliton radii to $r_\mathrm{sol} < 13$--$112\,\mathrm{pc}$ (68\% confidence level).
        We find that the galaxies are consistent with not having been significantly tidally stripped within their half-light radii.
        The virial masses and concentrations are sensitive to the choice of dynamical modelling tool: GravSphere produces results consistent with $M_{200} \sim 10^9\,M_\sun$, as expected from models in which ultra-faint dwarf galaxies are re-ionization fossils, while CJAM prefers haloes that are less massive.
    }{}
    \maketitle

\section{Introduction}
\label{sec:introduction}
    The smallest galaxies promise unparalleled constraints on cosmological models, on the nature of dark matter, and on how feedback processes regulate galaxy formation~\citep[e.g.][]{Battaglia-2013-NewAR-57-52, Bullock-2017-ARA&A-55-343, Simon-2019-ARA&A-57-375, Agertz-2020-MNRAS-491-1656}.
    They have become a particular focus due to several tensions at the scale of dwarf galaxies between observations and the prevailing $\Lambda$CDM cosmological model~\citep{Bullock-2017-ARA&A-55-343}.
    Two well-known issues are the core--cusp problem and the missing satellites problem.
    In the first, dark matter--only simulations predicted steep central density profiles for galaxies of any mass, while observations of several classical dwarf galaxies have shown evidence of constant-density cores~\citep{Flores-1994-ApJ-427-L1, Moore-1994-Natur-370-629}.
    The second problem entails that there are far fewer observed satellite galaxies of the Milky Way than that there are subhaloes in dark matter--only simulations~\citep{Klypin-1999-ApJ-522-82, Moore-1999-ApJ-524-L19}.
    Both problems could be addressed without abandoning $\Lambda$CDM by accounting for baryonic physics.

    Modern hydrodynamical simulations~\citep[e.g.][]{Mashchenko-2008-Sci-319-174, Governato-2010-Natur-463-203, Teyssier-2013-MNRAS-429-3068, DiCintio-2014-MNRAS-437-415, Brooks-2014-ApJ-786-87, Onnorbe-2015-MNRAS-454-2092, Read-2016-MNRAS-459-2573} are able to reproduce cores in dwarf galaxies of mass $M_{200} \sim 10^{10}\,M_\sun$ through repeated cycles of supernova feedback, a dynamical mechanism that has become known as dark-matter heating~\citep[e.g.][]{Navarro-1996-MNRAS-283-L72, Read-2005-MNRAS-356-107, Pontzen-2012-MNRAS-421-3464, Pontzen-2014-Natur-506-171}.
    This process requires sufficient star formation to take place over an extended period such that dwarf galaxies with only old stars are expected to have smaller, denser, dark-matter cores, or no cores at all~\citep[e.g.][]{Brook-2015-MNRAS-450-3920, Read-2016-MNRAS-459-2573, BermejoCliment-2018-MNRAS-479-1514}.
    Indeed, there seems to be mounting observational evidence for this scenario~\citep[e.g.][]{Read-2019-MNRAS-484-1401, 2021arXiv210907545B, 2021arXiv210914224S}.

    It remains unclear, however, whether the very smallest and faintest galaxies -- ultra-faint dwarf~(UFD) galaxies~\citetext{$M_V > -7.7$; \citealp{Simon-2019-ARA&A-57-375}} -- have undergone sufficient star formation to excite dark-matter cusp--core transformations.
    Models agree that such transformations are likely to be incomplete, yielding in some cases a lowering of the inner density from a pristine dark-matter cusp, but unlikely to form a large, low-density core as found in higher-mass dwarf galaxies~\citep[e.g.][]{Read-2019-MNRAS-484-1401}.
    The issue is made more complex, however, by a competing dynamical mechanism for core formation, dynamical friction~\citep[e.g.][]{ElZant-2001-ApJ-560-636, Nipoti-2015-MNRAS-446-1820}, that can act to lower the inner dark-matter density of UFDs, even when they have experienced little star formation~\citep{2021MNRAS.504.3509O}.

    The missing satellites problem can also be solved by baryonic effects, as it depends on which dark matter halos become luminous due to star formation.
    The latest abundance-matching models suggest that the most massive subhaloes are occupied by the Milky Way's classical dwarf galaxies.
    These models favour pre-infall masses for UFDs of ${\lesssim}10^9\,M_\sun$~\citep{Jethwa-2018-MNRAS-473-2060, Kim-2018-PhRvL-121-211302, Read-2019-MNRAS-487-5799, DESCollaboration-2020-ApJ-893-48}, consistent with the latest high-resolution numerical models~\citep[e.g.][]{2017arXiv170506286M, Agertz-2020-MNRAS-491-1656, Applebaum-2021-ApJ-906-96}.
    However, direct dynamical estimates of UFD pre-infall masses, while consistent with the abundance-matching methods and numerical models, are inconclusive due to their order-of-magnitude uncertainties~\citep[e.g.][]{Errani-2018-MNRAS-481-5073, Forbes-2018-MNRAS-481-5592, Read-2019-MNRAS-487-5799}.

    The above solutions to the cusp--core and missing satellite problems yield two key predictions that remain untested: (i)~UFDs should have cuspier inner dark-matter density profiles than nearby dwarf irregular galaxies that experience significantly more star formation; and (ii)~UFDs should inhabit dark-matter haloes of mass $M_{200} \sim 10^9\,M_\sun$ or lower.
    To test these predictions, we require accurate constraints on the dark-matter density profiles of UFDs.
    Such profiles, when measured close enough to the centre, will directly address the first prediction, while integration of the profile to larger radii will yield the halo masses of UFDs.

    With MUSE-Faint~\citetext{\citealp{Zoutendijk-2020-A&A-635-A107}; hereafter \citetalias{Zoutendijk-2020-A&A-635-A107}}, a survey of UFDs with MUSE~\citep{Bacon-2010-SPIE-7735-773508}, we are observing the spectroscopically unexplored centres of faint and ultra-faint satellites of the Milky Way.
    Previously, we have derived constraints on the dark-matter density profile of \object{Eridanus~2} (Eri~2) and on the properties of self-interacting and fuzzy dark matter~\citetext{\citealp{Zoutendijk-2021-A&A-651-A80}; hereafter \citetalias{Zoutendijk-2021-A&A-651-A80}}.
    While we could not resolve a core, we found that a core in Eri~2 must be smaller than ${\approx}100\,\mathrm{pc}$ (95\% confidence level), which is smaller than the cores found in larger dwarf galaxies.
    To start addressing the above predictions on UFD core size and mass more rigorously, in this paper we expand our analysis to the four additional dwarf galaxies with completed MUSE-Faint observations; in decreasing order of luminosity: \object{Antlia~B} (Ant~B), \object{Leo~T}, \object{Hydra~II} (Hya~II), and \object{Grus~1} (Gru~1).
    We use different density profile models to test whether UFDs have cores and how large these cores can be, and to calculate halo properties.
    We focus on testing the predicted UFD properties in the $\Lambda$CDM paradigm and will therefore model each galaxy individually, without assuming a particular core formation mechanism.
    We defer a joint analysis of the density profiles, in which we will constrain the nature and properties of dark matter through the core formation prescribed by the dark-matter physics of each model, to a follow-up paper.

    By expanding the number of dark-matter profiles from MUSE-Faint galaxies from one to five, we are able to compare the properties of a small population of galaxies with scaling relations and other expectations.
    First, we introduce our observations and data reduction process~(Sect.~\ref{sec:obsred}) and our methods~(Sect.~\ref{sec:methods}).
    We then constrain the density profiles of the new galaxies using three different profile models (core, cusp, and soliton), derive constraints on halo properties, and compare the evidence for the different models~(Sects.~\ref{ssec:param}--\ref{ssec:modcomp}).
    Additionally, we test whether the four new galaxies, supplemented with Eri~2, could have undergone tidal stripping of dark matter, and examine how their galaxy and halo properties compare to mass--concentration relations and stellar-to-halo mass ratios determined from simulations and observations of more massive galaxies~(Sects.~\ref{ssec:stripping} and~\ref{ssec:scaling}).
    We end with a discussion of our results~(Sect.~\ref{sec:discussion}) and a summary of our conclusions~(Sect.~\ref{sec:conclusions}).
    In Appendix~\ref{app:gravsphere} we test the robustness of our results by repeating our analyses with a different dynamical modelling tool, and we provide figures of the constraints on the profile model parameters, including the core size, in Appendix~\ref{app:corner}.

    In keeping with the previous papers in this series, we adopt the \emph{Planck} 2015 cosmological parameters~\citep{PlanckCollaboration-2016-A&A-594-A13}.
    For convenience, we consistently show results for the different profile models with the same colour in every figure.

\section{Observations and data reduction}
\label{sec:obsred}
    We describe here the galaxies in our sample~(Sect.~\ref{ssec:sample}) and our observations and general process of data reduction~(Sect.~\ref{ssec:obsred}), followed by details about the individual galaxies~(Sects.~\ref{ssec:obsredAntB}--\ref{ssec:obsredGru1}).

\subsection{Sample}
\label{ssec:sample}
    Ant~B is part of the nearby \object{NGC~3109} association and was discovered by \citet{Sand-2015-ApJL-812-L13} using Blanco/DECam imaging.
    These data indicate a distance of $1.29 \pm 0.10\,\mathrm{Mpc}$, an absolute $V$-band magnitude of $-9.7 \pm 0.6\,\mathrm{mag}$, a half-light radius of $0.72 \pm 0.7\,\mathrm{arcmin}$ ($273 \pm 29\,\mathrm{pc}$), and the presence of two stellar populations: one with an age ${>}10\,\mathrm{Gyr}$ and metallicity $[\element{Fe}/\element{H}] \approx -2$, the other ${\approx}200$--$400\,\mathrm{Myr}$ old and with $[\element{Fe}/\element{H}] \approx -1$.
    No H$\alpha$ was detected with SOAR/Goodman spectroscopy, which is consistent with the lack of a population ${<}100\,\mathrm{Myr}$ old~\citep{Sand-2015-ApJL-812-L13}.
    Radio spectroscopy with GBT/VEGAS revealed $(2.8 \pm 0.2) \times 10^5\,M_\sun$ of \ion{H}{i} with a line-of-sight velocity of ${\approx}375\,\mathrm{km}\,\mathrm{s}^{-1}$.
    \citet{Hargis-2020-ApJ-888-31} followed up with \emph{HST}/ACS photometry, revising the distance to $1.35 \pm 0.06\,\mathrm{Mpc}$.
    They also constrained the star-formation history of Ant~B using this higher-resolution photometry, which shows that Ant~B started forming stars ${\approx}13\,\mathrm{Gyr}$ ago and had its last episode of star formation ${\approx}2$--$3\,\mathrm{Gyr}$ ago.

    \begin{table*}
        \centering
        \caption{Structural parameters adopted from the literature.}
        \label{tab:adopted}
        \begin{tabular}{lcccccccccc}
            \hline
            \hline
                   & RA               & Dec.             & $D$                & $\theta$         & $\varepsilon$ & $R_{1/2}$           & $M_V$                        & $\mu_{V,0}$                            & $m_V - M_V$      & References \\
                   & ($\mathrm{deg}$) & ($\mathrm{deg}$) & ($\mathrm{kpc}$)   & ($\mathrm{deg}$) &               & ($\mathrm{arcmin}$) & ($\mathrm{mag}$)             & ($\mathrm{mag}\,\mathrm{arcsec}^{-2}$) & ($\mathrm{mag}$) &            \\
            \hline
            Ant~B  & $147.2337$       & $-25.9900$       & $1.35 \times 10^3$ &   $+4.0$         & $0.30$        & $0.72$              & $-9.7$                       &                                        & $25.65$          & 1, 2       \\
            Leo~T  & $143.7292$       & $+17.0482$       & $417.0$            & $-104$           & $0.24$        & $1.27$              & $-7.60$                      & $25.42$                                &                  & 3, 4, 5    \\
            Hya~II & $185.4251$       & $-31.9860$       & $151$              &  $+13$           & $0.25$        & $1.65$              & $-4.86$~~\tablefootmark{(a)} & $26.14$                                &                  & 3, 6       \\
            Gru~1  & $344.166$        & $-50.168$        & $125$              & $+153$           & $0.44$        & $4.16$              & $-4.1$                       &                                        & $20.48$          & 7          \\
            \hline
        \end{tabular}
        \tablefoot{%
            RA: central right ascension;
            Dec.: central declination;
            $D$: heliocentric distance;
            $\theta$: position angle;
            $\varepsilon$: ellipticity;
            $R_{1/2}$: angular half-light radius;
            $M_V$: absolute $V$-band magnitude;
            $\mu_{V,0}$: central $V$-band surface brightness;
            $m_V-M_V$: $V$-band distance modulus.
            \tablefoottext{a}{Adjusted for the difference in adopted distance.}%
        }
        \tablebib{%
            (1)~\citet{Sand-2015-ApJL-812-L13};
            (2)~\citet{Hargis-2020-ApJ-888-31};
            (3)~\citet{Munnoz-2018-ApJ-860-66};
            (4)~\citet{Irwin-2007-ApJ-656-L13};
            (5)~\citet{Simon-2007-ApJ-670-313};
            (6)~\citet{Vivas-2016-AJ-151-118};
            (7)~\citet{2021ApJ...916...81C}.%
        }
    \end{table*}

    The earliest-discovered galaxy in our sample, Leo~T, was found by \citet{Irwin-2007-ApJ-656-L13} in SDSS DR5 images.
    Follow-up with INT/WFC revealed two stellar populations, one ${<}1\,\mathrm{Gyr}$ old, the other $6$--$8\,\mathrm{Gyr}$ old with $[\element{Fe}/\element{H}] \approx -1.6$.
    The galaxy was found to have a total absolute magnitude of $-7.1\,\mathrm{mag}$ and a half-light radius of $1.4\,\mathrm{arcmin}$ ($170 \pm 15\,\mathrm{pc}$; \citealp{Simon-2007-ApJ-670-313}).
    An \ion{H}{i} component of ${\approx}2 \times 10^5\,M_\sun$ was found in HIPASS with a line-of-sight velocity of $35\,\mathrm{km}\,\mathrm{s}^{-1}$~\citep{Irwin-2007-ApJ-656-L13}, at a distance of $417^{+20}_{-19}\,\mathrm{kpc}$~\citep{Irwin-2007-ApJ-656-L13, Simon-2007-ApJ-670-313}.
    \citet{Simon-2007-ApJ-670-313} observed Leo~T with Keck/DEIMOS and determined the stars to have a mean line-of-sight velocity of $38.1 \pm 2.0\,\mathrm{km}\,\mathrm{s}^{-1}$ with a dispersion of $7.5\,\mathrm{km}\,\mathrm{s}^{-1}$.
    Their spectroscopy indicates a much lower metallicity, $[\element{Fe}/\element{H}] = -2.29 \pm 0.10$, than the photometry.
    Using GMRT and WSRT, \citet{RyanWeber-2008-MNRAS-384-535} find that the \ion{H}{i} gas consists of two components with different temperatures, ${\approx}500\,\mathrm{K}$ and ${\approx}6000\,\mathrm{K}$, where the cold component is more centrally concentrated, as is the younger stellar population.
    This is confirmed by \citet{Adams-2018-A&A-612-A26} with deeper WSRT data, resulting in an increased \ion{H}{i} mass of $4.1 \pm 0.4 \times 10^5\,M_\sun$, of which ${\approx}10\%$ is a cold neutral medium with a velocity dispersion of $2.5 \pm 0.1\,\mathrm{km}\,\mathrm{s}^{-1}$, while the remaining part is a warm neutral medium with a velocity dispersion of $7.1 \pm 0.4\,\mathrm{km}\,\mathrm{s}^{-1}$.
    Further photometric studies with LBT/LBC~\citep{deJong-2008-ApJ-680-1112} and HST/WFPC2~\citep{Weisz-2012-ApJ-748-88} and re-analyses of photometric \citep{Clementini-2012-ApJ-756-108} and spectroscopic \citep{Kirby-2008-ApJL-685-L43, Kirby-2013-ApJ-779-102} data confirm the existence of two stellar populations and report metallicities ranging from $[\element{Fe}/\element{H}] = -2.02 \pm 0.05$ to ${\approx}{-1.5}$.
    The LBT/LBC data of \citet{deJong-2008-ApJ-680-1112} indicates an absolute magnitude of $-8.0\,\mathrm{mag}$ and a half-light radius of $0.99 \pm 0.06\,\mathrm{arcmin}$ ($120 \pm 7\,\mathrm{pc}$), while \citet{Munnoz-2018-ApJ-860-66} find $-7.60 \pm 0.14\,\mathrm{mag}$ and $1.27 \pm 0.13\,\mathrm{arcmin}$ ($154 \pm 16\,\mathrm{pc}$) from new Magellan/Megacam imaging.

    Hya~II was discovered by \citet{Martin-2015-ApJ-804-L5} in the Survey of the MAgellanic Stellar History~(SMASH; \citealp{Nidever-2017-AJ-154-199}), consisting of Blanco/DECam imaging.
    From these data \citeauthor{Martin-2015-ApJ-804-L5} estimate that Hya~II is $13\,\mathrm{Gyr}$ old, has a metallicity of $[\element{Fe}/\element{H}] = -2.2$, has an absolute $V$-band magnitude of $-4.8 \pm 0.3\,\mathrm{mag}$, has a half-light radius of $1.7^{+0.3}_{-0.2}\,\mathrm{arcmin}$ ($68 \pm 11\,\mathrm{pc}$), and is located at a distance of $134 \pm 10\,\mathrm{kpc}$.
    \citet{Kirby-2015-ApJ-810-56} have obtained Keck/DEIMOS spectroscopy and determined a metallicity of $[\element{Fe}/\element{H}] = -2.02 \pm 0.08$.
    The intrinsic velocity dispersion could not be resolved.
    \citet{Vivas-2016-AJ-151-118} find a distance of $151 \pm 8\,\mathrm{kpc}$, based on time series observations of an RR~Lyrae star.
    \citet{Munnoz-2018-ApJ-860-66} re-analyse the data of \citet{Martin-2015-ApJ-804-L5} and find an absolute $V$-band magnitude of $-4.60 \pm 0.37\,\mathrm{mag}$ and a half-light radius of $1.65 \pm 0.39\,\mathrm{arcmin}$ ($64.3 \pm 15.2\,\mathrm{pc}$).
    An analysis of photometry from \emph{HST} by \citet{2021ApJ...920L..19S} indicates that Hya~II reached half of its cumulative star formation $13.21 \pm 0.32\,\mathrm{Gyr}$ ago.

    \citet{Koposov-2015-ApJ-805-130} discovered Gru~1 in public DES data, with an absolute $V$-band magnitude of $-3.4 \pm 0.3\,\mathrm{mag}$ and a half-light radius of $1.77^{+0.85}_{-0.39}\,\mathrm{arcmin}$ ($62^{+29.8}_{-13.6}\,\mathrm{pc}$), at a distance of $120~\mathrm{kpc}$.
    No \ion{H}{i} was detected in its direction \citep{Westmeier-2015-MNRAS-453-338}.
    Though the sky position of Gru~1 was not part of the DES Y1A1 catalogue analysed by the \citet{DESCollaboration-2015-ApJ-807-50} simultaneously to and independent of the study of \citet{Koposov-2015-ApJ-805-130}, its presence was confirmed using data of the DES Y2 catalogue \citep{DESCollaboration-2015-ApJ-813-109}.
    \citet{Walker-2016-ApJ-819-53} studied Gru~1 with Magellan/M2FS spectroscopy and find a mean line-of-sight velocity of $-140.5^{+2.4}_{-1.6}\,\mathrm{km}\,\mathrm{s}^{-1}$, but could not resolve the dispersion.
    In the same study, the metallicity was determined to be $[\element{Fe}/\element{H}] = -1.42^{+0.55}_{-0.42}$.
    \citet{Munnoz-2018-ApJ-860-66} re-analysed archival DES data and find an absolute $V$-band magnitude of $-3.47 \pm 0.59\,\mathrm{mag}$ and a half-light radius of $1.50 \pm 0.68\,\mathrm{arcmin}$ ($52.4 \pm 23.8\,\mathrm{pc}$).
    \citet{2018arXiv180902259J} find a much more metal-poor result, $[\element{Fe}/\element{H}] = -2.5 \pm 0.3$, using Gemini/GMOS-S photometry, and find a distance of $115 \pm 6\,\mathrm{kpc}$.
    \citet{Ji-2019-ApJ-870-83}, on the other hand, find a metallicity of $[\element{Fe}/\element{H}] \approx -2.5$ for two stars with high-resolution Magellan/MIKE spectroscopy, and suggest the difference with the result of \citet{Walker-2016-ApJ-819-53} is due to the presence of low--signal-to-noise spectra in the original spectroscopic sample.
    With SOAR/Goodman and Blanco/DECam, the \citet{DESCollaboration-2019-MNRAS-490-2183} find two RLLs in Gru~1, which lead to a distance measurement of $127 \pm 6\,\mathrm{kpc}$.
    Reanalysing DES data, \citet{Moskowitz-2020-ApJ-892-27} find a significantly larger half-light radius of $2.84^{+0.35}_{-0.28}\,\mathrm{arcmin}$.
    The \citet{2021ApJ...916...81C} performed a deeper photometric study with Magellan/Megacam and find Gru~1 consists of a single stellar population at a distance of $125^{+6}_{-12}\,\mathrm{kpc}$, with an age of $13.26^{+0.18}_{-0.25}\,\mathrm{Gyr}$, a metallicity of $[\element{Fe}/\element{H}] \approx -1.88^{+0.09}_{-0.03}$, an absolute $V$-band magnitude of $-4.1 \pm 0.3\,\mathrm{mag}$, and a half-light radius of $4.16^{+0.54}_{-0.84}\,\mathrm{arcmin}$ ($151^{+21}_{-31}\,\mathrm{pc}$).

    For consistency, when available and reasonably up-to-date, we adopted the same positions and distances as in the homogeneous study by \citet{Munnoz-2018-ApJ-860-66} and the same photometric and structural parameters as they derive.
    Though \citet{Sersic-1963-BAAA-6-41, Sersic-1968-AGA-OA-1} profiles provide the best fits, exponential profiles are consistent and have fewer parameters; therefore we adopt the latter.
    Ant~B, not being a satellite of the Milky Way, was not included in the study of \citet{Munnoz-2018-ApJ-860-66}.
    We therefore use the distance of \citet{Hargis-2020-ApJ-888-31} and the position, absolute magnitude, and structural parameters of \citet{Sand-2015-ApJL-812-L13}.
    For Hya~II, we found that the distance assumed by \citet{Munnoz-2018-ApJ-860-66} leads to isochrones that do not fit the horizontal branch.
    The other distance determination, from \citet{Vivas-2016-AJ-151-118}, did fit.
    Though we retain the angular structural parameters and surface brightness from \citet{Munnoz-2018-ApJ-860-66}, we adjust their absolute magnitude in accordance with the change in distance.
    Lastly, a recent, deeper study of Gru~1~\citep{2021ApJ...916...81C} revealed a substantially larger half-light radius and brighter magnitude, therefore we adopt the parameters found in this study.
    We display the list of adopted parameters in Table~\ref{tab:adopted}.

\subsection{Overview of observations and data reduction}
\label{ssec:obsred}
    In addition to $21.5\,\mathrm{h}$ of data in five fields towards Eri~2, reduced and presented in \citetalias{Zoutendijk-2020-A&A-635-A107} and \citetalias{Zoutendijk-2021-A&A-651-A80}, in this paper we used $4.5\,\mathrm{h}$ on Ant~B (one field), $4\,\mathrm{h}$ on Leo~T (one field), $14.75\,\mathrm{h}$ on Hya~II (four fields), and $4\,\mathrm{h}$ on Gru~1 (one field) from the MUSE-Faint survey.
    These data were taken between February 2018 and February 2020 during MUSE Collaboration guaranteed-time observing runs.
    The natural seeing varied between $0.6$ and $1.2\,\mathrm{arcsec}$, with median values between $0.7$ and $0.9\,\mathrm{arcsec}$ for each dwarf galaxy.
    The adaptive-optics system indicated corrected median seeings between $0.5$ and $0.6\,\mathrm{arcsec}$.
    After the data reduction we measured full widths at half-maximum of $0.55$, $0.61$, $0.40$, and $0.67~\mathrm{arcsec}$ for Ant~B, Leo~T, Hya~II, and Gru~1, respectively, at $7000\,\AA$, by fitting a Moffat function to the point spread functions of the brightest stars.

\begin{figure*}
        \centering
        \includegraphics[scale=0.45]{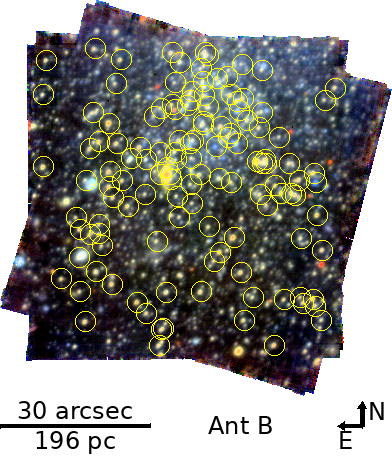}\hfill
        \includegraphics[scale=0.45]{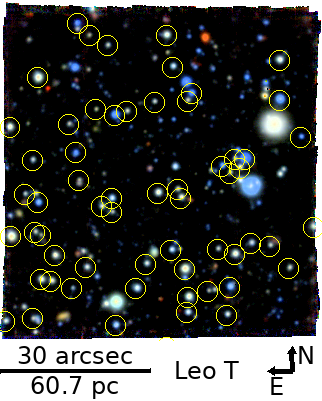}\hfill
        \includegraphics[scale=0.45]{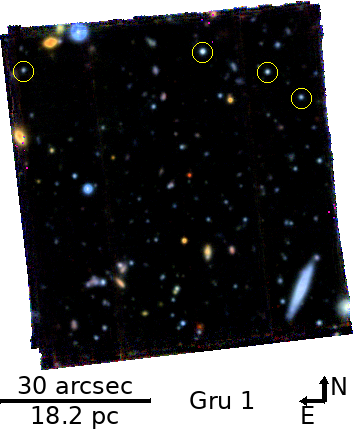}
        \includegraphics[scale=0.45]{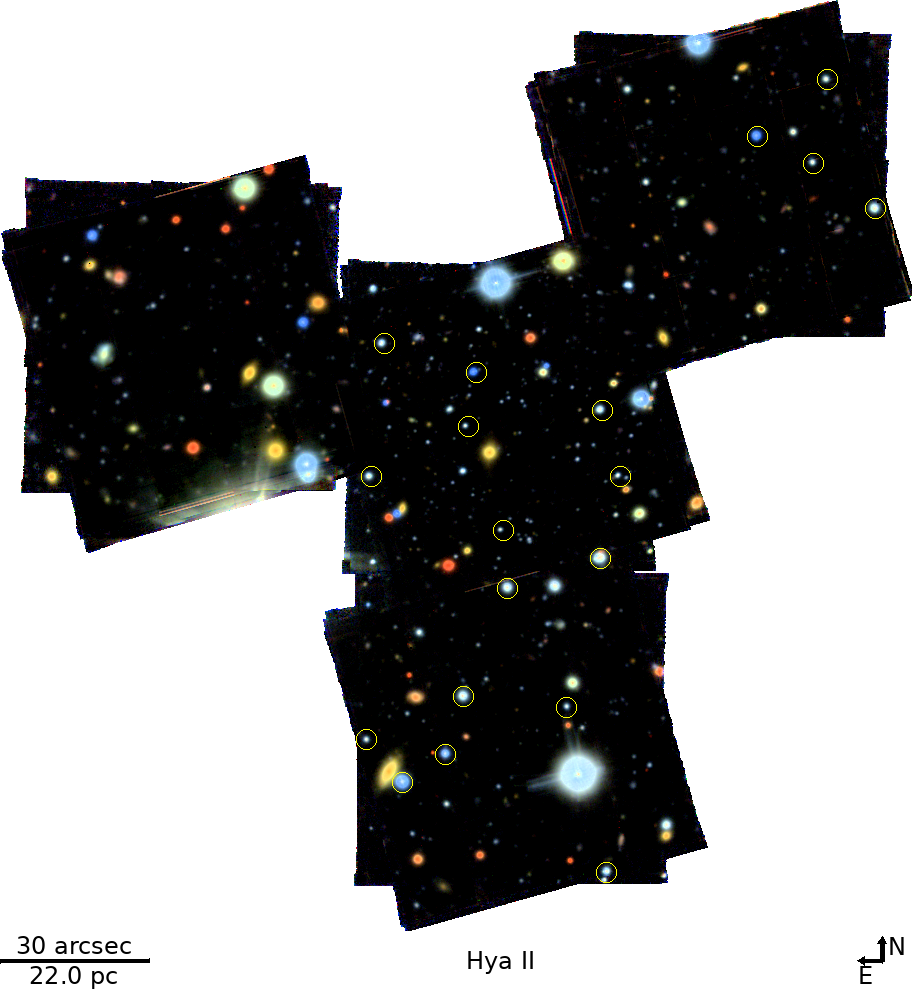}%
        \caption{%
            Composite-colour images of the targeted galaxies based on data from the MUSE-Faint survey.
            The SDSS filters $g$, $r$, and $i$ were used to create the blue, red, and green channels, respectively, of the images.
            Stars identified in this paper as members of these galaxies that are located within the bounds of these MUSE-Faint images are indicated with yellow circles.
            The angular and physical scales of the images are indicated at their lower left corners.
            The directions north and east are indicated in the lower right corners of the images.%
        }
        \label{fig:im}
    \end{figure*}

    We used the same data reduction procedure as for Eri~2 in \citetalias{Zoutendijk-2020-A&A-635-A107} and \citetalias{Zoutendijk-2021-A&A-651-A80}, which we summarize here for convenience.
    We used the standard procedure for MUSE data reduction using the MUSE Data Reduction Software~(DRS; version~2.6; \citealp{Weilbacher-2020-A&A-641-A28}), supplemented with a bad-pixel table from \citet{Bacon-2017-A&A-608-A1} and an auto-calibration step, when possible.
    The field of Ant~B was too crowded, therefore the auto-calibration was skipped for this target.
    Contrary to the reduction for Eri~2, it was not necessary to build a source mask from a source catalogue during the auto-calibration of Leo~T, Hya~II, and Gru~1.
    Instead we relied on the automatic masking based on pixel brightness.
    These same three galaxies were reduced with an updated overscan setting that became the default in DRS version~2.8~\citep{Weilbacher-2020-A&A-641-A28} after this was found to improve the reduction quality of the MUSE Extremely Deep Field~(MXDF; Bacon et al.\ in prep.).
    For Leo~T a satellite trail had to be masked as well.
    The data cubes reduced with the DRS were post-processed with the Zurich Atmosphere Purge (ZAP; version~2.0; \citealp{Soto-2016-MNRAS-458-3210}) to remove residual sky signatures.
    Composite-colour images of the post-processed cubes created using the method of \citet{Lupton-2004-PASP-116-133}, using SDSS filters $g$, $r$, and $i$ for blue, green, and red, respectively, are shown in Fig.~\ref{fig:im}.
    Spectral extraction was performed with PampelMuse (version~1.0rc2; \citealp{Kamann-2013-A&A-549-A71}) using source catalogues produced from public \emph{Hubble Space Telescope}~(\emph{HST}) data (see Sects.~\ref{ssec:obsredAntB}--\ref{ssec:obsredGru1}).
    In the case of Hya~II, the source catalogue included a number of spurious detections around a bright star; these we removed manually from the catalogue.
    The extracted spectra were then fit with spexxy (version~2.5; \citealp{Husser-2012-3DSDSP-UG-0}) in combination with the PHOENIX synthetic stellar spectral library, resulting in line-of-sight velocities.

    We used publicly available \emph{HST} photometry in the F606W and F814W bands, which we analysed with SExtractor~\citep{Bertin-1996-A&AS-117-393} to construct colour--magnitude diagrams for each dwarf galaxy.
    We compared the \emph{HST} photometry to PARSEC~\citep{Bressan-2012-MNRAS-427-127, Chen-2014-MNRAS-444-2525, Chen-2015-MNRAS-452-1068, Tang-2014-MNRAS-445-4287, Marigo-2017-ApJ-835-77, Pastorelli-2019-MNRAS-485-5666, Pastorelli-2020-MNRAS-498-3283} isochrones, assuming a Kroupa IMF~\citep{Kroupa-2001-MNRAS-322-231, Kroupa-2002-Sci-295-82, 2013pss5.book..115K} corrected for unresolved binaries.
    We converted the isochrones from Vega to AB magnitudes using the solar magnitudes of \citet{Willmer-2018-ApJS-236-47} and applied Galactic dust extinction~\citetext{\citealp{Schlegel-1998-ApJ-500-525}; \citealp{Schlafly-2011-ApJ-737-103} recalibration}.
    We used metallicity, age, and distance measurements from the literature as starting points to select one or more isochrones.
    As the goal here was to determine source membership and not a detailed star-formation history, we did not fine tune the isochrone parameters but accepted a combination that gave a good match by eye to the colour-magnitude diagram.
    To determine which stars are photometrically consistent with membership, we adopted an uncertainty of $0.1\,\mathrm{mag}$ on the colour values of the isochrone.
    This is to compensate for the coarse isochrone parameter selection, the discontinuous nature of the isochrone samples, the possibility of variations in stellar parameters (e.g.\ metallicity dispersion), and any error in the theoretical isochrones.
    We combined this uncertainty with the stellar measurement uncertainties on both colours, and rejected stars offset more than two combined standard deviations from the isochrone.
    From the fitted isochrone and the accompanying PARSEC simple stellar population simulation, we calculated a stellar mass-to-light ratio for each dwarf galaxy by adding stellar remnant masses to the active stellar masses using the recipes from \citet{Renzini-1993-ApJ-416-L49}.

    We queried the astrometry and kinematics of sources in \emph{Gaia} Early Data Release~3~\citetext{EDR3; \citealp{GaiaCollaboration-2016-A&A-595-A1, 2021A&A...649A...1G, 2021A&A...649A...2L}} within a radius from the centre of each dwarf galaxy sufficiently large to encapsulate the sky coverage of both the MUSE and the \emph{HST} observations.
    We considered a given parallax measurement reliable if it is positive and if the renormalized unit-weight error~(RUWE) is less than $1.4$~\citep{ruwe}.
    If a source had a reliable parallax measurement that was inconsistent with zero at a level of at least $2\sigma$, we rejected the source.
    When available, we used \emph{Gaia} line-of-sight velocity measurements carried over from Data Release~2; however, no sources with \emph{Gaia} line-of-sight velocities made it into the final source selections.

    Not all spectra extracted with PampelMuse have a converged spexxy fit, which naturally limits the number of sources we can include in our sample.
    Below a signal-to-noise ratio~(S/N) of 5, spexxy may incorrectly estimate uncertainties \citep{Kamann-2018-MNRAS-473-5591}, which would influence the intrinsic velocity dispersion estimate.
    However, the faintness of these galaxies makes it challenging to assemble a large enough sample of sources to trace the density profile.
    We have therefore relaxed the S/N criterion to 3 in the three faintest galaxies Leo~T, Hya~II, and Gru~1.
    To safeguard ourselves from introducing a bias into the kinematics in this way, we have confirmed that the bulk intrinsic velocity and bulk intrinsic velocity dispersion are consistent at both S/N criteria.

    When available, we added line-of-sight velocity measurements from the literature to our sample.
    These generally come from high-resolution spectroscopic measurements, which typically have smaller velocity uncertainties, but the lower spatial resolution of these instruments introduces a lack of measurements in the dense centres of faint dwarf galaxies; a lack we can address with the higher spatial resolution and smaller field of view of MUSE.

    \begin{table*}
        \centering
        \caption{Derived properties of the stellar populations and the kinematic samples.}
        \label{tab:derived}
        \begin{tabular}{lccccc}
            \hline
            \hline
                   & $\log_{10}(M_*/M_\sun)$ & $N_\mathrm{mem}^\mathrm{MUSE}$ & $N_\mathrm{mem}^\mathrm{tot}$ & $\mu$                            & $\sigma_\mathrm{int}$            \\
                   &                         &                                &                               & ($\mathrm{km}\,\mathrm{s}^{-1}$) & ($\mathrm{km}\,\mathrm{s}^{-1}$) \\
            \hline
            Ant~B  & $5.88$                  & 127                            & 127                           &  $375.5 \pm 1.5$                 &  $8.0^{+1.6}_{-1.4}$             \\
            Leo~T  & $5.16$                  &  55                            &  75                           &   $39.5 \pm 2.1$                 &  $7.6^{+2.3}_{-1.7}$             \\
            Hya~II & $4.30$                  &  15                            &  28                           &  $299.5^{+4.5}_{-4.7}$           & $12.0^{+5.0}_{-3.5}$             \\
            Gru~1  & $2.80$                  &   4                            &  14                           & $-139.2^{+6.1}_{-5.2}$           & $10.4^{+9.3}_{-5.1}$             \\
            \hline
        \end{tabular}
        \tablefoot{%
            $M_*$: stellar mass;
            $N_\mathrm{mem}^\mathrm{MUSE}$: number of member stars in the sample with a MUSE-Faint velocity measurement;
            $N_\mathrm{mem}^\mathrm{tot}$: total number of member stars in the sample;
            $\mu$: mean velocity;
            $\sigma_\mathrm{int}$: intrinsic velocity dispersion.%
        }
    \end{table*}
    We matched the celestial positions of entries in the \emph{HST} F814W, \emph{Gaia} EDR3, and any high-resolution spectroscopy catalogues using TOPCAT~\citep{2005ASPC..347...29T}.
    We remind the reader that the spexxy velocities are already linked to the \emph{HST} positions through the use of the \emph{HST} catalogue in the extraction of spectra with PampelMuse.
    We calibrated the positions of \emph{HST} and high-resolution spectroscopy entries to \emph{Gaia} EDR3 using the sources in common.
    We then built a unified catalogue of positions and velocities:
    We used the \emph{Gaia} EDR3 position if available for a source, otherwise we take the average of any other (\emph{Gaia}-calibrated) positions available.
    We combined velocity measurements by taking the average weighted by the inverse variance and propagated the measurement uncertainties accordingly.
    To filter out galaxies, we reject sources for which SExtractor determined $\verb|CLASS_STAR| < 0.8$ in both the F606W and the F814W \emph{HST} data.
    The remaining sources we considered stars.
    We removed foreground stars using our parallax criterion and applied the isochrone criterion.

    Finally, we used the following procedure to identify kinematic outliers:
    We modelled the distribution underlying the velocity measurements~$v_i$ and their uncertainties~$\varepsilon_i$ as a normal distribution of member stars, with mean~$\mu$ and intrinsic dispersion~$\sigma$, and a uniform distribution of contaminants within the velocity range considered taking up a fraction~$\eta$ of the sample.
    The likelihood that the measurements are generated by this model with a particular combination of parameters is
    \begin{equation}
        \begin{aligned}
        &\mathcal{L}(\mu, \sigma, \eta | v_i, \varepsilon_i) =\\
        &\quad\prod_i \Biggl(\frac{1-\eta}{\sqrt{2\pi\bigl(\sigma^2+\varepsilon_i^2\bigr)}} \exp\Biggl(-\frac{(v_i-\mu)^2}{2(\sigma^2+\varepsilon_i^2)}\Biggr) + \frac{\eta}{\max(v_i)-\min(v_i)}\Biggr),
        \end{aligned}
        \label{eq:sel-like}
    \end{equation}
    We used emcee~\citep{ForemanMackey-2013-PASP-125-306} to find the posterior distributions of $\mu$, $\sigma$, and $\eta$.
    We assumed uniform priors of $\min(v_i)$ to $\max(v_i)$ for $\mu$, $0$ to $50\,\mathrm{km}\,\mathrm{s}^{-1}$ for $\sigma$, and $0$ to $1$ for $\eta$.
    For the median values $\hat{\mu}$ and $\hat{\sigma}$ of the posteriors of $\mu$ and $\sigma$, respectively, we calculated the membership likelihood of each star:
    \begin{equation}
        \mathcal{L}_\mathrm{mem}(v_i, \varepsilon_i | \hat{\mu}, \hat{\sigma}) = \frac{1}{\sqrt{2\pi(\hat{\sigma}^2+\varepsilon_i^2)}} \exp\Biggr(-\frac{(v_i-\hat{\mu})^2}{2(\hat{\sigma}^2+\varepsilon_i^2)}\Biggr).
        \label{eq:pmem}
    \end{equation}
    When ordered according to decreasing membership likelihood, we found that the membership likelihoods gradually decrease down to a certain point after which the membership likelihoods of remaining stars rapidly decrease.
    This break was usually located at a likelihood value between the equivalent of a $2\sigma$ and a $5\sigma$ outlier of a normal distribution with standard deviation~$\sigma$.
    These breaks in the membership likelihood distributions seemed an appropriate choice to divide the stars into members and non-members.
    We also compared the membership likelihoods against the expected values for normally distributed measurements, using a quantile--quantile plot, to ensure that the proposed selection of members and non-members would not overly curtail the distribution.
    After finalizing the selection, we removed the non-members.
    The kinematic selection procedure was repeated on the remaining stars until we found no more non-members.
    Finally, we set $\eta = 0$ and ran emcee once again.
    We compared the resulting intrinsic distribution to the one reported in the literature, if available (see Sects.~\ref{ssec:obsredAntB}--\ref{ssec:obsredGru1}).

    In the following sections we discuss the sample selection procedure for each galaxy.
    We provide a summary of the derived sample and stellar properties in Table~\ref{tab:derived}.

\subsection{Antlia~B}
\label{ssec:obsredAntB}
    J\'ulio et al.\ \citetext{in prep.} describe the sample of selected member stars in Ant~B.
    For the sake of completeness, we summarize the procedure here.
    An extraction catalogue was created by running SExtractor on publicly available \emph{HST}/ACS photometry\footnote{\emph{Hubble Space Telescope} Proposal 14078, principal investigator Jonathan Hargis.}.
    Using this catalogue, 141 spectra with a S/N of ${>}5$ were extracted from the MUSE-Faint data and 131 of these were successfully fitted.
    A visual inspection of the \emph{HST} photometry and the extracted spectra revealed no galaxies.
    The extracted stars had a very clean velocity distribution, with four clear outliers more than twice the standard deviation away from the mean.
    This lead to a sample of 127~stars.
    The Ant~B colour--magnitude diagram shows a very clear separation between photometrically consistent and inconsistent stars; we could verify by eye that no photometric outliers were present in this sample.

    Out to a radius of $5\,\mathrm{arcmin}$ from the centre of Ant~B, we found 305~\emph{Gaia} EDR3 sources.
    Seven sources had a DR2 line-of-sight velocity, of which six also had a good RUWE.
    These six were however all foreground stars, leaving our sample at 127~sources.

    Though a kinematic cut was already made, we applied our kinematic outlier procedure to test for remaining contaminants.
    We did not find any additional outliers.
    We also inspected the \verb|CLASS_STAR| values returned by SExtractor as an extra precaution, but found no additional galaxies in the sample.
    The intrinsic mean line-of-sight velocity of the final sample is $375.5 \pm 1.5\,\mathrm{km}\,\mathrm{s}^{-1}$, with an intrinsic dispersion of $8.0^{+1.6}_{-1.4}\,\mathrm{km}\,\mathrm{s}^{-1}$.

    Ant~B reached half its present cumulative star formation ${\approx}6\,\mathrm{Gyr}$ ago~\citep{Hargis-2020-ApJ-888-31}.
    For the purpose of obtaining stellar mass-to-light ratios, we fitted a PARSEC isochrone with distance $D = 1.35\,\mathrm{Mpc}$, age $t = 6.0\,\mathrm{Gyr}$, and metallicity $[\element{M}/\element{H}] = -2.0$.
    This isochrone led to a stellar mass-to-light ratio of $1.19$.
    Adopting an absolute magnitude $M_V = -9.7\,\mathrm{mag}$, the resulting stellar mass is $M_* = 10^{5.88}\,M_\sun = 7.59 \times 10^5\,M_\sun$.

\subsection{Leo~T}
\label{ssec:obsredLeoT}
    The selection of member stars in Leo~T is described in more detail by Vaz et al.\ \citetext{in prep.}, but we summarize it here for completeness.
    With spexxy we were able to fit 130 out of the 252 extracted spectra with a line-of-sight velocity.
    For Leo~T a S/N criterion of ${>}3$ was used, which yielded 56~sources.
    Three emission-line stars were found, the spectra of which spexxy was not able to fit due to the absence of a suitable template in PHOENIX; for these three stars a velocity was determined using ULySS~\citep{Koleva-2009-A&A-501-1269} with the MIUSCAT library~\citep{Vazdekis-2012-MNRAS-424-157, Ricciardelli-2012-MNRAS-424-172}.
    Isochrones of various ages, reflecting the extended star-formation history of Leo~T~\citep{Weisz-2012-ApJ-748-88}, were fitted to the colour-magitude diagram of public\footnote{\emph{Hubble Space Telescope} Proposals 12914, pincipal investigator Tuan Do, and 14224, principal investigator Carme Gallart.} \emph{HST}/ACS data, adopting a distance $D = 417\,\mathrm{kpc}$ and a metallicity $[\element{M}/\element{H}] = -1.6$.
    One of the 59~sources was inconsistent with the isochrones and was removed.
    The 19 member stars of \citet{Simon-2007-ApJ-670-313} were added to the MUSE-Faint sample.
    Two stars were in common between \citet{Simon-2007-ApJ-670-313} and the MUSE-Faint observations.
    No kinematic outliers were found in this sample of 75~stars by Vaz et al.\ \citetext{in prep.}.

    For consistency with the analysis used on other sources in this paper, we additionally queried \emph{Gaia} EDR3 and found 110~sources, of which three had a DR2 velocity measurement, two of which with a good RUWE.
    These, however, were foreground sources according to their parallax measurements.
    Our other selection criteria did not reveal additional interlopers.
    We therefore arrived at the same sample of 75~stars as Vaz et al.\ \citetext{in prep.}.
    Of these stars, 55 have a velocity measurement from MUSE-Faint.
    We measured an intrinsic mean line-of-sight velocity of $39.5 \pm 2.1\,\mathrm{km}\,\mathrm{s}^{-1}$ and an intrinsic line-of-sight velocity dispersion of $7.6^{+2.3}_{-1.7}\,\mathrm{km}\,\mathrm{s}^{-1}$.
    The inferred kinematics in the final sample did not significantly differ when we tested with a S/N criterion of ${>}5$, but this would have resulted in 14~fewer stars; therefore we found the lower S/N cut to be justified.

    Considering half of the stellar mass of Leo~T was assembled $7.6\,\mathrm{Gyr}$ ago~\citep{Weisz-2012-ApJ-748-88}, we use this age and $[\element{M}/\element{H}] = -1.7$ to derive a stellar mass-to-light ratio of $1.57$.
    This resulted in a stellar mass of $M_* = 10^{5.16}\,M_\sun = 1.44 \times 10^5\,M_\sun$, somewhat larger than the $1.05^{+0.27}_{-0.23} \times 10^5\,M_\sun$ found by \citet{Weisz-2012-ApJ-748-88}.

\subsection{Hydra~II}
\label{ssec:obsredHyaII}
    We used publicly available \emph{HST}/ACS photometry\footnote{\emph{Hubble Space Telescope} Proposal~14734, principal investigator Nitya Kallivayalil.} to construct a colour--magnitude diagram for Hya~II.
    We compared the \emph{HST} photometry to PARSEC isochrones.
    We assumed an age of $t = 13.2\,\mathrm{Gyr}$ based on the star-formation history.
    We tried metallicities of $[\element{M}/\element{H}] = -2.2$ and $-2.0$ at both distance measurements, and found the combination of metallicity $[\element{M}/\element{H}] = -2.2$ and distance $D = 151\,\mathrm{kpc}$ fits the horizontal branch the best.
    From the fitted isochrone, we obtained a stellar mass-to-light ratio of $2.70$.
    Combined with the luminosity, we calculated a stellar mass of $M_* = 10^{4.30}\,M_\sun = 2.00 \times 10^4\,M_\sun$.

    Out of 132 extracted spectra, spexxy was able to converge to a spectral fit for 80 sources.
    Of the extracted spectra, 79 had a S/N of at least $3$, and 48 of these also had a converged spexxy fit.

    Through the Keck/DEIMOS spectroscopy of \citet{Kirby-2015-ApJ-810-56}, we have access to an additional 31~measurements of line-of-sight velocities in the direction of Hya~II.
    Of these sources, 13 are considered members by \citet{Kirby-2015-ApJ-810-56}.
    We included all 31~measurements in our preliminary sample to have a unified selection procedure.

    We queried the astrometry and kinematics of sources in \emph{Gaia} EDR3 within $5\,\mathrm{arcmin}$ from the centre of Hya~II.
    Six of the 1076~sources in the queried catalogue had DR2 line-of-sight velocities, of which five had a RUWE of less than $1.4$.
    We added these five measurements to our preliminary selection.

    Out of 84 velocity measurements, we built a catalogue of 81 unique sources with a (combined) velocity measurement.
    Nine sources were not sufficiently star-like according to SExtractor, the remaining 72~sources we considered stars.
    We removed foreground stars using our parallax criterion and found 59 possibly distant stars.
    Of these stars, 46 were consistent with the isochrone criterion.

    After removing kinematic outliers, we obtained a final selection of 28 stars with a mean line-of-sight velocity of $299.5^{+4.5}_{-4.7}\,\mathrm{km}\,\mathrm{s}^{-1}$ and an intrinsic dispersion of $12.0^{+5.0}_{-3.5}\,\mathrm{km}\,\mathrm{s}^{-1}$.
    Out of this sample, 15~stars have a velocity measurement from MUSE-Faint.
    We confirmed that a S/N criterion of ${>}5$ yielded consistent kinematics.
    The dispersion is over two standard deviations higher than the \citet{Kirby-2015-ApJ-810-56} 95\% upper limit of $4.5\,\mathrm{km}\,\mathrm{s}^{-1}$, but we find no further kinematic outliers, and the posterior distribution is well-shaped.
    There is no obvious cause of this discrepancy, though \citet{Kirby-2015-ApJ-810-56} note that their upper limit is sensitive to the in- or exclusion of two less likely member stars.

\subsection{Grus~1}
\label{ssec:obsredGru1}
    Using publicly available \emph{HST}/ACS photometry\footnote{\emph{Hubble Space Telescope} Proposal 14734, principal investigator Nitya Kallivayalil.} of Gru~1, we constructed a colour--magnitude diagram.
    Guided by the literature reviewed in Sect.~\ref{ssec:sample}, we found a good match by eye using PARSEC isochrones with age $t = 13.5\,\mathrm{Gyr}$, metallicity $[\element{M}/\element{H}] = -1.9$, and distance $D = 120\,\mathrm{kpc}$.
    The adopted parameters correspond to a stellar mass-to-light ratio of $1.72$.
    This resulted in a stellar mass of $M_* = 10^{3.80}\,M_\sun = 6.31 \times 10^3\,M_\sun$.

    We extracted 309~spectra, of which we successfully fitted 194 with spexxy.
    Only 11 extracted sources had a S/N of at least $5$, and only seven of these had a velocity fit.
    When we relaxed the S/N criterion to ${>}3$, we had 28 extracted spectra, with 18 velocities.
    Given the low number of sources even before considering membership, we chose to use a S/N cut of ${>}3$.
    We repeated the selection with a cut of ${>}5$ as well and found no significant difference in the inferred bulk dynamics.

    We supplemented our sample with that of \citet{Walker-2016-ApJ-819-53}, which consists of 133~sources in the direction of Gru~1, though only seven are considered members.
    We limited the \citet{Walker-2016-ApJ-819-53} sample to 64~sources with a velocity error ${\lesssim}20\,\mathrm{km}\,\mathrm{s}^{-1}$ to avoid poor-quality velocity measurements.
    For the sake of a uniform selection procedure, we did not adopt the membership probabilities determined by \citet{Walker-2016-ApJ-819-53}, but performed our own membership analysis.
    We further added the two velocity measurements of \citet{Ji-2019-ApJ-870-83}, whose sources are also included in the \citet{Walker-2016-ApJ-819-53} sample.

    We queried \emph{Gaia} EDR3 in a radius of $20\,\mathrm{arcmin}$ around the position of Gru~1 and found 1733~sources.
    There were 15 sources with a DR2 velocity measurement, of which 13 had a good RUWE.
    These 13 sources were added to our sample, though we remind the reader that for none of our galaxies did a source with a \emph{Gaia} velocity end up in our final selection of member stars.

    Our preliminary sample consisted of 97~velocity measurements of 95~sources.
    Consecutively applying membership criteria, we found 86~sources were stars, of which 58 were not nearby, of which 54 matched to the PARSEC isochrone.

    This sample was still heavily contaminated with kinematic outliers.
    We applied our iterative kinematic member selection and arrived at a final sample of 14~members.
    There are four stars in this sample with a velocity measurement from MUSE-Faint.
    This sample includes the seven members of \citet{Walker-2016-ApJ-819-53}, but also three stars that \citet{Walker-2016-ApJ-819-53} did not consider members.
    We find this sample has a mean line-of-sight velocity of $-139.2^{+6.1}_{-5.2}\,\mathrm{km}\,\mathrm{s}^{-1}$ and a velocity dispersion of $10.4^{+9.3}_{-5.1}\,\mathrm{km}\,\mathrm{s}^{-1}$.
    The velocity dispersion is marginally resolved and is about one standard deviation away from the median of  the unresolved posterior of \citet{Walker-2016-ApJ-819-53}.
    The dispersion measurement supports the classification of Gru~1 as a galaxy.

\section{Methods}
\label{sec:methods}
    In \citetalias{Zoutendijk-2021-A&A-651-A80} we used two dynamical modelling tools, CJAM~\citep{Watkins-2013-MNRAS-436-2598} and pyGravSphere~\citep{Read-2017-MNRAS-471-4541, Genina-2020-MNRAS-498-144} in combination with hkbin\footnote{\url{https://github.com/slzoutendijk/hkbin}}~\citepalias{Zoutendijk-2021-A&A-651-A80}, which is an alternative velocity binning algorithm based on that of pyGravSphere, in our analysis of the dark-matter profiles.
    We found consistent results between the two tools for our galaxy Eri~2.
    In this paper, we use CJAM together with the updated GravSphere\footnote{\url{https://github.com/justinread/gravsphere}}~\citep{Read-2017-MNRAS-471-4541, Read-2018-MNRAS-481-860, Genina-2020-MNRAS-498-144, 2021MNRAS.505.5686C}, which we describe in Sects.~\ref{ssec:cjam} and~\ref{ssec:gsmeth}, respectively.
    When the two tools are in agreement, we show the results from one of the tools in the main body of this paper and the results from the other in Appendix~\ref{app:gravsphere}, whereas in the case of disagreement, we present the results of both in the main body of the paper, without declaring a preference.
    We choose CJAM as our fiducial tool in case of agreement, because its simpler implementation of a fully cored profile allows us to address our hypothesis about the cuspiness of UFDs, while we have too few measurements to constrain the more complex implementation of cores in GravSphere, which allows partial core formation.

\subsection{CJAM}
\label{ssec:cjam}
    In our CJAM analysis we used three density profile models from \citetalias{Zoutendijk-2021-A&A-651-A80}.
    The diversity of profile shapes between the three models makes them well-suited to explore the properties of individual galaxies.
    This choice of models has the additional advantages of enabling the joint analysis of dark-matter properties using multiple galaxies in future work, and of already having been implemented.
    We will be brief in the description of these profiles and refer the reader to \citetalias{Zoutendijk-2021-A&A-651-A80} for details on the associated dark-matter physics.

    Firstly, we use the cuspy Navarro--Frenk--White (NFW; \citealp{Navarro-1996-ApJ-462-563}) profile, determined by a characteristic density~$\rho_0$ and a scale radius~$r_\mathrm{s}$:
    \begin{equation}
        \rho_\mathrm{cusp}(r; \rho_0, r_\mathrm{s}) = \frac{\rho_0}{(r/r_\mathrm{s}) (1+r/r_\mathrm{s})^2}
        \label{eq:cusp}
    \end{equation}
    This profile is divergent ($\rho_\mathrm{cusp} \propto r^{-1}$) at small radii~$r$ and transitions to a steeper slope ($\rho_\mathrm{cusp} \propto r^{-3}$) at radii larger than $r_\mathrm{s}$.
    The NFW profile is found in simulations of cold dark matter without baryonic interactions, but is also expected for UFDs due to their scarcity of baryons~\citep{Pennarrubia-2012-ApJ-759-L42, Onnorbe-2015-MNRAS-454-2092, Wheeler-2019-MNRAS-490-4447}.

    An alternative to cuspy profiles are cored profiles.
    Here we use the profile of \citet{Lin-2016-JCAP-03-009}:
    \begin{equation}
        \rho_\mathrm{core}(r; \rho_0, r_\mathrm{s}, r_\mathrm{c}) = \frac{\rho_0}{r_\mathrm{c}/r_\mathrm{s} + (r/r_\mathrm{s})(1+r/r_\mathrm{s})^2}.
        \label{eq:core}
    \end{equation}
    This model introduces a core radius~$r_\mathrm{c}$; at radii much smaller than this scale, the profile becomes flat ($\rho_\mathrm{core} \propto r^0$).
    At larger radii the profile behaves like the NFW profile, and if $r_\mathrm{c} = 0$ the two are identical at every radius.
    Though this profile is designed to describe self-interacting dark matter, it is flexible enough to fit to other core-producing mechanisms, such as baryonic feedback~\citep{Brooks-2014-ApJ-786-87, DiCintio-2014-MNRAS-437-415}.

    As our final model, we have a profile that is steeper than the NFW profile at intermediate radii.
    This profile, derived for fuzzy dark matter, is parametrized by \citet{Marsh-2015-MNRAS-451-2479} as
    \begin{equation}
        \rho_\mathrm{sol}(r; \rho_{\mathrm{sol},0}, r_\mathrm{sol}, \rho_{\mathrm{cusp},0}, r_\mathrm{s}) =
        \begin{cases}
            \frac{\rho_{\mathrm{sol},0}}{(1+(r/r_\mathrm{sol})^2)^8}, & (r < r_\mathrm{t}),\\
            \frac{\rho_{\mathrm{cusp},0}}{(r/r_\mathrm{s})(1+r/r_\mathrm{s})^2}, & (r \geq r_\mathrm{t}).\\
        \end{cases}
    \label{eq:sol}
    \end{equation}
    This is identical to the NFW profile at radii larger than the transition radius $r_\mathrm{t}$, but then steeply rises (up to $\rho_\mathrm{sol} \propto r^{-16}$) until becoming flat at the centre.
    The inner part deviating from the NFW profile is known as the soliton.

    Though a full explanation of our CJAM modelling and the implementation of our density profiles is given in \citetalias{Zoutendijk-2021-A&A-651-A80}, we recapitulate here.
    Given a density profile, characterized by parameters $\boldsymbol{\theta}$ and modelled with a multi-Gaussian expansion~(MGE; \citealp{Emsellem-1994-A&A-285-723}), CJAM calculates expected line-of-sight velocity dispersion~$\sigma_i(\boldsymbol{\theta})$ at the projected radius of each tracer~$i$ (i.e.\ each star).
    It is then up to the user to implement a strategy to find the best model and the uncertainties thereon.
    We make use of Equations~(7), (9), (11), and~(15) provided by \citet{Graham-2005-PASA-22-118} and the absolute solar magnitude $M_{V,\sun} = 4.81$ of \citet{Willmer-2018-ApJS-236-47} to transform the adopted magnitudes and central surface brightness parameters in Table~\ref{tab:adopted} to the surface brightness at the half-light radius that is required by CJAM.
    We construct an axisymmetric luminosity profile with the adopted position angles and ellipticities, and the assumption that the system is seen face-on.
    To avoid the model from becoming computationally too expensive, we assume an isotropic velocity distribution and a spherical dark-matter halo.
    We shall see later that GravSphere, which does not assume isotropy, indicates that anisotropy has no significant influence on our results.

    The likelihood that a model with profile parameters~$\boldsymbol{\theta}$ and systemic line-of-sight velocity~$v_0$ describes the observed velocities~$v_i$ and their measurement uncertainties~$\varepsilon_i$, is
    \begin{equation}
        \mathcal{L}(\boldsymbol{\theta}, v_0 | v_i, \varepsilon_i) = \prod_i \frac{1}{\sqrt{2\pi\bigl(\sigma_i^2(\boldsymbol{\theta})+\varepsilon_i^2\bigr)}} \exp\Biggl(-\frac{(v_i-v_0)^2}{2(\sigma_i^2(\boldsymbol{\theta})+\varepsilon_i^2)}\Biggr).
        \label{eq:like}
    \end{equation}
    The model is varied by sampling different model parameters $\boldsymbol{\theta}$ and $v_0$.
    Using MultiNest~\citep{Feroz-2008-MNRAS-384-449, Feroz-2009-MNRAS-398-1601, Feroz-2019-OJAp-2-10} through the pyMultiNest~\citep{Buchner-2014-A&A-564-A125} interface, the parameters are constrained.
    MultiNest calculates the Bayesian evidence for each model, with which the models can be compared.

    We found that a parametrization of the cored profile in the densities $\rho_1 = \rho(50\,\mathrm{pc})$, $\rho_2 = \rho(100\,\mathrm{pc})$, and $\rho_3 = \rho(150\,\mathrm{pc})$ gives the best constraints on the posterior distribution.
    These three radii were chosen because the greatest density of tracers is found within this radial range.
    As the cuspy profile is a special case of the cored profile with $r_\mathrm{c} = 0$, for this model we parametrize only with $\rho_2$ and $\rho_3$.
    The soliton profile we parametrize with the value $\rho_{\mathrm{cusp},100} = \rho_\mathrm{cusp}(100\,\mathrm{pc})$ of the outer (cuspy) density profile at $100\,\mathrm{pc}$, the logarithmic slope $\alpha_{\mathrm{cusp},100} \coloneqq (\mathrm{d}\ln\rho_\mathrm{cusp}/\mathrm{d}\ln r)(100\,\mathrm{pc})$ of the outer density profile at $100\,\mathrm{pc}$, the ratio $r_\mathrm{sol}/r_\mathrm{s}$ between soliton and scale radius, and the ratio $\varepsilon = \rho_\mathrm{sol}(r_\mathrm{t})/\rho_{\mathrm{sol},0}$ between the density at the transition radius and at the centre.

    We used the same priors as in \citetalias{Zoutendijk-2021-A&A-651-A80}, indicated in Table~\ref{tab:priors}.
    \begin{table}
        \caption{Priors on parameters explored with MultiNest.}
        \label{tab:priors}
        \centering
        \begin{tabular}{lc@{\hspace{3pt}}cc}
            \hline
            \hline
            Prior                                                            & Min.   & Max.            & Profiles         \\
            \hline
            $\log_{10}(\rho_1/M_\sun\,\mathrm{kpc}^{-3})$\tablefootmark{(a)} &    $6$ &            $12$ & core             \\
            $\log_{10}(\rho_2/M_\sun\,\mathrm{kpc}^{-3})$\tablefootmark{(a)} &    $6$ &            $12$ & cusp, core       \\
            $\log_{10}(\rho_3/M_\sun\,\mathrm{kpc}^{-3})$\tablefootmark{(a)} &    $6$ &            $12$ & cusp, core       \\
            $\log_{10}(\rho_{\mathrm{cusp},100}/M_\sun\,\mathrm{kpc}^{-3})$  &    $6$ &            $10$ & sol.             \\
            $\alpha_{\mathrm{cusp},100}$                                     &   $-3$ &            $-1$ & sol.             \\
            $\log_{10}(r_\mathrm{sol}/r_\mathrm{s})$                         &   $-3$ &             $0$ & sol.             \\
            $\log_{10} \varepsilon$                                          &   $-5$ & $\log_{10} 1/2$ & sol.             \\
            $v_0/\mathrm{km}\,\mathrm{s}^{-1}$ (Ant~B)                       &  $365$ &           $385$ & cusp, core, sol. \\
            $v_0/\mathrm{km}\,\mathrm{s}^{-1}$ (Leo~T)                       &   $25$ &            $45$ & cusp, core, sol. \\
            $v_0/\mathrm{km}\,\mathrm{s}^{-1}$ (Hya~II)                      &  $280$ &           $330$ & cusp, core, sol. \\
            $v_0/\mathrm{km}\,\mathrm{s}^{-1}$ (Gru~1)                       & $-165$ &          $-115$ & cusp, core, sol. \\
            \hline
        \end{tabular}
        \tablefoot{%
            All priors are uniform between the indicated minima and maxima and apply to the indicated profiles: core, cusp, or soliton.
            $\rho_1$, $\rho_2$, $\rho_3$: dark-matter density at $50$, $100$, and $150\,\mathrm{pc}$ from the centre, respectively;
            $\rho_{\mathrm{cusp},100}$: density of the large-scale, cusp-like component of the soliton profile, measured at $100\,\mathrm{pc}$;
            $\alpha_{\mathrm{cusp},100}$: power-law slope of the large-scale, cusp-like component of the soliton profile, measured at $100\,\mathrm{pc}$;
            $r_\mathrm{sol}/r_\mathrm{s}$: ratio of the soliton radius to the scale radius of the large-scale, cusp-like component of the soliton profile;
            $\varepsilon$: fraction of the dark-matter density at the soliton profile's transition radius versus the central density;
            $v_0$: systemic velocity.
            \tablefoottext{a}{Within the indicated priors, $\rho_i \geq \rho_{i+1}$.}%
        }
    \end{table}
    Additionally, as in \citetalias{Zoutendijk-2021-A&A-651-A80}\footnote{\citetalias{Zoutendijk-2021-A&A-651-A80} describes these conditions incorrectly. The conditions given in that paper should read identical to those given here.}, the following conditions were imposed to ensure physically plausible profiles with a well-fit MGE: $r_\mathrm{c} \leq r_\mathrm{s}$ and $1.5\,\mathrm{pc} \leq r_\mathrm{s} \leq 50\,\mathrm{kpc}$.

\subsection{GravSphere}
\label{ssec:gsmeth}
    In our GravSphere~\citep{Read-2017-MNRAS-471-4541, Read-2018-MNRAS-481-860, Genina-2020-MNRAS-498-144, 2021MNRAS.505.5686C} analysis we use two profile models.
    The first is a cuspy model using the same NFW profile as CJAM.
    However, GravSphere parametrizes this profile with the virial mass $M_{200}$ and concentration $c_{200}$.
    These are related to the conventional~$\rho_0$ and~$r_\mathrm{s}$ through
    \begin{align}
        \rho_0 &= \frac{200\rho_\mathrm{crit}c_{200}^3}{3\bigl(\ln(1+c_{200}) - c_{200}/(1+c_{200})\bigr)},
        \label{eq:gsrho0}\\
        r_\mathrm{s} &= r_{200}/c_{200},
        \label{eq:gsrs}
    \end{align}
    where
    \begin{equation}
        r_{200} = \Biggl(\frac{3M_{200}}{800\pi\rho_\mathrm{crit}}\Biggr)^{1/3}
        \label{eq:gsr200}
    \end{equation}
    and $\rho_\mathrm{crit}$ is the critical density of the Universe.

    The second profile model we will refer to as core+tides, as it modifies the cusp model by allowing for a (partial) central core and a lowering of the density beyond a tidal radius.
    This is described by the coreNFWtides profile~\citep{Read-2016-MNRAS-459-2573, Read-2018-MNRAS-481-860}.
    Within the tidal radius~$r_\mathrm{t}$, this profile redistributes of mass from the central cusp to larger radii, as is expected to take place during core formation due to star-formation feedback:
    \begin{equation}
        M_\mathrm{cNFW}(<r) = M_\mathrm{NFW}(<r)f^n,
    \end{equation}
    with
    \begin{equation}
        f^n = \Bigl(\tanh(r/r_\mathrm{c})\Bigr)^n,
    \end{equation}
    where $r_\mathrm{c}$ is the core radius and $0 \leq n \leq 1$ the completeness of core formation.
    In terms of density, this is
    \begin{equation}
        \rho_\mathrm{cNFW}(r) = f^n\rho_\mathrm{NFW}(r) + \frac{nf^{n-1} \Bigl(1-f^2\Bigr)}{4\pi r^2r_\mathrm{c}} M_\mathrm{NFW}(<r),
    \end{equation}
    where
    \begin{equation}
        M_\mathrm{NFW}(<r) = \frac{M_{200} \Bigl(\ln(1+r/r_\mathrm{s}) - (r/r_\mathrm{s})(1+r/r_\mathrm{s})^{-1}\Bigr)}{\ln(1+c_{200}) - c_{200}/(1+c_{200})}
    \end{equation}
    is the NFW cumulative mass profile.
    Outside of $r_\mathrm{t}$, the density profile is changed into a power law with negative slope $\delta \geq 3$:
    \begin{equation}
        \rho_\mathrm{cNFWt}(r) = \rho_\mathrm{cNFW}(r_\mathrm{t})\,(r/r_\mathrm{t})^{-\delta}.
    \end{equation}
    The complete coreNFWtides mass profile is then
    \begin{equation}
        M_\mathrm{cNFWt}(<r) =
        \begin{cases}
            M_\mathrm{cNFW}(<r), & r < r_\mathrm{t}, \\
            M_\mathrm{cNFW}(<r_\mathrm{t}) + {}\\
                \quad 4\pi\rho_\mathrm{cNFW}(r_\mathrm{t}) \frac{r_\mathrm{r}^3}{3-\delta} \Bigl((r/r_\mathrm{t})^{3-\delta} - 1\Bigr), & r > r_\mathrm{t}.
        \end{cases}
    \end{equation}
    Because the coreNFWtides profile is parametrized with the virial parameters of its progenitor NFW profile, we will refer to these parameters as $M_{200}^\mathrm{prog}$ and $c_{200}^\mathrm{prog}$ when discussing the core+tides model, to avoid confusion with the actual $M_{200}$ and $c_{200}$ of the present-day profile.

    Unlike for our CJAM modelling, where we have assumed an isotropic velocity distribution, our GravSphere modelling uses a \citet{BAes-2007-A&A-471-419} velocity anisotropy profile,
    \begin{equation}
        \beta(r) = \beta_0 + (\beta_\infty - \beta_0) \frac{1}{1+(r_0/r)^\eta}.
    \end{equation}
    GravSphere uses a symmetrized anisotropy,
    \begin{equation}
        \tilde\beta = \frac{\beta}{2-\beta},
    \end{equation}
    and equivalently $\beta_0 \to \tilde\beta_0$ and $\beta_\infty \to \tilde\beta_\infty$, to avoid infinities.
    The velocity anisotropy is fully tangential for $\tilde\beta = -1$ and fully radial for $\tilde\beta = +1$, while $\tilde\beta = 0$ means isotropy.

    GravSphere uses binning to compare the velocity moments of the profiles with the measurements.
    This binning is done through an algorithm called the binulator.
    The intrinsic velocity distribution function is modelled with a generalized Gaussian.
    The mean line-of-sight velocity~$\mu_v$, intrinsic line-of-sight dispersion $\sigma_\mathrm{los}$, and kurtosis~$\kappa$ are optimized to fit the generalized Gaussian probability density function to the velocity measurements in each bin, taking into account the observational broadening by the measurement uncertainties.
    To break the degeneracy between density and velocity anisotropy, two virial shape parameters~\citep{Merrifield-1990-AJ-99-1548} are then calculated from the fourth velocity moments $\langle v_\mathrm{los}^3 \rangle = \kappa\sigma_\mathrm{los}^4$ by integration over the projected radius~$R$:
    \begin{align}
        v_\mathrm{s1} &= \int_0^\infty \Sigma\langle v_\mathrm{los}^4 \rangle R\,\mathrm{d}R,\\
        v_\mathrm{s2} &= \int_0^\infty \Sigma\langle v_\mathrm{los}^4 \rangle R^3\,\mathrm{d}R,
    \end{align}
    where $\Sigma$ is the surface brightness profile of the galaxy.
    The fourth velocity moment is assumed to be constant beyond the outermost bin.
    The uncertainty of the virial shape parameters is determined by random sampling of $\langle v_\mathrm{los}^4 \rangle$ from the posterior of the generalized Gaussian fits, generating a probability density function.
    GravSphere will then compare the models to the data using the velocity dispersion of each bin, including the Gaussian uncertainties, and the two global virial shape parameters with their probability density functions.

    Due to the small number of stars in our samples and the large measurement uncertainties on the stellar velocities, the fourth velocity moments are unconstrained.
    The result is that the virial shape parameters have little constraining power.
    We have tested GravSphere on Eri 2 both with and without using the virial shape parameters, and found no significant difference in the results.
    We have chosen to use the virial shape parameters for all our galaxies, per the default setting of GravSphere.

    We fit the exponential surface-brightness profiles assumed for CJAM with a superposition of three \citet{Plummer-1911-MNRAS-71-460} spheres and adopt the same stellar mass as for CJAM.
    GravSphere uses emcee~\citep{ForemanMackey-2013-PASP-125-306} to explore the parameter space made up of the dark-matter profile parameters, anisotropy parameters, and small variations around the adopted stellar mass and surface-brightness profile parameters.
    The piors are uniform; we show the ranges of the priors in Table~\ref{tab:gspriors}.
    \begin{table}
        \caption{Priors on parameters explored by GravSphere with emcee.}
        \label{tab:gspriors}
        \centering
        \begin{tabular}{lc@{\hspace{3pt}}cc}
            \hline
            \hline
            Prior                                  & Min.   & Max.              & Profiles         \\
            \hline
            $\log_{10}(M_{200}/M_\sun)$            &  $7.5$  &          $11.5$  & cusp             \\
            $\log_{10}(M_{200}^\mathrm{prog}/M_\sun)$          &  $7.5$  &          $11.5$  & core+tides       \\
            $c_{200}$                              &  $1.0$  &          $50.0$  & cusp             \\
            $c_{200}^\mathrm{prog}$                            &  $1.0$  &          $50.0$  & core+tides       \\
            $\log_{10}(r_\mathrm{c}/\mathrm{kpc})$ & $-3$    &           $1$    & core+tides       \\
            $n$                                    &  $0$    &           $1$    & core+tides       \\
            $\log_{10}(r_\mathrm{t}/\mathrm{kpc})$ &  $0$    & $\log_{10}(20)$  & core+tides       \\
            $\delta$                               &  $3.01$ &           $5.0$  & core+tides       \\
            $\tilde\beta_0$                        & $-1$    &           $1$    & cusp, core+tides \\
            $\tilde\beta_\infty$                   & $-1$    &           $1$    & cusp, core+tides \\
            $\log_{10}(r_0/\mathrm{kpc})$          & $-0.9$  &           $0$    & cusp, core+tides \\
            $\eta$                                 &  $1.0$  &           $3.0$  & cusp, core+tides \\
            $M_*/M_{*,\mathrm{nom}}$               &  $0.75$ &           $1.25$ & cusp, core+tides \\
            $M_i/M_{i,\mathrm{best}}$              &  $0.9$  &           $1.1$  & cusp, core+tides \\
            $a_i/a_{i,\mathrm{best}}$              &  $0.9$  &           $1.1$  & cusp, core+tides \\
            \hline
        \end{tabular}
        \tablefoot{%
            All priors are uniform between the indicated minima and maxima and apply to the indicated profiles.
            $M_{200}$: virial mass of the cusp model;
            $M_{200}^\mathrm{prog}$: virial mass of the cuspy progenitor to the core+tides model;
            $c_{200}$: concentration of the cusp model;
            $c_{200}^\mathrm{prog}$: concentration of the cuspy progentor to the core+tides model;
            $r_\mathrm{c}$: core radius;
            $n$: coredness parameter;
            $r_\mathrm{t}$: tidal radius;
            $\delta$: negative logarithmic density slope beyond $r_\mathrm{t}$;
            $\tilde\beta_0$: inner symmetrized ansisotropy;
            $\tilde\beta_\infty$: outer symmetrized anisotropy;
            $r_0$: anisotropy transition radius;
            $\eta$: anistotropy transition rapidity;
            $M_*/M_{*,\mathrm{nom}}$: stellar mass, relative to the nominal stellar mass from Table~\ref{tab:derived};
            $M_i/M_{i,\mathrm{best}}$: surface-brightness profile multi-Plummer amplitude parameters, relative to the best-fit values;
            $a_i/a_{i,\mathrm{best}}$: surface brightness profile multi-Plummer scale parameters, relative to the best-fit values.%
        }
    \end{table}
    We use 250~walkers with 25000~steps and discard the first 75\% of the steps as burn-in, which are the default settings of GravSphere.

\section{Results}
\label{sec:results}
    We present our results in the following sections, broken down into the estimation of model parameters~(Sect.~\ref{ssec:param}), the determination of the constraints on the density profile and derived quantities~(Sect.~\ref{ssec:recovery}), the comparison of the different models~(Sect.~\ref{ssec:modcomp}), a test for the effects of tidal stripping~(Sect.~\ref{ssec:stripping}), and the comparison of our results to the expected scaling relations~(Sect.~\ref{ssec:scaling}).
    Most of these results are determined with the fiducial CJAM analysis.
    To evaluate the robustness of the CJAM results, we have also analysed our galaxies with GravSphere.
    Where CJAM and GravSphere agree, we present only CJAM here and refer to the GravSphere results in Appendix~\ref{app:gravsphere}.
    If the two tools are not in agreement, the results of both are shown here.

\subsection{Parameter estimation}
\label{ssec:param}
    We sampled the parameter spaces of our models with MultiNest, using the parametrizations described above.
    We then converted the posterior distributions to a different set of parameters, those that occur in Equations \eqref{eq:cusp}--\eqref{eq:sol}: $\rho_0$ and $r_\mathrm{s}$ for the cusp model, additionally $r_\mathrm{c}$ for the core model, and $\rho_{\mathrm{sol},0}$, $r_\mathrm{sol}$, $\rho_{\mathrm{cusp},0}$, and $r_\mathrm{s}$ for the soliton model.
    These converted posteriors are shown in Figs.~\ref{fig:cdmaltcorner}--\ref{fig:fdm4altcorner} in Appendix~\ref{app:corner}, including the confidence intervals or upper and lower limits.
    Here we describe and compare the constraints on the parameters.

    Comparing the galaxies among themselves (Figs.~\ref{fig:cdmaltcorner}--\ref{fig:fdm4altcorner}), we see a general trend with Ant~B and Hya~II usually having a larger $r_\mathrm{s}$ than Leo~T and Gru~1, while Leo~T and Hya~II usually have a larger $\rho_0$ or $\rho_{\mathrm{cusp},0}$ than Ant~B and Gru~1.
    These differences are, however, not significant due to the large uncertainties on these parameters.

    When we look at the posteriors of $r_\mathrm{c}$ in Fig.~\ref{fig:sidmaltcorner}, we see a clear peak.
    However, these distributions have tails extending to values smaller than the projected radii of the innermost tracers.
    As was the case with Eri~2 in \citetalias{Zoutendijk-2021-A&A-651-A80}, the posterior distributions on $r_\mathrm{c}$ therefore do not constitute a detection of a core, but should instead be regarded as upper limits.
    All four galaxies have a core radius $r_\mathrm{c} < 66$--$95\,\mathrm{pc}$ at the 68\% confidence level.
    For Leo~T, the 95\% upper limit is $r_\mathrm{c} < 200\,\mathrm{pc}$.
    The other three galaxies have $r_\mathrm{c} < 295$--$309\,\mathrm{pc}$ at the 95\% confidence level.

    \begin{figure*}
        \includegraphics[width=\linewidth]{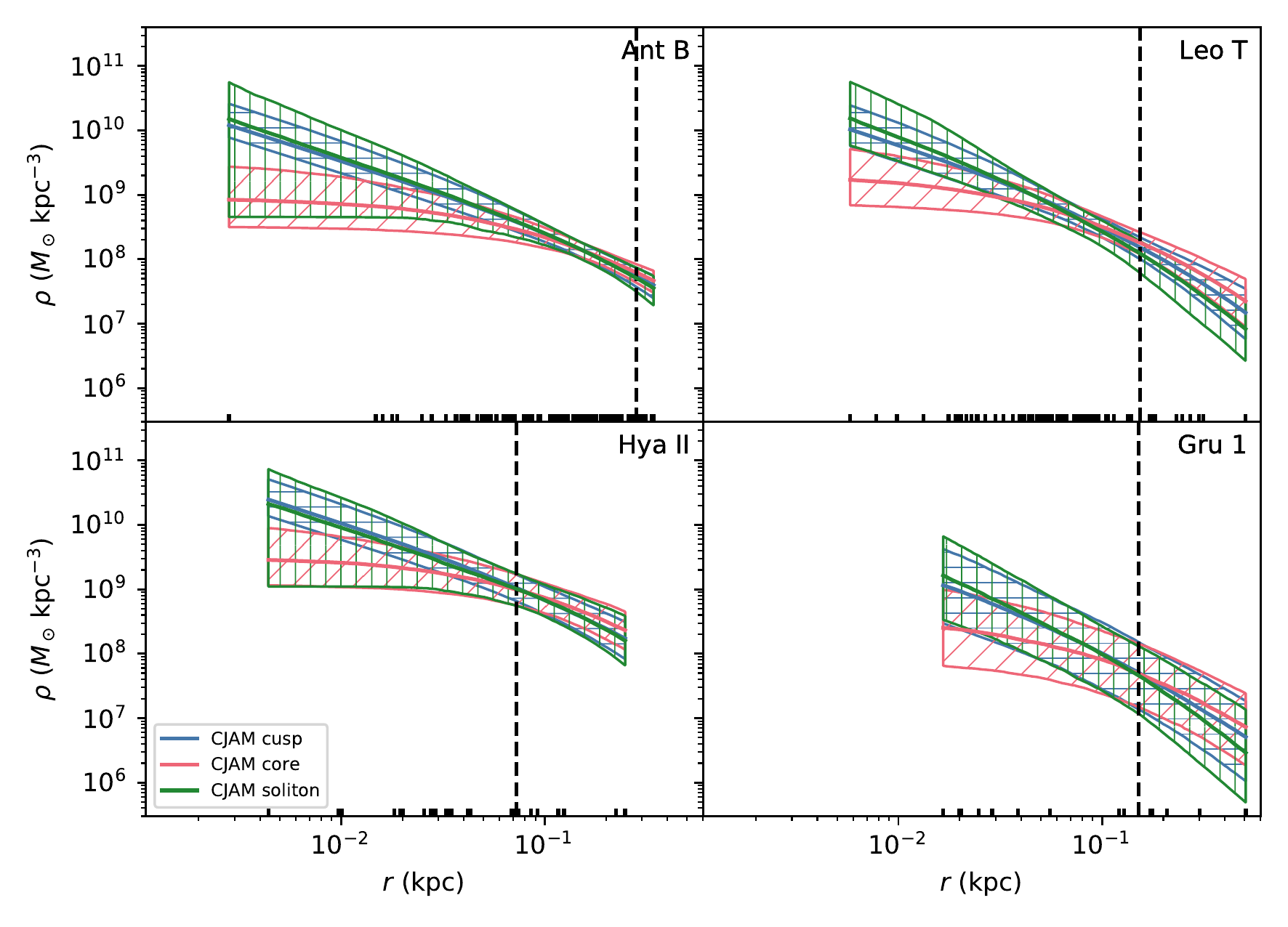}
        \caption{%
            Recovered dark-matter density profiles of the four (ultra-)faint dwarf galaxies, for cusp, core, and soliton profiles, modelled with CJAM.
            The hatched bands indicate the 68\% confidence interval on the density at each radius.
            The central, thicker line is the median density at each radius.
            The galaxy's projected half-light radii are indicated with the vertical dashed lines.
            The markers along the bottom of each panel indicate the projected radii of the kinematic tracers.%
        }
        \label{fig:recovery}
    \end{figure*}

    Interpreting the posterior distributions of the soliton parameters $\rho_{\mathrm{sol},0}$ and $r_\mathrm{sol}$, shown in Fig.~\ref{fig:fdm4altcorner}, is more complex.
    Ant~B and Hya~II show multi-modal distributions and there may be a hint of multi-modality in the other galaxies.
    Upon further inspection, we see that the mode with large $r_\mathrm{sol}$ and small $\rho_{\mathrm{sol},0}$ lies along $r_\mathrm{sol} = r_\mathrm{s}$, which is the edge of the prior, and is accompanied with a sudden drop in $\rho_{\mathrm{cusp},0}$.
    It therefore seems that this mode corresponds to a scenario where the soliton part of the profile is used to fit the change in slope at $r_\mathrm{s}$, a feature of all three of our models, instead of an actual soliton.
    If we consider the modes with large $r_\mathrm{sol}$ spurious and concentrate on the other modes and their tails, we see again that the posterior distributions are consistent with radii smaller than the projected radii of the innermost tracers.
    We therefore interpret the posterior distributions of $r_\mathrm{sol}$ as upper limits.
    Due to the degeneracy between $r_\mathrm{sol}$ and $\rho_{\mathrm{sol},0}$, we consider the posteriors of the latter as lower limits.
    The strength of the limit correlates with the strength of the multi-modality; for Leo~T and Gru~1 we can constrain $r_\mathrm{sol} < 13$–$14\,\mathrm{pc}$ at the 68\% confidence level, whereas the opposite extreme is Hya~II with $r_\mathrm{sol} < 112\,\mathrm{pc}$.
    Ant~B is in the middle with $r_\mathrm{sol} < 43\,\mathrm{pc}$ at the 68\% confidence level.
    The 95\% confidence levels are more divergent: $r_\mathrm{sol} < 251\,\mathrm{pc}$ for Leo~T, ${<}724\,\mathrm{pc}$ for Gru~1, ${<}1.20\,\mathrm{kpc}$ for Ant~B, and ${<}1.58\,\mathrm{kpc}$ for Hya~II.
    The large difference between the 68\% and 95\% confidence limits is a consequence of the spurious large-$r_\mathrm{sol}$ modes, which pull the 95\% confidence limits to larger values of $r_\mathrm{sol}$.

    Due to the differences between the implemented models, the parameter constrains from CJAM and GravSphere are not directly comparable.
    However, we can point out that the core radius~$r_\mathrm{c}$ of the GravSphere core+tides model, as discussed in Appendix~\ref{app:gravsphere} and shown in Figs.~\ref{fig:gscorenfwcornerAntB}--\ref{fig:gscorenfwcornerGru1}, is not constrained and includes the possibility of cores larger than $1\,\mathrm{kpc}$.
    This lack of constraints is due to a degeneracy between the core radius~$r_\mathrm{c}$ and the completeness of core formation, $n$: for partially formed cores, larger core radii are allowed than for the fully formed cores modelled with CJAM.

\subsection{Profile recovery}
\label{ssec:recovery}
    We take the posterior distributions of model parameters, evaluate the density profile, and marginalize over the model parameters to calculate constraints on the density profiles of our dwarf galaxies.
    These profiles are displayed in Fig.~\ref{fig:recovery}.
    For each galaxy, we see that the different models agree within the uncertainties at the larger radii.
    The constraints are the tightest and best-agreeing at the radii with the largest concentration of tracers.
    The profiles start to diverge more as one goes to smaller radii, particularly the core model.
    The agreement between the cusp and soliton models is very good at all radii.

    We further note that in the centres the soliton model prefers the highest density for three out of four galaxies and the core model the lowest for all galaxies.
    This can be understood as a consequence of the formulations of these models.
    At intermediate radii the soliton profile rises steeply, only to flatten out at scales around the soliton radius.
    For equal $\rho_0$ ($\rho_{\mathrm{cusp},0}$) and $r_\mathrm{s}$, this forces the density of the soliton model to be higher than that of the cusp model for radii inside the transition radius, which is reflected in the marginalized density profiles.
    Similarly, the core model always has a lower central density than the cusp model due to the presence of a core, or at most an equal density when $r_\mathrm{c} = 0$, for which the two models produce identical profiles.
    This has as a side effect that the median density profile will be skewed away from the mode.
    For example, if we were to fit a core model to a finite number of measurements with non-zero uncertainties, drawn from a cuspy profile, the mode of the resulting density profile should be at $r_\mathrm{c} = 0$, the truth, but because of the finite information there will be an asymmetric tail towards $r_\mathrm{c} > 0$.
    Therefore the median recovered $r_\mathrm{c}$ will always be larger than the mode or the truth, which translates into a lower central density.

    By comparing the CJAM profiles with those derived using GravSphere in Appendix~\ref{app:gravsphere} and shown in Fig.~\ref{fig:gsrecovery}, we see that they are generally in good agreement, especially near the half-light radius due to the high concentration of tracers.
    We do see that the CJAM profiles are systematically steeper.
    This could mean the profile is sensitive to bias when extrapolated beyond the range containing tracers.
    The GravSphere results also indicate that the profile parameters are not correlated with the velocity anisotropy, which means that the assumed isotropy of our CJAM model should not have influenced its results.

    From the density profiles we can calculate constraints on additional quantities, such as the virial radius~$r_{200}$, inside which the average density is $200$ times the critical density of the Universe, and the associated concentration $c_{200} \coloneqq r_{200}/r_\mathrm{s}$ and virial mass $M_{200}$.
    As the virial mass of satellite galaxies is hard to determine in simulations, we also calculate $V_\mathrm{max}$, the maximum of the circular velocity profile.
    We further provide circular velocities and masses for the projected half-light radius $R_{1/2}$ and the three-dimensional half-light radius $r_{1/2} \simeq (4/3)R_{1/2}$~\citep{Wolf-2010-MNRAS-406-1220}.
    The mass-to-light ratios $\Upsilon(r_{200}) \coloneqq M_{200}/L_V$ and $\Upsilon(r_{1/2}) \coloneqq M(r_{1/2})/(L_V/2)$ are an indicator of how dark matter--dominated the galaxies are.
    Finally, the astrophysical $J$ and $D$ factors are measures of the potential to detect signals from dark-matter annihilation and decay, respectively, from a galaxy.
    These factors are integrals of the density profile (for decay) or the square of the density profile (for annihilation, a two-particle process) over the line of sight and a solid angle on the sky, and are therefore proportional to the reaction rate in a field of view.
    To calculate the expected signal flux from a certain source for a certain proposed kind of dark-matter particle, the source-specific astrophysical factors need to be combined with a particle-specific factor.
    We have determined the astrophysical factors inside the critical integration angles, which maximize the signal.
    The critical integration angle for decay, $\alpha_\mathrm{c}^D$, is the angle subtended by the projected half-light radius, while the critical integration angle for annihilation, $\alpha_\mathrm{c}^J$, corresponds to twice the projected half-light radius~\citep{Walker-2011-ApJ-742-20, Bonnivard-2015-MNRAS-446-3002}.
    Constraints on all these quantities are given per galaxy and model in Table~\ref{tab:m200}.
    \begin{sidewaystable}
        \caption{Parameters derived from the recovered dark-matter density profiles from CJAM.}
        \label{tab:m200}
        \centering
        \begin{tabular}{lc@{\hspace{3pt}}c@{\hspace{3pt}}cc@{\hspace{3pt}}c@{\hspace{3pt}}cc@{\hspace{3pt}}c@{\hspace{3pt}}cc@{\hspace{3pt}}c@{\hspace{3pt}}c}
            \hline
            \hline
                                                                             & \multicolumn{3}{c}{Ant~B}                                                & \multicolumn{3}{c}{Leo~T}                                                   & \multicolumn{3}{c}{Hya~II}                                                  & \multicolumn{3}{c}{Gru~1}                                                   \\
                                                                             & cusp                   & core                   & soliton                & cusp                    & core                    & soliton                 & cusp                    & core                    & soliton                 & cusp                    & core                    & soliton                 \\
            \hline
            $\log_{10}(V_\mathrm{max}/\mathrm{km}\,\mathrm{s}^{-1})$         & $1.20^{+0.21}_{-0.10}$ & $1.24^{+0.21}_{-0.13}$ & $1.17^{+0.15}_{-0.08}$ &  $1.17^{+0.18}_{-0.11}$ &  $1.23^{+0.23}_{-0.14}$ &  $1.14^{+0.13}_{-0.10}$ &  $1.41^{+0.25}_{-0.18}$ &  $1.49^{+0.29}_{-0.21}$ &  $1.38^{+0.24}_{-0.16}$ &  $0.98^{+0.29}_{-0.29}$ &  $1.02^{+0.30}_{-0.30}$ &  $0.99^{+0.24}_{-0.26}$ \\
            $\log_{10}(r_{200}/\mathrm{kpc})$                                & $1.18^{+0.30}_{-0.19}$ & $1.22^{+0.29}_{-0.18}$ & $1.09^{+0.23}_{-0.17}$ &  $1.07^{+0.29}_{-0.19}$ &  $1.15^{+0.32}_{-0.20}$ &  $0.95^{+0.25}_{-0.18}$ &  $1.31^{+0.33}_{-0.24}$ &  $1.39^{+0.36}_{-0.25}$ &  $1.20^{+0.32}_{-0.22}$ &  $0.92^{+0.37}_{-0.30}$ &  $1.01^{+0.34}_{-0.29}$ &  $0.80^{+0.32}_{-0.27}$ \\
            $\log_{10}(c_{200})$                                             & $1.39^{+0.32}_{-0.32}$ & $1.46^{+0.25}_{-0.30}$ & ---                    &  $1.61^{+0.35}_{-0.32}$ &  $1.63^{+0.24}_{-0.32}$ & ---                     &  $1.61^{+0.31}_{-0.29}$ &  $1.70^{+0.25}_{-0.29}$ & ---                     &  $1.41^{+0.49}_{-0.45}$ &  $1.32^{+0.32}_{-0.36}$ & ---                     \\
            $\log_{10}(M_{200}/M_\sun)$                                     & $8.57^{+0.89}_{-0.58}$ & $8.67^{+0.87}_{-0.55}$ & $8.29^{+0.71}_{-0.53}$ &  $8.22^{+0.87}_{-0.56}$ &  $8.46^{+0.95}_{-0.59}$ &  $7.88^{+0.73}_{-0.54}$ &  $8.94^{+0.99}_{-0.73}$ &  $9.19^{+1.08}_{-0.75}$ &  $8.62^{+0.96}_{-0.68}$ &  $7.79^{+1.11}_{-0.90}$ &  $8.06^{+1.03}_{-0.87}$ &  $7.43^{+0.96}_{-0.81}$ \\
            $\log_{10}(\Upsilon(r_{200})/M_\sun\,L_\sun^{-1})$             & $2.76^{+0.89}_{-0.58}$ & $2.87^{+0.87}_{-0.55}$ & $2.49^{+0.71}_{-0.53}$ &  $3.26^{+0.87}_{-0.56}$ &  $3.50^{+0.95}_{-0.59}$ &  $2.92^{+0.73}_{-0.54}$ &  $5.07^{+0.99}_{-0.73}$ &  $5.32^{+1.08}_{-0.75}$ &  $4.76^{+0.96}_{-0.68}$ &  $4.22^{+1.11}_{-0.90}$ &  $4.50^{+1.03}_{-0.87}$ &  $3.87^{+0.96}_{-0.81}$ \\
            $\log_{10}(V(R_{1/2})/\mathrm{km}\,\mathrm{s}^{-1})$             & $1.10^{+0.05}_{-0.06}$ & $1.09^{+0.05}_{-0.05}$ & $1.10^{+0.05}_{-0.06}$ &  $1.07^{+0.07}_{-0.07}$ &  $1.06^{+0.08}_{-0.08}$ &  $1.05^{+0.08}_{-0.09}$ &  $1.12^{+0.12}_{-0.11}$ &  $1.05^{+0.13}_{-0.13}$ &  $1.11^{+0.13}_{-0.16}$ &  $0.83^{+0.24}_{-0.28}$ &  $0.75^{+0.23}_{-0.26}$ &  $0.85^{+0.22}_{-0.25}$ \\
            $\log_{10}(M(R_{1/2})/M_\sun)$                                  & $7.01^{+0.10}_{-0.12}$ & $7.00^{+0.10}_{-0.11}$ & $7.01^{+0.11}_{-0.13}$ &  $6.68^{+0.15}_{-0.14}$ &  $6.67^{+0.16}_{-0.17}$ &  $6.65^{+0.16}_{-0.17}$ &  $6.47^{+0.23}_{-0.21}$ &  $6.33^{+0.25}_{-0.26}$ &  $6.44^{+0.26}_{-0.32}$ &  $6.20^{+0.48}_{-0.55}$ &  $6.04^{+0.46}_{-0.52}$ &  $6.24^{+0.43}_{-0.50}$ \\
            $\log_{10}(\Upsilon(R_{1/2})/M_\sun\,L_\sun^{-1})$             & $1.51^{+0.10}_{-0.12}$ & $1.50^{+0.10}_{-0.11}$ & $1.51^{+0.11}_{-0.13}$ &  $2.02^{+0.15}_{-0.14}$ &  $2.01^{+0.16}_{-0.17}$ &  $1.99^{+0.16}_{-0.17}$ &  $2.90^{+0.23}_{-0.21}$ &  $2.76^{+0.25}_{-0.26}$ &  $2.87^{+0.26}_{-0.32}$ &  $2.93^{+0.48}_{-0.55}$ &  $2.78^{+0.46}_{-0.52}$ &  $2.97^{+0.43}_{-0.50}$ \\
            $\log_{10}(V(r_{1/2})/\mathrm{km}\,\mathrm{s}^{-1})$             & $1.12^{+0.06}_{-0.06}$ & $1.13^{+0.06}_{-0.06}$ & $1.12^{+0.06}_{-0.07}$ &  $1.09^{+0.07}_{-0.07}$ &  $1.11^{+0.08}_{-0.08}$ &  $1.06^{+0.08}_{-0.09}$ &  $1.17^{+0.10}_{-0.10}$ &  $1.13^{+0.12}_{-0.13}$ &  $1.15^{+0.12}_{-0.13}$ &  $0.85^{+0.24}_{-0.27}$ &  $0.81^{+0.23}_{-0.26}$ &  $0.86^{+0.21}_{-0.25}$ \\
            $\log_{10}(M(r_{1/2})/M_\sun)$                                  & $7.19^{+0.11}_{-0.12}$ & $7.21^{+0.11}_{-0.12}$ & $7.18^{+0.12}_{-0.13}$ &  $6.87^{+0.14}_{-0.15}$ &  $6.89^{+0.16}_{-0.16}$ &  $6.81^{+0.16}_{-0.19}$ &  $6.68^{+0.21}_{-0.21}$ &  $6.60^{+0.24}_{-0.25}$ &  $6.66^{+0.24}_{-0.26}$ &  $6.38^{+0.48}_{-0.55}$ &  $6.04^{+0.46}_{-0.52}$ &  $6.40^{+0.41}_{-0.51}$ \\
            $\log_{10}(J(\alpha_\mathrm{c}^J)/M_\sun^2\,\mathrm{kpc}^{-5})$ & $9.29^{+0.26}_{-0.24}$ & $9.23^{+0.29}_{-0.26}$ & $9.31^{+0.36}_{-0.26}$ & $10.44^{+0.34}_{-0.32}$ & $10.40^{+0.36}_{-0.34}$ & $10.48^{+0.49}_{-0.36}$ & $11.91^{+0.44}_{-0.42}$ & $11.95^{+0.51}_{-0.48}$ & $12.02^{+0.50}_{-0.49}$ & $10.47^{+0.93}_{-1.13}$ & $10.32^{+0.90}_{-1.08}$ & $10.55^{+0.88}_{-1.04}$ \\
            $\log_{10}(D(\alpha_\mathrm{c}^D)/M_\sun\,\mathrm{kpc}^{-2})$   & $1.52^{+0.49}_{-0.32}$ & $1.60^{+0.49}_{-0.32}$ & $1.39^{+0.39}_{-0.30}$ &  $2.19^{+0.46}_{-0.31}$ &  $2.34^{+0.52}_{-0.34}$ &  $1.99^{+0.38}_{-0.33}$ &  $3.31^{+0.53}_{-0.42}$ &  $3.45^{+0.60}_{-0.44}$ &  $3.17^{+0.51}_{-0.37}$ &  $2.82^{+0.65}_{-0.66}$ &  $2.97^{+0.65}_{-0.65}$ &  $2.65^{+0.60}_{-0.65}$ \\
            \hline
        \end{tabular}
        \tablefoot{%
            $V_\mathrm{max}$: maximum circular velocity;
            $r_{200}$: virial radius;
            $c_{200}$: concentration parameter;
            $M_{200}$: virial mass;
            $\Upsilon(r)$: mass-to-light ratio integrated within radius~$r$;
            $V(r)$: circular velocity at radius~$r$;
            $R_{1/2}$: projected half-light radius;
            $M(r)$: mass within radius~$r$;
            $r_{1/2}$: three-dimensional half-light radius;
            $J(\alpha_\mathrm{c}^J)$: astrophysical $J$ factor within its critical angle;
            $D(\alpha_\mathrm{c}^D)$: astrophysical $D$ factor within its critical angle.%
        }
    \end{sidewaystable}

    \begin{table*}
        \caption{Bayesian evidence and Bayes factors for CJAM.}
        \label{tab:modcomp}
        \centering
        \begin{tabular}{lc@{\hspace{3pt}}c@{\hspace{3pt}}cc@{\hspace{3pt}}c@{\hspace{3pt}}cc@{\hspace{3pt}}c@{\hspace{3pt}}cc@{\hspace{3pt}}c@{\hspace{3pt}}c}
            \hline
            \hline
                                 & \multicolumn{3}{c}{Ant~B}         & \multicolumn{3}{c}{Leo~T}         & \multicolumn{3}{c}{Hya~II}        & \multicolumn{3}{c}{Gru~1}      \\
                                 & cusp      & core      & soliton   & cusp      & core      & soliton   & cusp      & core      & soliton   & cusp     & core     & soliton  \\
            \hline
            $\ln(Z)$             & $-499.39$ & $-499.76$ & $-499.37$ & $-312.01$ & $-312.86$ & $-311.65$ & $-118.28$ & $-118.46$ & $-117.84$ & $-58.49$ & $-58.62$ & $-58.24$ \\
            $\Delta\log_{10}(Z)$ & $-0.01$   & $-0.17$   & $0$       & $-0.16$   & $-0.53$   & $0$       & $-0.19$   & $-0.27$   & $0$       & $-0.11$  & $-0.16$  & $0$      \\
            \hline
        \end{tabular}
        \tablefoot{%
            $\ln(Z)$: natural logarithm of the Bayesian evidence~$Z$;
            $\Delta\log_{10}(Z)$: decimal logarithm of the Bayes factor, or difference of the decimal logarithms of Bayesian evidence, relative to the best model of each galaxy.%
        }
    \end{table*}

    The differences in constraints between different models of the same galaxy are generally small, certainly when considering the size of the uncertainties.
    The differences between the galaxies are larger.
    We note that, although Hya~II is the third most luminous galaxy, its $M_{200}$ and $V_\mathrm{max}$ are the highest.
    Within the half-light radius, the ordering of the masses is in line with that of the luminosities.
    Consequently, Hya~II has high mass-to-light ratios and astrophysical $J$ and $D$ factors.
    Lastly, we note that Leo~T and Hya~II tend to higher concentrations than Ant~B and Gru~1.

    We repeated the calculation of the same quantities on the GravSphere profiles, the results of which we list in Table~\ref{tab:gsm200} and present in Appendix~\ref{app:gravsphere}.
    We see good agreement between the two tools for quantities determined within the half-light radius, as expected from the agreement on the profiles.
    However, at the larger scales, we see systematic differences, for example in $M_{200}$.
    This is likely the result of the aforementioned systematic differences in the density slopes.
    We will discuss the implications of these differences in Sect.~\ref{ssec:scaling}, where we compare our results against theoretical scaling relations.

\subsection{Model comparison}
\label{ssec:modcomp}
    Having appreciated the results of our models and their differences, we now turn to establishing which of these models is the best description of each galaxy.
    MultiNest provides several estimators of Bayesian evidence.
    As in \citetalias{Zoutendijk-2021-A&A-651-A80}, we use the nested sampling global log-evidence estimator.
    The Bayesian evidence is the posterior likelihood of observing the data given the model and its priors.
    Because this likelihood has been integrated over the entire parameter space, the Bayesian evidence takes into account not only how well the model fits the data for the best parameter values, but how well the model does in general.
    This penalizes models with larger parameter spaces and avoids over-fitting.
    We display the natural logarithm of the Bayesian evidence, $\ln(Z)$, in Table~\ref{tab:modcomp}.

    Different models for the same galaxy can be compared by computing the posterior odds ratios.
    We assume no prior preference for certain models and take the prior odds ratios to be $1$.
    Therefore the posterior odds ratios are equal to the ratios of the Bayesian evidence, known as the Bayes factors.
    According to the scale of \citet[their Appendix~B]{Jeffreys-1961-TP-C-3}, an odds ratio of $10^{-2}$ is required to rule out a model decisively.
    For each galaxy, we calculate the Bayes factors relative to the model with the largest Bayesian evidence, and display the decimal logarithm of the Bayes factor, $\Delta\log_{10}(Z)$,  in Table~\ref{tab:modcomp}.
    The only model with substantial evidence against it ($\Delta\log_{10}(Z) < 10^{-0.5}$) is the core model for Leo~T, though substantial evidence is still far from significant.
    For all other models the evidence is weak.
    We cannot exclude the cusp, core, or soliton model for any of our galaxies.
    For each galaxy individually, the soliton model is the most preferred and the core model the least.
    We defer a joint analysis, in which all galaxies are expected to follow the same profile model with the same dark-matter properties, to a follow-up paper.

\subsection{Tidal stripping}
\label{ssec:stripping}
    With the newly presented galaxies in this paper, we have increased the number of dark-matter density profiles from the MUSE-Faint survey from one to five.
    Now we are going to compare the measured properties of all five galaxies against theoretical expectations.
    From the comparison between the CJAM and GravSphere profiles it is clear that the profiles are best determined, with the smallest uncertainties and the least bias, around the half-light radii.
    We therefore begin our comparison between the derived profiles and theoretical expectations with a model based on quantities measured around the half-light radius.

    \citet{Fattahi-2018-MNRAS-476-3816} present a theoretical relation, based on the APOSTLE zoom-in simulations of Local Group-like environments, between the stellar mass $M_*^\mathrm{prog}$ and maximum circular velocity $V_\mathrm{max}^\mathrm{prog}$ of a satellite galaxy's progenitor, when the stellar mass and maximum circular velocity where at their peak, before the progenitor started falling into the host galaxy.
    This theoretical relation is combined with a procedure to relate $M_*^\mathrm{prog}$ and $V_\mathrm{max}^\mathrm{prog}$ to the satellite's present-day stellar mass $M_*$, present-day three-dimensional half-light radius $r_{1/2}$, and present-day circular velocity $V(r_{1/2})$ at that radius, by accounting for tidal stripping.
    This stripping procedure is based on the assumption that the progenitors at peak stellar mass and peak maximum circular velocity have an NFW profile, therefore it only applies to the cusp models in this paper.
    The procedure has three theoretical ingredients: the aforementioned $M_*^\mathrm{prog}$--$V_\mathrm{max}^\mathrm{prog}$ relation; the mass--concentration relation of \citet{Ludlow-2016-MNRAS-460-1214}, applied to the progenitors ($M_{200}^\mathrm{prog}$--$c_{200}^\mathrm{prog}$); and the tidal tracks of \citet{Errani-2015-MNRAS-449-L46} describing the coupled evolution of $M_*$, $r_{1/2}$, and $V(r_{1/2})$.
    These tidal tracks are a function of $x \coloneqq M(r_{1/2}^\mathrm{prog})/M_\mathrm{prog}(r_{1/2}^\mathrm{prog})$, which is the fraction of present-day versus peak total mass remaining inside $r_{1/2}^\mathrm{prog}$, the three-dimensional half-light radius of the progenitor.
    There should be one value of $x$ for which a progenitor satisfying the $M_*^\mathrm{prog}$--$V_\mathrm{max}^\mathrm{prog}$ and $M_{200}^\mathrm{prog}$--$c_{200}^\mathrm{prog}$ scaling relations, evolving along the tidal tracks, produces the observed galaxy.

\begin{figure*}
        \includegraphics[width=\linewidth]{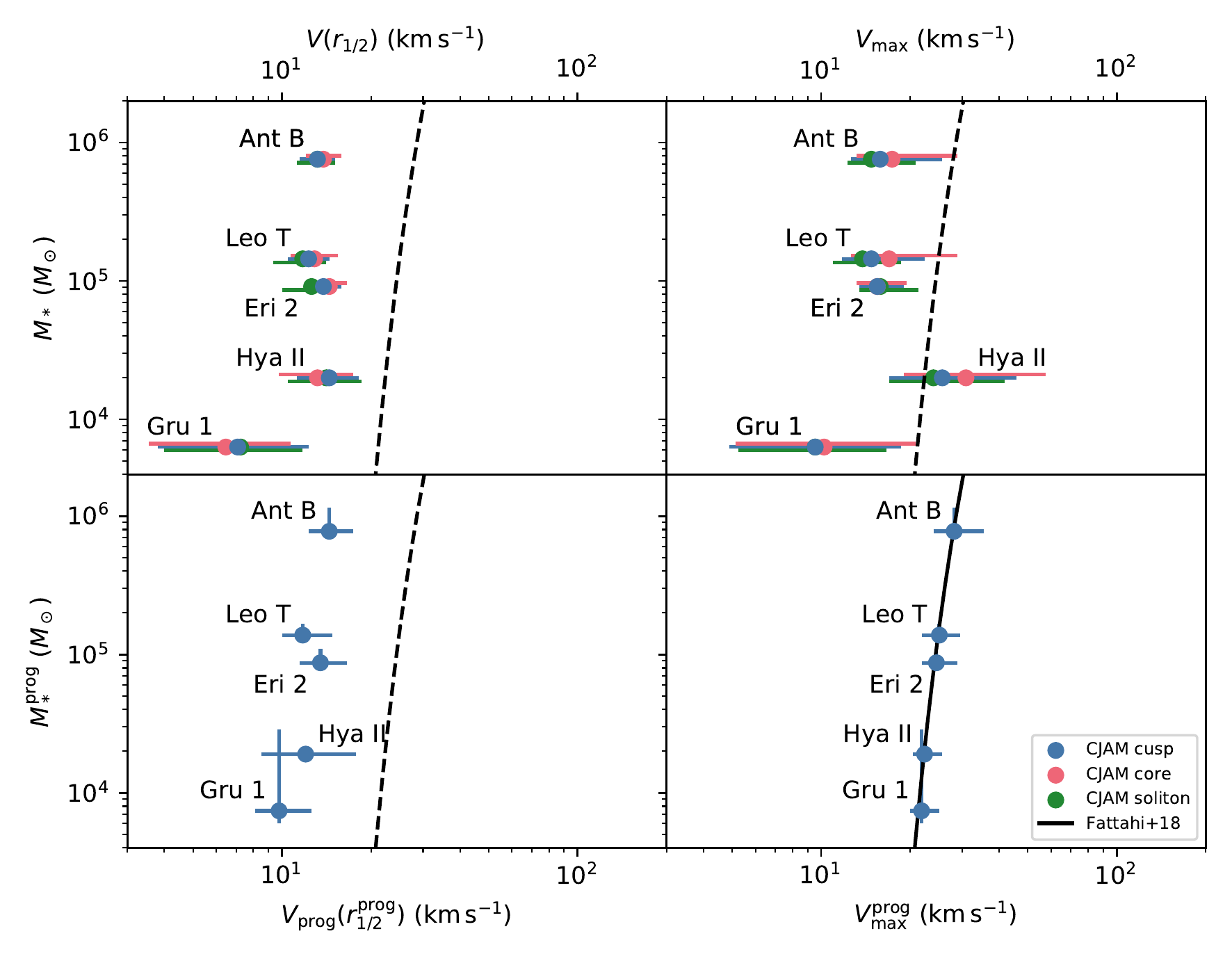}
        \caption{%
            Observed galaxy properties (\emph{top row}), as determined with CJAM, matched with predicted progenitor properties (\emph{bottom row}) according to the method of \citet{Fattahi-2018-MNRAS-476-3816}.
            \emph{Top left}: Observed circular velocities~$V(r_{1/2})$ at the three-dimensional half-light radius against observed stellar masses~$M_*$.
            \emph{Top right}: Observed maximum circular velocities~$V_\mathrm{max}$ against observed stellar masses~$M_*$.
            \emph{Bottom left}: Predicted progenitor circular velocities~$V_\mathrm{prog}(r_{1/2}^\mathrm{prog})$ at the three-dimensional half-light radius against predicted progenitor stellar masses~$M_*^\mathrm{prog}$.
            \emph{Bottom right}: Predicted progenitor maximum circular velocities~$V_\mathrm{max}^\mathrm{prog}$ against predicted progenitor stellar masses~$M_*^\mathrm{prog}$.
            The observed galaxies have properties measured under the assumption of three different density profile models, cusp, core, and soliton, while the progenitor properties can only be predicted for the cusp model.
            The progenitors are assumed to obey theoretical scaling relations, including the $M_*^\mathrm{prog}$--$V_\mathrm{max}^\mathrm{prog}$ scaling relation of \citet{Fattahi-2018-MNRAS-476-3816}, which is displayed in the \emph{bottom-right} panel with a solid line and repeated for reference in the other panels with a dashed line.
            We devolve the observed galaxy properties in the \emph{top-left} panel along theoretical tidal tracks to obtain the predicted progenitor properties in the \emph{bottom} row.
            The observations in the \emph{top-right} panel are not used in this procedure.
            The observed galaxies are tidally stripped of dark matter or stars if the observed circular velocity or stellar mass, respectively, indicated in the \emph{top-left} panel, are smaller than those of the progenitor in the \emph{bottom-left} panel.
        }
        \label{fig:MstarVprog}
    \end{figure*}

    Because in \citetalias{Zoutendijk-2021-A&A-651-A80} we did not compute quantities associated with the three-dimensional half-light radius, we provide in Table~\ref{tab:m200eri2} the circular velocity at, and mass within, this radius, based on the original profiles.
    \begin{table}
        \caption{Supplementary parameters derived from the CJAM density profiles of Eridanus~2 presented in \citetalias{Zoutendijk-2021-A&A-651-A80}.}
        \label{tab:m200eri2}
        \centering
        \begin{tabular}{lc@{\hspace{3pt}}c@{\hspace{3pt}}c}
            \hline
            \hline
                                                                 & \multicolumn{3}{c}{Eri~2}                                                \\
                                                                 & cusp                   & core                   & soliton                \\
            \hline
            $\log_{10}(V(r_{1/2})/\mathrm{km}\,\mathrm{s}^{-1})$ & $1.14^{+0.06}_{-0.06}$ & $1.16^{+0.06}_{-0.06}$ & $1.10^{+0.07}_{-0.11}$ \\
            $\log_{10}(M(r_{1/2})/\mathrm{km}\,\mathrm{s}^{-1})$ & $7.16^{+0.12}_{-0.12}$ & $7.20^{+0.13}_{-0.11}$ & $7.08^{+0.14}_{-0.21}$ \\
            \hline
        \end{tabular}
        \tablefoot{%
            The cusp, core, and soliton models are identified in \citetalias{Zoutendijk-2021-A&A-651-A80} as cold dark matter, self-interacting dark matter, and fuzzy dark matter, respectively.
            $V(r_{1/2})$: circular velocity at the three-dimensional half-light radius;
            $M(r_{1/2})$: mass within the three-dimensional half-light radius.%
        }
    \end{table}
    In the top row of Fig.~\ref{fig:MstarVprog} we show our galaxies in the $M_*$--$V(r_{1/2})$ and $M_*$--$V_\mathrm{max}$ planes as well as the $M_*^\mathrm{prog}$--$V_\mathrm{max}^\mathrm{prog}$ relation of \citet{Fattahi-2018-MNRAS-476-3816} that is repeated in all panels.
    If the galaxies have not experienced any tidal stripping, we expect that they lie on the indicated relation in the top right panel.
    If stripping did occur, they should be located to the left of the relation, having lost dark matter more heavily than stars.
    Meanwhile, in the top left panel, the galaxies are expected to always lie to the left of the relation, because by definition $V_\mathrm{max} > V(r_{1/2})$.
    However, we need to take into account that $V_\mathrm{max}$ is determined at a larger radius than $r_{1/2}$ and could therefore be biased by the systematic difference in density profile slope that we noticed in our comparison between CJAM and GravSphere.
    Therefore, the fact that our measurements do not follow the relation in the top right panel is not solid evidence of stripping.
    As the stripping calculations are based on $V(r_{1/2})$, not $V_\mathrm{max}$, this possible bias will not affect the results of these calculations.

    Here we describe the determination of the progenitor properties and the amount of tidal stripping in more detail.
    Given a value of the stripping parameter $x$, the observed $M_*$, $V(r_{1/2})$, and $r_{1/2}$ can be converted to the progenitor's $M_*^\mathrm{prog}$, $V_\mathrm{prog}(r_{1/2}^\mathrm{prog})$, and $r_{1/2}^\mathrm{prog}$ using the tidal tracks.
    This $x$ is the sole degree of freedom in the tidal tracks, which we must constrain.
    Together, $V_\mathrm{prog}(r_{1/2}^\mathrm{prog})$ and $r_{1/2}^\mathrm{prog}$ allow us to calculate $M_\mathrm{prog}(r_{1/2}^\mathrm{prog})$, the mass of the progenitor inside its half-light radius.
    We will constrain the degree of freedom by additionally requiring that the progenitor obeys the $M_*^\mathrm{prog}$--$V_\mathrm{max}^\mathrm{prog}$ and $M_{200}^\mathrm{prog}$--$c_{200}^\mathrm{prog}$ scaling relations.
    Every choice of $x$ then corresponds to an $M_{200}^\mathrm{prog}$ and $c_{200}^\mathrm{prog}$, enough to fully specify the two-parameter NFW profile of the progenitor.
    From this density profile we can also calculate $M_\mathrm{prog}(r_{1/2}^\mathrm{prog})$.
    We can therefore constrain our one degree of freedom by demanding that the results of these two different $M_\mathrm{prog}(r_{1/2}^\mathrm{prog})$ calculations are identical.
    Thus we find $x$ as well as $M_*^\mathrm{prog}$, $V_\mathrm{prog}(r_{1/2}^\mathrm{prog})$, and $r_{1/2}^\mathrm{prog}$.

    For computational ease, we search for the solution of the above problem by varying $M_{200}^\mathrm{prog}$ instead of $x$.
    In this way, we save ourselves from inverting the $M_{200}^\mathrm{prog}$--$c_{200}^\mathrm{prog}$ relation.
    We proceed as follows:
    Given an initial guess for the virial mass $M_{200}^\mathrm{prog}$ of the progenitor, we calculate its concentration $c_{200}^\mathrm{prog}$ according to \citet{Ludlow-2016-MNRAS-460-1214}.
    As we know the progenitor follows an NFW profile, we can use both of these parameters to calculate $V_\mathrm{max}^\mathrm{prog}$.
    The application of the \citet{Fattahi-2018-MNRAS-476-3816} $M_*^\mathrm{prog}$--$V_\mathrm{max}^\mathrm{prog}$ scaling relation then leads to $M_*^\mathrm{prog}$.
    We numerically solve the tidal track for the evolution of stellar mass to find the $x$ for which the $M_*^\mathrm{prog}$ progenitor produces an $M_*$ satellite.
    This $x$ is then substituted into the other tidal tracks to find $r_{1/2}^\mathrm{prog}$ from $r_{1/2}$ and $V_\mathrm{prog}(r_{1/2}^\mathrm{prog})$ from $V(r_{1/2})$.
    Together, $V_\mathrm{prog}(r_{1/2}^\mathrm{prog})$ and $r_{1/2}^\mathrm{prog}$ allow us to calculate $M_\mathrm{prog}(r_{1/2}^\mathrm{prog})$.
    We calculate the same from $M_{200}^\mathrm{prog}$ and $c_{200}^\mathrm{prog}$, which should give the same result if our guess of $M_{200}^\mathrm{prog}$ is correct.
    We numerically minimize the difference of the two calculations of $M_\mathrm{prog}(r_{1/2}^\mathrm{prog})$ by varying $M_{200}^\mathrm{prog}$ until a match is found.

    \begin{table*}
        \caption{Progenitor properties based on CJAM, according to the method of \citet{Fattahi-2018-MNRAS-476-3816}.}
        \label{tab:prog}
        \centering
        \begin{tabular}{lccccc}
            \hline
            \hline
                                                                         & Ant~B                   & Leo~T                   & Eri~2                   & Hya~II                  & Gru~1                   \\
            \hline
            $\log_{10}(M_*^\mathrm{prog}/M_\sun)$                                 &  $5.89^{+0.17}_{-0.03}$ &  $5.14^{+0.08}_{-0.00}$ &  $4.94^{+0.10}_{-0.01}$ &  $4.28^{+0.00}_{-0.00}$ &  $3.87^{+0.59}_{-0.09}$ \\
            $\log_{10}(V_\mathrm{max}^\mathrm{prog}/\mathrm{km}\,\mathrm{s}^{-1})$ &  $1.45^{+0.10}_{-0.07}$ &  $1.40^{+0.07}_{-0.06}$ &  $1.39^{+0.07}_{-0.05}$ &  $1.35^{+0.06}_{-0.04}$ &  $1.34^{+0.06}_{-0.04}$ \\
            $\log_{10}(r_{1/2}^\mathrm{prog}/\mathrm{kpc})$                          & $-0.44^{+0.06}_{-0.05}$ & $-0.68^{+0.06}_{-0.06}$ & $-0.48^{+0.06}_{-0.06}$ & $-0.97^{+0.08}_{-0.08}$ & $-0.71^{+0.08}_{-0.06}$ \\
            $\log_{10}(V_\mathrm{prog}(r_{1/2}^\mathrm{prog})/\mathrm{km}\,\mathrm{s}^{-1})$     &  $1.16^{+0.08}_{-0.07}$ &  $1.07^{+0.10}_{-0.07}$ &  $1.13^{+0.09}_{-0.07}$ &  $1.08^{+0.17}_{-0.15}$ &  $0.99^{+0.11}_{-0.08}$ \\
            $\log_{10}(x)$                                               & $-0.07^{+0.13}_{-0.16}$ &  $0.03^{+0.13}_{-0.16}$ &  $0.00^{+0.13}_{-0.15}$ &  $0.12^{+0.17}_{-0.19}$ & $-0.17^{+0.26}_{-0.50}$ \\
            $\log_{10}(\mu_L)$                                           & $-0.01^{+0.03}_{-0.17}$ &  $0.02^{+0.00}_{-0.08}$ &  $0.01^{+0.01}_{-0.10}$ &  $0.02^{+0.00}_{-0.00}$ & $-0.07^{+0.09}_{-0.59}$ \\
            \hline
        \end{tabular}
        \tablefoot{%
            $M_*^\mathrm{prog}$: progenitor stellar mass;
            $V_\mathrm{max}^\mathrm{prog}$: progenitor maximum circular velocity;
            $r_{1/2}^\mathrm{prog}$: progenitor three-dimensional half-light radius;
            $V_\mathrm{prog}(r_{1/2}^\mathrm{prog})$: progenitor circular velocity within the progenitor three-dimensional half-light radius;
            $x$: $M(r_{1/2}^\mathrm{prog})/M_\mathrm{prog}(r_{1/2}^\mathrm{prog})$, fraction of the total mass within the progenitor's three-dimensional half-light radius that remains after tidal stripping;
            $\mu_L$: $M_*/M_*^\mathrm{prog}$, fraction of the stellar mass that remains after tidal stripping.%
        }
    \end{table*}

    We propagate the measurement uncertainties on $M_*$, $r_{1/2}$, and $V(r_{1/2})$ in a Monte Carlo fashion by taking random samples from their posterior distributions.
    To account for scatter in the scaling relations, we vary the nominal values from the several scaling relations with a log-normally distributed factor based on the scatter visible in the figures of \citet{Ludlow-2016-MNRAS-460-1214}, \citet{Fattahi-2018-MNRAS-476-3816}, and \citet{Errani-2015-MNRAS-449-L46}.
    This scatter amounts to $0.1\,\mathrm{dex}$ on $c_{200}^\mathrm{prog}$, $1\,\mathrm{dex}$ on $M_*^\mathrm{prog}$, $0.2\,\mathrm{dex}$ on $x$, $0.05\,\mathrm{dex}$ on $r_{1/2}^\mathrm{prog}$, and $0.05\,\mathrm{dex}$ on $V(r_{1/2}^\mathrm{prog})$.
    We also vary the redshift at which we evaluate the mass--concentration relation uniformly between $0$--$1$.
    The added scatter and redshift variation increase the uncertainties on our results, but we find that our conclusions would not have been different if we had opted to not propagate these theoretical uncertainties.

    \begin{figure*}
        \includegraphics[width=\linewidth]{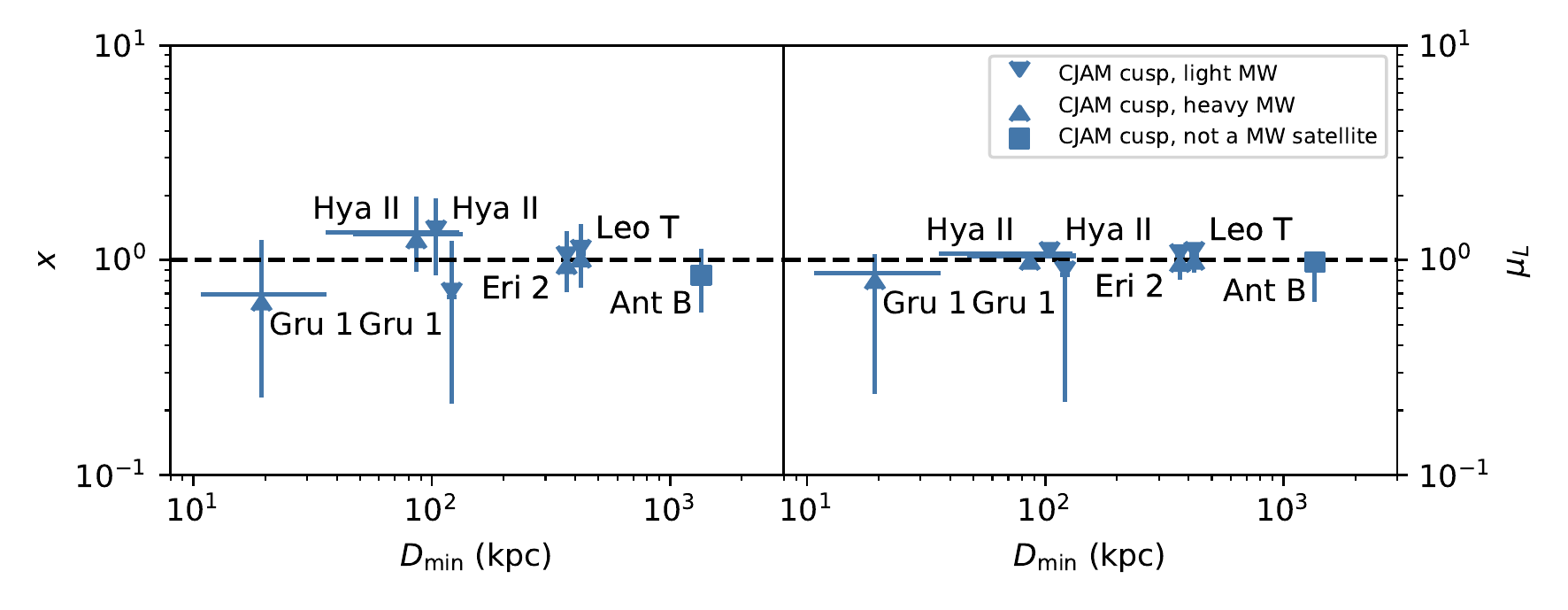}
        \caption{%
            Stripping parameters of the galaxies, as determined with CJAM and comparison to theoretical models, versus their closest distance to the Milky Way ever reached.
            The stripping parameters $x$ and $\mu_L$ are the fractions of total mass within the half-light radius and of the total stellar mass, respectively, remaining after tidal stripping; lower values indicate more stripping has occurred.
            We show results for two models of the Milky Way that differ in the halo mass.
            In most cases, the closest distance is not the pericentre distance but the current distance, as most of these galaxies are still in their first infall.
            For Hya~II in both Milky Way models and for Gru~1 in the heavy Milky Way model, the closest distance is the pericentre.
            For Gru~1 in the light Milky Way models and for Leo~T and Eri~2 in both Milky Way models, the closest distance is the present-day distance, because in these models these galaxies are still on their first infall.
            We note that Antlia~B is a satellite of NGC~3109, not of the Milky Way, and that its distance to the Milky Way is therefore not relevant for its stripping parameters; its display in this figure is merely illustrative.%
        }
        \label{fig:stripdist}
    \end{figure*}

    For all five of our galaxies we find a progenitor that solves all constraints.
    The progenitors' properties, as well as the stripping parameters $x$ and $\mu_L \coloneqq M_*/M_*^\mathrm{prog}$ are presented in Table~\ref{tab:prog} and their locations in the $M_*^\mathrm{prog}$--$V_\mathrm{prog}(r_{1/2}^\mathrm{prog})$ and $M_*^\mathrm{prog}$--$V_\mathrm{max}^\mathrm{prog}$ planes are shown in the bottom row of Fig.~\ref{fig:MstarVprog}.
    By design, we see that the maximum circular velocities of the progenitors follow the \citet{Fattahi-2018-MNRAS-476-3816} relation.
    The circular velocity at the three-dimensional half-light radius being naturally smaller than the maximum circular velocity, the galaxies lie a little to the left in the bottom-left panel.
    By eye, the locations of the progenitor galaxies in the bottom-left panel does not differ significantly from the locations of the observed galaxies in the top-left panel.
    This indicates that the total mass within the half-light radius and the stellar mass have not significantly changed during tidal stripping.

    Indeed, from Table~\ref{tab:prog} we can see that all galaxies are consistent with having undergone no stripping, as they include $x = 1$ in their confidence intervals.
    However, the large uncertainties on Gru~1 still include $x = 10^{-0.67} \approx 0.2$, which means up to ${\approx}80\%$ of its original mass could have been stripped.
    Given its present-day distance of ${>}100\,\mathrm{kpc}$ from the Milky Way, such significant stripping seems surprising.
    However, there may have been a previous time when these galaxies were closer to the Milky Way, at which time they could have been tidally stripped more significantly.
    \citet{2021arXiv210608819B} provide pericentre distances for our four Milky Way satellites, based on \emph{Gaia} EDR3 and two different Milky Way mass models: a light model ($M_{200} = 8.8 \times 10^{11}\,M_\sun)$ and a heavy model ($M_{200} = 1.6 \times 10^{12}\,M_\sun$).
    We remind the reader that Ant~B is not a satellite of the Milky Way, but rather of NGC~3109.
    In the light Milky Way model, only Hya~II has reached its pericentre in the past; the other satellites are making their first infall.
    According to the heavy Milky Way model, both Hya~II and Gru~1 have reached their pericentres in the past.

    \begin{figure*}
        \includegraphics[width=\linewidth]{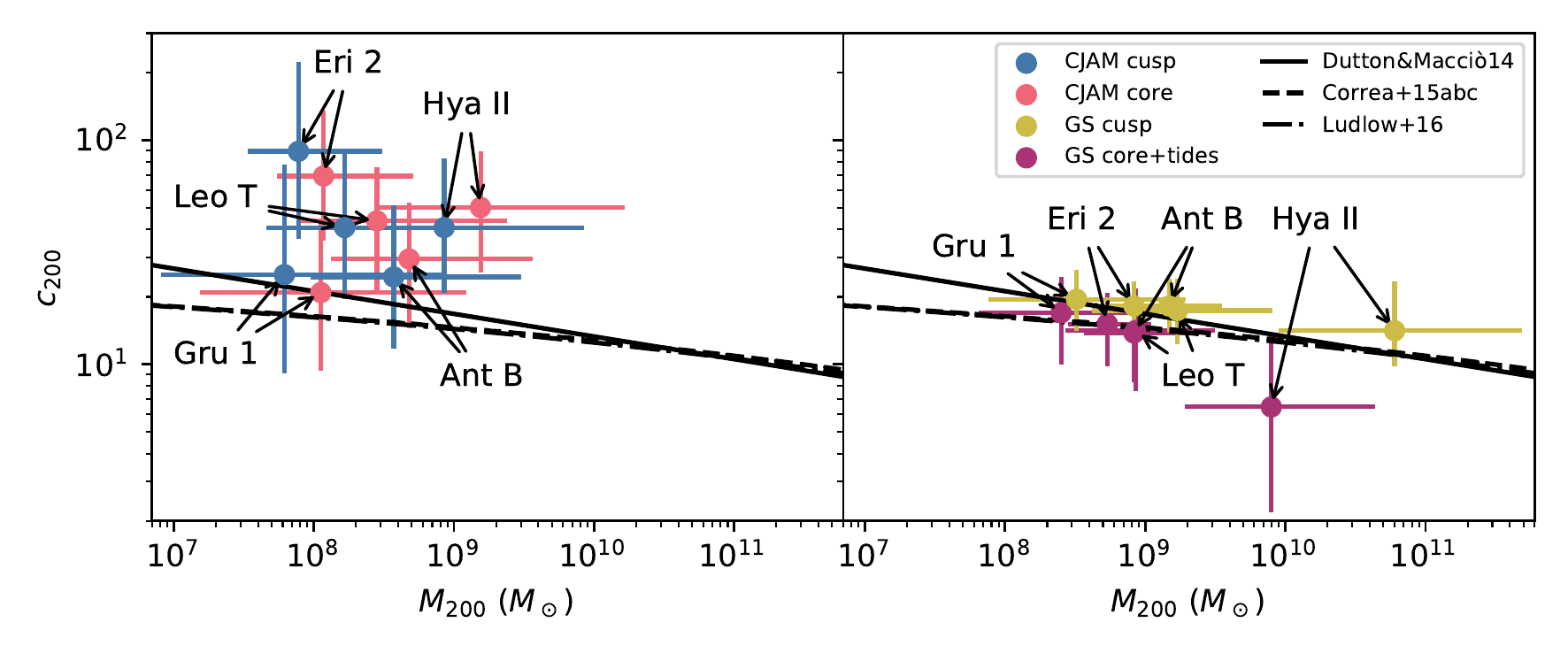}
        \caption{%
            Comparison of the galaxies' relation between virial masses~$M_{200}$ and concentrations~$c_{200}$, as determined with CJAM~(\emph{left}) and GravSphere~(\emph{right}), to expectations from models.
            The model of \citet{Dutton-2014-MNRAS-441-3359} is empirically fitted to simulations, while the models of \citet{Correa-2015-MNRAS-450-1514, Correa-2015-MNRAS-450-1521, Correa-2015-MNRAS-452-1217} and \citet{Ludlow-2016-MNRAS-460-1214} are semi-analytical.
            The latter two models produce nearly indistinguishable results.%
        }
        \label{fig:c200M200}
    \end{figure*}

    In Fig.~\ref{fig:stripdist} we display $x$ and $\mu_L$ versus the closest distances to the Milky Way that the galaxies have ever reached, which is a past pericentre or the present-day distance in the case of first infall.
    Depending on the Milky Way model, Gru~1 may have made a close approach to the Milky Way, in which case significant stripping may indeed have occurred.
    The consistency of the more distant galaxies Leo~T, Eri~2, and Hya~II with no stripping indicates that our $V(r_{1/2})$ measurements match the theoretical expectations.

    When we repeat the stripping procedure with the GravSphere results, as presented in Appendix~\ref{app:gravsphere}, we find consistent results, listed in Table~\ref{tab:gsprog} and shown in Fig.~\ref{fig:gsstripdist}.
    This confirms the robustness of our no-stripping conclusions and the $V(r_{1/2})$ measurements.
    However, the smaller uncertainties on the GravSphere profiles result in Gru~1 no longer being consistent with a significant amount of stripping.

\subsection{Galaxy--halo scaling relations}
\label{ssec:scaling}
    We continue to test our results against theoretical expectations with a comparison of our measurements at larger radii against theoretical scaling relations.
    It is important to remember that these measurements at the virial radius may be biased due to the tendency of our two dynamical modelling tools to fits systematically different profile slopes.
    Though within the half-light radius, where we have kinematic tracers, the profiles agree within the measurement uncertainties, the difference in slope will make the profiles diverge as one goes to larger radii.
    Considering the virial radii~$r_{200}$ are over an order of magnitude larger than the half-light radii, the virial parameters may be significantly biased.
    Indeed, the lack of tracers far outside the half-light radii make the fitted profiles insensitive to deviations from the profile models, such as tidal stripping, if they occur at large radii.
    As we found no evidence of stripping inside the half-light radii and are incapable of detecting stripping outside, the virial parameters that we calculate would be those of our galaxies' progenitors, aside from the possible bias.
    We therefore do not have to concern ourselves with taking into account the effects of stripping on the scaling relations and can instead use the relations derived for isolated galaxies.

    First we perform a comparison of virial masses and concentrations.
    In the left panel of Fig.~\ref{fig:c200M200} we display the values derived with CJAM for the four galaxies presented in this paper, supplemented with those of Eri~2 from \citetalias{Zoutendijk-2021-A&A-651-A80}.
    We compare these galaxies to a selection of mass--concentration relations.
    The models should be evaluated at the redshift at which the galaxies became satellites of the Milky Way.
    Lacking this information, we take a redshift of zero.
    The relation of \citet{Dutton-2014-MNRAS-441-3359} is fitted to simulations, though these fits are limited to $M_{200} \gtrsim 10^{10}\,M_\sun$.
    The application of this fitted relation to our galaxies is therefore an extrapolation.
    \citet{Correa-2015-MNRAS-450-1514, Correa-2015-MNRAS-450-1521, Correa-2015-MNRAS-452-1217} and \citet{Ludlow-2016-MNRAS-460-1214} present semi-analytical models, which are physically motivated but still calibrated against simulations.
    Although the simulations against which the latter two models are calibrated do not reach UFD masses either, their physical backing makes the models more constrained and therefore these models should withstand extrapolation better.

    The comparison in the left panel of Fig.~\ref{fig:c200M200} shows that all three models systematically underpredict the observations.
    The two semi-analytical models produce results very close to each other, while the model of \citet{Dutton-2014-MNRAS-441-3359} predicts somewhat higher concentrations that are slightly more consistent with the observations.
    Assuming a higher redshift will make the difference between theory and observations larger.
    The mismatch between the \citet{Ludlow-2016-MNRAS-460-1214} relation and the observations is on first thought remarkable because the stripping analysis indicated consistency with this relation.

    We repeat the comparison with the GravSphere results presented in Appendix~\ref{app:gravsphere} and show this in the right panel of Fig.~\ref{fig:c200M200}.
    Now we do see an overall agreement between observation and theory, except for Hya~II.
    The cusp model produces results close to the \citet{Dutton-2014-MNRAS-441-3359} scaling relations, while the lower concentrations of the core+tides model are closer to the predictions from the semi-analytical scaling relations.

\begin{figure*}
        \includegraphics[width=\linewidth]{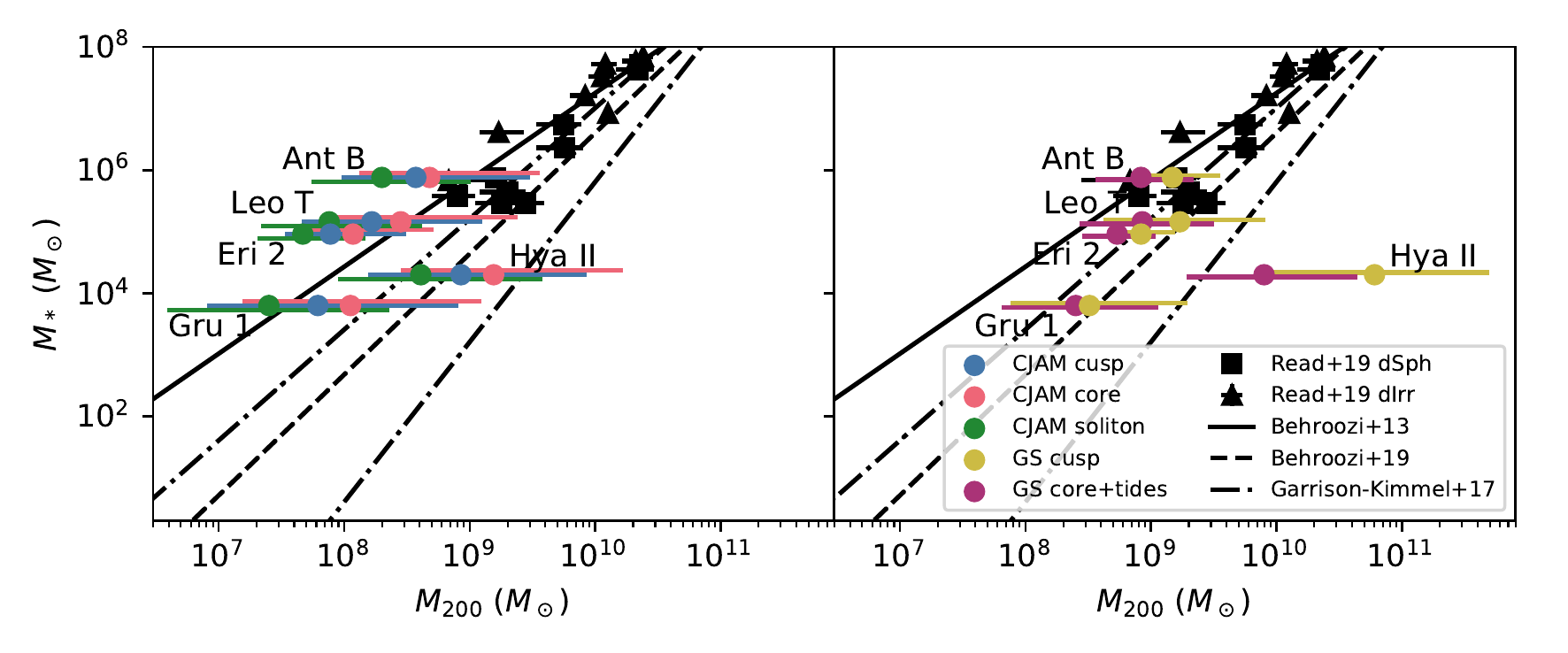}
        \caption{%
            Comparison of the galaxies' stellar masses~$M_*$ and virial masses~$M_{200}$, as determined with CJAM~(\emph{left}) and GravSphere~(\emph{right}), to expectations from models.
            The models of \citet{Behroozi-2013-ApJ-770-57, Behroozi-2019-MNRAS-488-3143} are determined with UniverseMachine.
            The model of \citet{GarrisonKimmel-2017-MNRAS-464-3108} has the scatter in the relation as a free parameter; we show lines corresponding to zero scatter~(larger $M_*$) and $2.0\,\mathrm{dex}$ scatter~(smaller $M_*$).
            We additionally show measurements for the sample of dwarf spheroidal~(dSph) and dwarf irregular~(dIrr) galaxies studied by \citet{Read-2019-MNRAS-484-1401}.%
        }
        \label{fig:MstarM200}
    \end{figure*}

    In the left panel Fig.~\ref{fig:MstarM200} we compare the stellar and virial masses from CJAM.
    When plotted in this way, we clearly see that the core models systematically tend to higher virial masses than the cusp models, which in turn tend to higher virial masses than the soliton models.
    The high virial mass of Hya~II for its luminosity or stellar mass is also noticeable.
    We compare the measurements against a selection of stellar-to-halo mass scaling relations.
    These relations are determined applying a proposed scaling relation to haloes in a dark-matter simulation and comparing the observable properties of the simulated galaxies to observations.
    The relations of \citet{Behroozi-2013-ApJ-770-57, Behroozi-2019-MNRAS-488-3143} are based on the UniverseMachine models.
    Though these models have been calibrated to galaxies with $M_* \gtrsim 10^8\,M_\sun$, \citet{Wang-2021-ApJ-915-116} find that the extrapolated stellar-to-halo mass relation of the second iteration (UniverseMachine DR1; \citealp{Behroozi-2019-MNRAS-488-3143}) is consistent with simulations of ultra-faint dwarf galaxies around a Milky Way-like host.
    \citet{GarrisonKimmel-2017-MNRAS-464-3108} produce a model with a degeneracy between the low-mass slope of the scaling relation and the scatter in that relation, calibrated against observations of galaxies with $M_*$ down to ${\sim}10^5\,M_\sun$.
    We show the two extremes, $M_* \propto M_{200}^{1.8}$ with zero scatter and $M_* \propto M_{200}^{2.6}$ with $2.0\,\mathrm{dex}$ scatter.
    The range between these extremes includes the UniverseMachine DR1 relation.
    The stellar-to-halo mass relation that best describes our measurements seems to be the old UniverseMachine model, even though this model is the only one not calibrated against observations or simulations of dwarf galaxies.
    We additionally compare our measurements with those of \citet{Read-2019-MNRAS-484-1401} for more massive dwarf galaxies, and see that these galaxies do follow the expected scaling relations.

    We show the same comparison, now using the GravSphere results, in the right panel of Fig.~\ref{fig:MstarM200}.
    Again, GravSphere produces results consistent with the theoretical expectations, except for Hya~II.
    The GravSphere measurements of our five faint dwarf galaxies are also consistent with an extrapolation of the the trend followed by the more massive dwarf galaxy sample of \citet{Read-2019-MNRAS-484-1401}.

    The above comparisons of the CJAM and GravSphere results against each other and against theoretical scaling relations demonstrate that the virial parameters are indeed sensitive to the dynamical modelling.
    Gravsphere does produce results more in line with theoretical expectations, but given that the dynamical models are not well tested in this regime, this cannot been taken in support of these models without more extensive tailored simulated data, the construction of which is outside the scope of this paper.

\section{Discussion}
\label{sec:discussion}
    Here we place our results in the context of previous works~(Sect.~\ref{ssec:prev}), and discuss how reliable our results~(Sect.~\ref{ssec:relres}) and our dynamical analysis~(Sect.~\ref{ssec:reldyn}) are.

\subsection{Comparison to previous results}
\label{ssec:prev}
    Though we presented the first dark-matter density profiles for Ant~B, Leo~T, Hya~II, and Gru~1, dynamical masses and mass-to-light ratios have been measured before for some of these galaxies.
    For Ant~B we have presented the first kinematics, therefore no dynamical measurements were made before.

    \citet{Simon-2007-ApJ-670-313} measured a total mass of $M_{200} = (8.2 \pm 3.6) \times 10^6\,M_\sun$ for Leo~T based on its velocity dispersion and an associated mass-to-light ratio of $138 \pm 71 M_\sun\,L_\sun^{-1}$.
    This is inconsistent with our measurements of $M_{200} = 7.6^{+33.2}_{-5.4} \times 10^7$--$2.9^{+22.8}_{-2.1} \times 10^8\,M_\sun$ and $\Upsilon(r_{200}) = 8.3^{+36.4}_{-5.9} \times 10^2$--$3.2^{+28.2}_{-2.3} \times 10^3\,M_\sun\,L_\sun^{-1}$, which are over an order of magnitude larger.
    Several authors estimate the mass of Leo~T within $300\,\mathrm{pc}$ from stellar kinematics -- $M(300\,\mathrm{pc}) = 1.30^{+0.88}_{-0.42} \times 10^7\,M_\sun$~\citep{Strigari-2008-Natur-454-1096} -- or gas kinematics -- $M(300\,\mathrm{pc}) = (8.2 \pm 0.2) \times 10^6\,M_\sun$~\citep{Faerman-2013-ApJ-777-119} and $M(300\,\mathrm{pc}) = (3.7 \pm 0.7) \times 10^6\,M_\sun$~\citep{Patra-2018-MNRAS-480-4369}.
    From our three CJAM profiles for Leo~T, we calculate $M(300\,\mathrm{pc}) = 1.00^{+0.51}_{-0.42} \times 10^7$--$1.45^{+0.64}_{-0.51} \times 10^7\,M_\sun$.
    This is consistent with the results of \citet{Strigari-2008-Natur-454-1096} and \citet{Faerman-2013-ApJ-777-119}, but not with that of \citet{Patra-2018-MNRAS-480-4369}.

    For Hya~II, \citet{Kirby-2015-ApJ-810-56} constrain the mass within the half-light radius to ${<}10^{6.1}\,M_\sun$ at 95\% confidence and the mass-to-light ratio within that radius to ${<}315\,M_\sun\,L_\sun^{-1}$.
    Our measurements $M(R_{1/2}) = 2.1^{+1.7}_{-1.0} \times 10^6$--$3.0^{+2.1}_{-1.1} \times 10^6\,M_\sun$ and $\Upsilon(R_{1/2}) = 5.8^{+4.5}_{-2.6} \times 10^2$--$7.9^{+5.5}_{-3.0} \times 10^2\,M_\sun\,L_\sun^{-1}$ are about one standard deviation higher than this upper limit, which is likely related to the fact that we resolve an intrinsic velocity dispersion $\sigma_\mathrm{int} = 12.0^{+5.0}_{-3.5}\,\mathrm{km}\,\mathrm{s}^{-1}$ that is significantly larger than the upper limit $4.5\,\mathrm{km}\,\mathrm{s}^{-1}$ of \citet{Kirby-2015-ApJ-810-56}.

    Gru~1 has constraints from \citet{Walker-2016-ApJ-819-53} on the mass within the half-light radius, ${<}2.5 \times 10^6\,M_\sun$, and the total mass-to-light ratio, ${<}2645$.
    Our $M(R_{1/2}) = 1.1^{+2.1}_{-0.8} \times 10^6$--$1.7^{+2.9}_{-1.2} \times 10^6\,M_\sun$ is fully consistent with this, but our $\Upsilon(r_{200}) = 7.4^{+60.2}_{-6.3} \times 10^3$--$3.2^{+33.9}_{-2.7} \times 10^4\,M_\sun\,L_\sun^{-1}$ is an order of magnitude larger.
    A possible explanation for the significant differences in $M_{200}$ and $\Upsilon(r_{200})$ for Leo~T and Gru~1 is that the previous determinations were based on a bulk measurement of the intrinsic dispersion, while we determine a profile.
    As we discuss in Sect.~\ref{ssec:relres}, determining $M_{200}$ (and by extension $\Upsilon(r_{200})$) from our profiles is hard already, and this is even more difficult when only a single dispersion is available.

    \citet{Bonnivard-2015-MNRAS-453-849} determined astrophysical $J$ and $D$ factors --~which are used to calculate the expected dark-matter annihilation and decay signal strength, respectively, as discussed in Sect.~\ref{ssec:recovery}~-- for a number of dwarf galaxies.
    These galaxies include Leo~T, for which the calculated $\log_{10}(J(\alpha_\mathrm{c}^J)/M_\sun^2\,\mathrm{kpc}^{-5}) = 10.7^{+0.5(+1.1)}_{-0.4(-0.8)}$ and $\log_{10}(D(\alpha_\mathrm{c}^D)/M_\sun\,\mathrm{kpc}^{-2}) = 2.1^{+0.4(+0.8)}_{-0.3(-0.5)}$ are very similar to our own $\log_{10}(J(\alpha_\mathrm{c}^J)/M_\sun^2\,\mathrm{kpc}^{-5}) = 10.40^{+0.36}_{-0.34}$--$10.48^{+0.49}_{-0.36}$ and $\log_{10}(D(\alpha_\mathrm{c}^D)/M_\sun\,\mathrm{kpc}^{-2}) = 1.99^{+0.38}_{-0.33}$--$2.34^{+0.52}_{-0.34}$.
    The $\log_{10}(J(\alpha_\mathrm{c}^J)/M_\sun^2\,\mathrm{kpc}^{-5}) = 11.91^{+0.44}_{-0.42}$--$12.02^{+0.50}_{-0.49}$ and $\log_{10}(D(\alpha_\mathrm{c}^D)/M_\sun\,\mathrm{kpc}^{-2}) = 3.17^{+0.51}_{-0.37}$--$3.45^{+0.60}_{-0.44}$ of Hya~II are similar to the highest measured for dwarf galaxies, which makes Hya~II an interesting target for studies aiming to constrain dark-matter annihilation and decay~\citep{Bonnivard-2015-MNRAS-453-849, 2020arXiv200201229A}.
    There is, however, the possibility of a bias of a factor of a few on the astrophysical factors when an incorrect halo triaxiality is assumed~\citep{Bonnivard-2015-MNRAS-446-3002}.
    The assumption in this paper is that the dark-matter haloes are spherical.

    It is noticeable that, despite the generally weak evidence, the soliton is always the preferred model.
    Additionally, the cusp model is preferred over the core model in all four of the galaxies.
    For Eri~2 we found the same order of preference as for the galaxies in this paper.
    With $r_\mathrm{c} < 200$--$309\,\mathrm{pc}$ (95\% confidence levels), we rule out the large cores detected in some of the classical dwarf galaxies, for example $r_\mathrm{c} \approx 500\,\mathrm{pc}$ for the dwarf irregular galaxies in the sample of \citet{Read-2019-MNRAS-484-1401}.

\subsection{Reliability of inferred parameters}
\label{ssec:relres}
    Because $r_{200}$ is larger than the radii of our most distant tracers by more than an order of magnitude, $M_{200}$ is calculated through an intregral over a density profile that is extrapolated much beyond the measured range.
    If there is a bias in the profile slope, then the extrapolation can lead to a large bias in density, possibly becoming larger than the quoted random error.
    There is, therefore, an extra degree of uncertainty surrounding the calculated $M_{200}$, and it is therefore important to compare results obtained with different tools.
    We have seen that, while CJAM produces results that are inconsistent with the theoretical scaling relations involving $M_{200}$, the results of GravSphere are consistent.
    As we are in an as of yet unexplored mass regime, a difference between the observations and the extrapolations from more massive galaxies is however not necessarily alarming.
    The UFDs are deep into the territory of the missing satellites problem.
    It is therefore conceivable that the successful formation of a galaxy in such small haloes requires exceptional halo properties.
    On the other hand, our observations may be subject to selection effects, because the galaxies selected for MUSE-Faint are limited to those that can be covered in a few pointings with MUSE.
    It is not immediate how these selection effects influence our comparison to the theoretical scaling relations.
    To determine which values of $M_{200}$ are correct, the ability of CJAM and GravSphere to recover unbiased parameters for UFDs needs to be tested with mock galaxies of the appropriate mass and with realistic numbers of measurements and measurement uncertainties.
    Meanwhile, to properly compare the results against theory, the scaling relations need to be extended with the results from high-resolution zoom-in simulations and need to take into account the observational selection functions in this regime.
    Both endeavours are beyond the scope of this paper.

    The circular velocity $V(r_{1/2})$, by contrast, is derived from the mass inside the three-dimensional half-light radius, which is included in our data ranges.
    Therefore it should not be affected by a possible bias in the slope of the profile.
    We have confirmed this through the consistent circular velocities obtained with CJAM and GravSphere.
    As $V(r_{1/2})$, not $V_\mathrm{max}$, is used as input for the procedure of \citet{Fattahi-2018-MNRAS-476-3816} to determine the amount of tidal stripping, the conclusions about stripping are similarly not significantly dependent on the modelling tool.

    Because the assumed stellar mass-to-light ratios are derived from rough isochrone fits, there is an additional uncertainty for the stellar mass not included in the calculations.
    A change in stellar mass-to-light ratio by a factor of almost 2 did not significantly change the derived stripping parameters for Ant B, therefore this does not seem to be a major source of uncertainty for these results.
    This can be understood by the steepness of the $V_\mathrm{max}^\mathrm{prog}$--$M_*^\mathrm{prog}$ relation, which means that even large changes in $M_*$ do not bring the galaxy much closer to or further from the relation, therefore the inferred stripping will not be sensitive to $M_*$.

\subsection{Reliability of the dynamical analysis}
\label{ssec:reldyn}
    In the dynamical analysis we have implicitly assumed that the galaxies are in dynamical equilibrium.
    Though our results indicate that most, if not all, galaxies have not been significantly stripped of dark matter or stars, this does not prove dynamical equilibrium, as these results are obtained on the basis of this assumption.
    Regardless, it seems unlikely that galaxies on their first infall into the Milky Way, at ${>}100\,\mathrm{kpc}$ distance, should have undergone significant stripping.
    To be more concrete, we estimate what the Jacobi radius~\citep[e.g.][their Sect.~8.3.1]{Binney-2008-GD-PU-2} would be if Gru~1 would approach the Milky Way to $D_\mathrm{min} \approx 10\,\mathrm{kpc}$, which is the lower bound on the possible closest approach for any of our galaxies.
    Following \citet{Simon-2011-ApJ-733-46}, we can estimate the Jacobi radius by taking the circular velocity of the Milky Way as $V_\mathrm{MW} = 220\,\mathrm{km}\,\mathrm{s}^{-1}$.
    As the stripping parameter~$x$ measures whether the galaxy has been stripped of dark matter within its three-dimensional half-light radius, we take our Gru~1 measurement $M(r_{1/2}) \approx 10^{6.38}\,M_\sun$ as the satellite mass.
    The Jacobi radius is then
    \begin{equation}
        r_\mathrm{J} \approx \Biggl(\frac{M(r_{1/2})}{3V_\mathrm{MW}^2D_\mathrm{min}/G}\Biggr)^{1/3} D_\mathrm{min} \approx 200\,\mathrm{pc}.
    \end{equation}
    This is approximately equal to the three-dimensional half-light radius of Gru~1, $r_{1/2} = (4/3)R_{1/2} = 201^{+28}_{-41}\,\mathrm{pc}$~\citep{2021ApJ...916...81C}.
    It is therefore plausible that Gru~1 could have had its dark matter tidally stripped if the Milky Way follows the heavier model with $M_{200} = 1.6 \times 10^{12}\,M_\sun$.
    Its larger ellipticity may be a hint of past tidal interaction.
    \citet{2018arXiv180902259J} observed substructure in Gru~1, but this was aligned with the Large Magellanic Cloud.
    However, Gru~1 has not been close to the Large Magellanic Cloud  according to \citet{2021arXiv210608819B}.
    On the other hand, in case of the light Milky Way model, where $D_\mathrm{min} \approx 100\,\mathrm{kpc}$, $r_\mathrm{J} \approx 900\,\mathrm{pc}$, which clearly rules out a significant tidal stripping within $r_{1/2}$.
    As the other four galaxies have $D_\mathrm{min} \gg 10\,\mathrm{kpc}$, we do not expect to see significant tidal stripping within $r_{1/2}$ for any of these galaxies.

    The presence of binary star systems in the UFDs studied has the potential to bias the inferred intrinsic velocity dispersion.
    The barycentre of a binary star system traces the gravitational potential of the dark-matter halo of its host galaxy.
    However, when we measure the line-of-sight velocity of a star in a binary system, we see an additional contribution from the orbit of the star around the barycentre.
    By combining exposures from multiple epochs we effectively average over the velocity variation and obtain broadened spectral features with a mean at the barycentric velocity.
    For most of the galaxies presented here, we have collected data over periods of the order of a year or longer, which should mitigate the effects of binary stars with the shortest periods and consequently the largest velocities.
    Ant~B is an exception; the data pertaining to this system was obtained over a few months.
    Though through the multi-epoch observations we attempt to address the worst effects of binary stars on the velocity dispersion, there is still little known about the populations of binary stars in UFDs, and therefore these stars remain a source of uncertainty.

\section{Conclusions}
\label{sec:conclusions}
    We presented new observations of four (ultra-)faint dwarf galaxies, Ant~B, Leo~T, Hya~II, and Gru~1, from the MUSE-Faint survey.
    We extracted stellar line-of-sight velocities, supplemented by literature velocities when available, and separated member stars from other sources.
    We resolved the velocity dispersion of Gru~1, which supports its classification as a galaxy.
    Through dynamical modelling with CJAM and GravSphere we constrained the dark-matter density profiles for these four galaxies for the first time.
    We used and compared cusp, core, and soliton models with CJAM and compared these with the cusp and core+tides models of GravSphere to determine whether our results are sensitive to the choice of modelling tool.
    We derived dynamical masses, concentrations, circular velocities, and astrophysical $J$ and $D$ factors from the profiles.
    We supplemented our galaxy sample with Eri~2, for which the CJAM dynamical modelling was done in \citetalias{Zoutendijk-2021-A&A-651-A80} of this series.
    Using the full sample of five galaxies, we furthermore compared the derived galaxy properties to expectations from theoretical scaling relations and in that process determined how much tidal stripping these galaxies could have incurred.

    We constrained the core radii of the four newly presented galaxies to $r_\mathrm{c} < 66$--$95\,\mathrm{pc}$ (68\% confidence limit) or $r_\mathrm{c} < 200$--$309\,\mathrm{pc}$ (95\% confidence limit).
    We constrained their soliton radii to $r_\mathrm{sol} < 13$--$112\,\mathrm{pc}$ (68\% confidence limit) or $r_\mathrm{sol} < 0.25$--$1.58\,\mathrm{kpc}$ (95\% confidence limit).
    These limits rule out cores of the size $r_\mathrm{c} \approx 500\,\mathrm{pc}$ as observed in more massive dwarf galaxies.

    We find substantial Bayesian evidence (Bayes factor $10^{-0.53}$) against the core model for Leo~T.
    The most preferred model for all galaxies is the soliton model.
    However, the evidence against the other models is weak, except for the aforementioned Leo~T result, and we have insufficient evidence to decisively discriminate between the models for the analysed galaxies.

    We find the highest values of $M_{200} = 4.2^{+33.9}_{-3.3} \times 10^8$--$1.5^{+18.6}_{-1.3} \times 10^9\,M_\sun$ and $V_\mathrm{max} = 24^{+18}_{-11}$--$31^{+29}_{-12}\,\mathrm{km}\,\mathrm{s}^{-1}$ for Hya~II, while this is one of the fainter galaxies.
    The astrophysical factors $\log_{10}(J(\alpha_\mathrm{c}^J)/M_\sun^2\,\mathrm{kpc}^{-5}) = 11.91^{+0.44}_{-0.42}$--$12.02^{+0.50}_{-0.49}$ and $\log_{10}(D(\alpha_\mathrm{c}^D)/M_\sun\,\mathrm{kpc}^{-2}) = 3.17^{+0.51}_{-0.37}$--$3.45^{+0.60}_{-0.44}$ are equivalently high for Hya~II, which makes this an interesting target for studies that search for annihilation or decay signals of dark-matter particles.

    We find that according to the theoretical expectations for isolated galaxy evolution and evolution after infall into the Milky Way halo, the galaxies are consistent with not having been significantly tidally stripping of dark matter within their half-light radii.
    This is consistent with the large distance between these satellites and the Milky Way, now and in the past.
    Only Gru~1 may have come close to the Milky Way, and is also indicated as the most stripped galaxy.
    The $V(r_{1/2})$ that these calculations are based on, are consistent between CJAM and GravSphere, as are the derived stripping parameters.

    All five galaxies have a larger $c_{200}$ than predicted for their $M_{200}$ according to CJAM.
    The $M_{200}$ is lower than expected given the $M_*$ of these galaxies.
    GravSphere does produce results consistent with the theoretical scaling relations.
    In this case, the UFDs are consistent with $M_{200} \sim 10^9\,M_\sun$, which is expected from models in which the smallest dwarf galaxies are re-ionization fossils.
    The $M_{200}$ and $c_{200}$ are easily biased, because their calculations rely on a large extrapolation of the density profiles to larger radii.
    This is also apparent in the different results produced by CJAM and GravSphere.
    Expectations can, however, turn out to be false, therefore we cannot conclude from these results that GravSphere gives more correct results than CJAM.
    Further tests are required to determine which of the dynamical modelling tools provides the least biased results.

    The determination of dark-matter density profiles for UFDs has only just begun.
    MUSE-Faint has so far provided us with velocity measurements in the centres of five faint and ultra-faint systems.
    With the current generation of instruments, improving on the measurements used in this paper will be expensive.
    Increasing the number of stars in the centres could lead to stronger constraints on cores and solitons, but will require a significant time investment.
    Higher-resolution spectroscopy of the central stars is also impractical with current instruments because of the crowdedness of these fields.
    If a future instrument could marry the spatial resolution of an integral-field spectrograph with the specral resolution of a fibre spectrograph, this would greatly improve the measurement uncertainties on the line-of-sight velocities.
    On the other hand, larger samples of stars at larger radii, potentially improving the constraints on the virial mass, are difficult to realize because the stellar surface density rapidly decreases outside the half-light radius.
    There is, however, a large number of UFDs with centres still unexplored with spectroscopy.
    Increasing the sample of UFDs with dark-matter density profiles will offer a more complete view of the properties of the faintest galaxies and how these relate to theory.
    We also hope that, with the increasing resolution of galaxy formation simulations, the theoretical scaling relations can be expanded into the ultra-faint regime.
    To properly compare the virial parameters of UFDs against such scaling relations, tests of different dynamical modelling tools against mock UFDs are required to determine their biases.
    Finally, in a follow-up paper we will study the consequences of the density profiles presented here for the nature and properties of dark matter.

    \begin{acknowledgements}
        SLZ wishes to thank Azadeh Fattahi and Matthieu Schaller for helpful discussions.

        SLZ acknowledges support by The Netherlands Organisation for Scientific Research~(NWO) through a TOP Grant Module~1 under project number 614.001.652.
        JB acknowledges support by Funda\c{c}\~{a}o para a Ci\^{e}ncia e a Tecnologia~(FCT) through the research grants UIDB/04434/2020 and UIDP/04434/2020, through the Investigador FCT Contract No.\ IF/01654/2014/CP1215/CT0003, and through FCT project PTDC/FIS-AST/4862/2020.

        Based on observations made with ESO Telescopes at the La Silla Paranal Observatory under programme IDs 0100.D-0807, 0101.D-0300, 0102.D-0372, 0103.D-0705, and 0104.D-0199.

        This research has made use of Astropy~\citep{Robitaille-2013-A&A-558-A33, AstropyCollaboration-2018-AJ-156-123}, corner.py~\citep{ForemanMackey-2016-JOSS-1-24}, matplotlib~\citep{Hunter-2007-CSE-9-90}, NASA's Astrophysics Data System Bibliographic Services, NumPy~\citep{Harris-2020-Natur-585-357}, SciPy~\citep{2020NatMe..17..261V}, and the colour schemes of \citet{tolcolor}.

        This work has made use of data from the European Space Agency (ESA) mission
        {\it Gaia} (\url{https://www.cosmos.esa.int/gaia}), processed by the {\it Gaia}
        Data Processing and Analysis Consortium (DPAC,
        \url{https://www.cosmos.esa.int/web/gaia/dpac/consortium}). Funding for the DPAC
        has been provided by national institutions, in particular the institutions
        participating in the {\it Gaia} Multilateral Agreement.

        Based on observations made with the NASA/ESA Hubble Space Telescope, and obtained from the Hubble Legacy Archive, which is a collaboration between the Space Telescope Science Institute (STScI/NASA), the Space Telescope European Coordinating Facility (ST-ECF/ESA) and the Canadian Astronomy Data Centre (CADC/NRC/CSA).

        This research has made use of the NASA/IPAC Extragalactic Database~(NED), which is funded by the National Aeronautics and Space Administration and operated by the California Institute of Technology.
    \end{acknowledgements}

    \bibliographystyle{aa}
    \bibliography{Zoutendijk_AntB-LeoT-HyaII-Gru1-profiles}

\appendix
\section{GravSphere}
\label{app:gravsphere}
    To test the robustness of our fiducial CJAM results against the choice of dynamical modelling tool, we present here the results of an alternative tool, GravSphere.
    GravSphere differs from CJAM in that it bins the velocity measurements and breaks the resulting degeneracy between density and anisotropy by also constraining higher velocity moments in the form of the virial shape parameters.
    It also allows for a radially varying velocity anisotropy profile.

    The presentation of the GravSphere results follows a similar structure as for CJAM.
    In Figs.~\ref{fig:gsnfwcornerAntB}--\ref{fig:gscorenfwcornerGru1} we display the posteriors of the GravSphere models on the dark-matter parameters.
    We do not show the Plummer parameters of the stellar distribution to limit the size of the figures.
    $M_{200}$ and $c_{200}$, or $M_{200}^\mathrm{prog}$ and $c_{200}^\mathrm{prog}$, are well-constrained to values of ${\sim}10^9\,M_\sun$ and $17$--$19$, respectively, for most galaxies.
    Hya~II is noticeably more massive (${\sim}10^{11}\,M_\sun$) and less concentrated ($12$--$13$).
    The core and tide parameters $r_\mathrm{c}$, $n$, $r_\mathrm{t}$, and $\delta$ are unconstrained and consistent with a cuspy profile, though core sizes larger than $1\,\mathrm{kpc}$ are allowed by the 95\% upper limits for all galaxies.
    For most galaxies the anisotropy profile in unconstrained and consistent with isotropy.
    Ant~B has a significant tangential anisotropy in its centre and Hya~II has a non-significant preference for tangential anisotropy at all radii.
    There is no visible correlation between the anisotropy and the other parameters, therefore the anisotropy profiles should not influence our results.
    This also supports our choice for isotropy in the CJAM modelling: anisotropy is not expected to change the CJAM results.
    The central tangential anisotropy of Ant~B will be further studied by J\'ulio et~al.~\citetext{in prep.}.

\begin{figure*}
        \includegraphics[width=\linewidth]{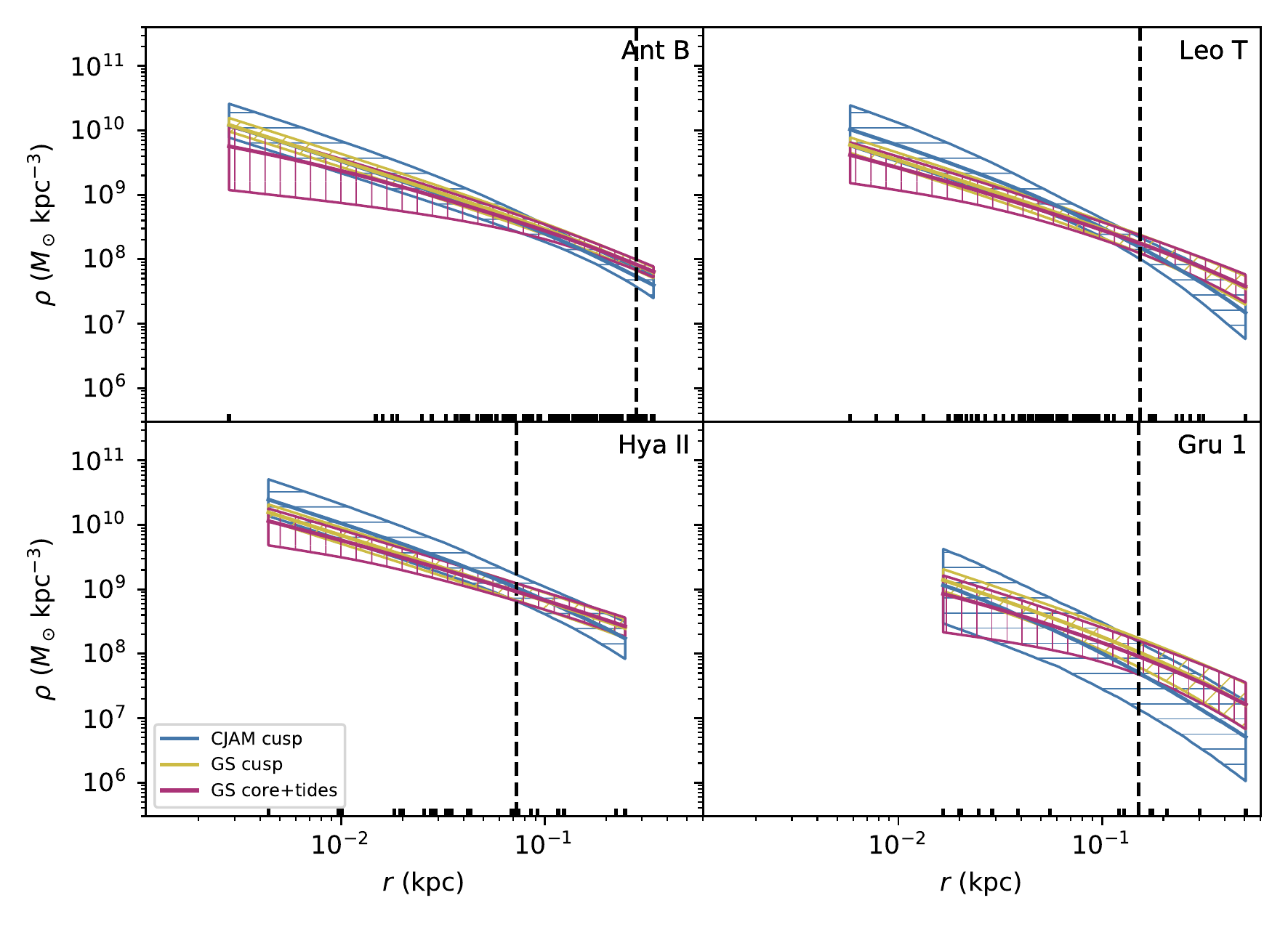}
        \caption{%
            Recovered dark-matter density profiles of the (ultra-)faint dwarf galaxies Antlia~B, Leo~T, Hydra~II, and Grus~1, for cusp and core+tides profiles, modelled with GravSphere, with the CJAM cusp profiles for reference.
            The hatched bands indicate the 68\% confidence interval on the density at each radius.
            The central, thicker line is the median density at each radius.
            The galaxy's projected half-light radii are indicated with the vertical dashed lines.
            The markers along the bottom of each panel indicate the projected radii of the kinematic tracers.%
        }
        \label{fig:gsrecovery}
    \end{figure*}

\begin{figure}
        \includegraphics[width=\linewidth]{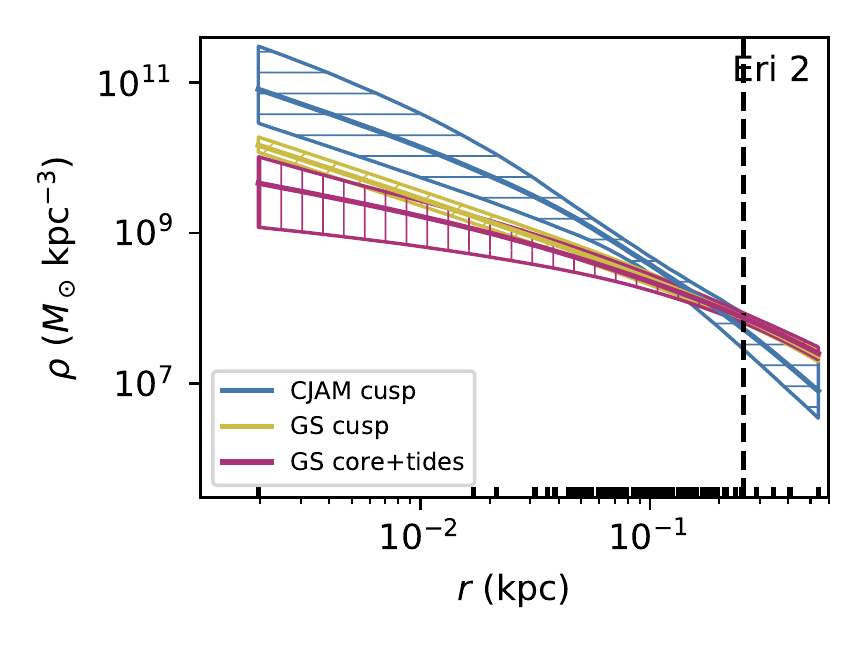}
        \caption{%
            Recovered dark-matter density profiles of the ultra-faint dwarf galaxy Eridanus~2, for cusp and core+tides profiles, modelled with GravSphere, with the CJAM cusp profile for reference.
            The hatched bands indicate the 68\% confidence interval on the density at each radius.
            The central, thicker line is the median density at each radius.
            The galaxy's projected half-light radii are indicated with the vertical dashed lines.
            The markers along the bottom of each panel indicate the projected radii of the kinematic tracers.%
        }
        \label{fig:gsrecoveryEri2}
    \end{figure}

    From the posteriors we derive the profiles shown in Figs.~\ref{fig:gsrecovery} and~\ref{fig:gsrecoveryEri2} together with the CJAM cuspy profile for reference.
    The GravSphere profiles show generally the same features: the profile uncertainties are smallest and the agreement between the profiles is the best where the tracer density is highest, and the divergence between the profiles increases towards the centre.
    There are, however, two noticeable differences.
    The GravSphere profiles have smaller uncertainties than the CJAM profiles, with the difference being particularly strong between the two cusp models.
    Secondly, GravSphere systematically prefers shallower profiles.

    We derive the same quantities from the GravSphere profiles as from the CJAM profiles and list these in Table~\ref{tab:gsm200}.
    \begin{sidewaystable}
        \caption{Parameters derived from the recovered dark-matter density profiles from GravSphere.}
        \label{tab:gsm200}
        \centering
        \begin{tabular}{lc@{\hspace{3pt}}cc@{\hspace{3pt}}cc@{\hspace{3pt}}cc@{\hspace{3pt}}cc@{\hspace{3pt}}c}
            \hline
            \hline
                                                                             & \multicolumn{2}{c}{Ant~B}                       & \multicolumn{2}{c}{Leo~T}                         & \multicolumn{2}{c}{Eri~2}                         & \multicolumn{2}{c}{Hya~II}                        & \multicolumn{2}{c}{Gru~1}                         \\
                                                                             & cusp                   & core+tides             & cusp                    & core+tides              & cusp                    & core+tides              & cusp                    & core+tides              & cusp                    & core+tides              \\
            \hline
            $\log_{10}(V_\mathrm{max}/\mathrm{km}\,\mathrm{s}^{-1})$         & $1.36^{+0.10}_{-0.08}$ & $1.38^{+0.10}_{-0.08}$ &  $1.38^{+0.19}_{-0.18}$ &  $1.39^{+0.17}_{-0.17}$ &  $1.28^{+0.07}_{-0.06}$ &  $1.30^{+0.07}_{-0.06}$ &  $1.85^{+0.27}_{-0.24}$ &  $1.78^{+0.20}_{-0.18}$ &  $1.15^{+0.23}_{-0.19}$ &  $1.17^{+0.25}_{-0.20}$ \\
            $\log_{10}(r_{200}/\mathrm{kpc})$                                & $1.38^{+0.13}_{-0.11}$ & $1.30^{+0.14}_{-0.12}$ &  $1.40^{+0.23}_{-0.20}$ &  $1.30^{+0.19}_{-0.17}$ &  $1.30^{+0.09}_{-0.08}$ &  $1.24^{+0.10}_{-0.10}$ &  $1.92^{+0.30}_{-0.28}$ &  $1.62^{+0.25}_{-0.21}$ &  $1.16^{+0.26}_{-0.21}$ &  $1.12^{+0.22}_{-0.20}$ \\
            $\log_{10}(c_{200})$                                             & $1.26^{+0.12}_{-0.12}$ & $1.14^{+0.17}_{-0.22}$ &  $1.24^{+0.14}_{-0.15}$ &  $1.15^{+0.19}_{-0.27}$ &  $1.26^{+0.11}_{-0.10}$ &  $1.18^{+0.14}_{-0.19}$ &  $1.15^{+0.22}_{-0.16}$ &  $0.81^{+0.32}_{-0.47}$ &  $1.29^{+0.13}_{-0.14}$ &  $1.23^{+0.16}_{-0.23}$ \\
            $\log_{10}(M_{200}/M_\sun)$                                     & $9.17^{+0.38}_{-0.33}$ & $8.92^{+0.42}_{-0.36}$ &  $9.23^{+0.68}_{-0.61}$ &  $8.93^{+0.57}_{-0.50}$ &  $8.92^{+0.27}_{-0.24}$ &  $8.73^{+0.31}_{-0.28}$ & $10.78^{+0.91}_{-0.83}$ &  $9.90^{+0.74}_{-0.62}$ &  $8.51^{+0.78}_{-0.62}$ &  $8.40^{+0.66}_{-0.59}$ \\
            $\log_{10}(\Upsilon(r_{200})/M_\sun\,L_\sun^{-1})$             & $3.37^{+0.38}_{-0.33}$ & $3.12^{+0.42}_{-0.36}$ &  $4.27^{+0.68}_{-0.61}$ &  $3.97^{+0.57}_{-0.50}$ &  $4.16^{+0.27}_{-0.24}$ &  $3.96^{+0.31}_{-0.28}$ &  $6.91^{+0.91}_{-0.83}$ &  $6.03^{+0.74}_{-0.62}$ &  $4.95^{+0.78}_{-0.62}$ &  $4.83^{+0.66}_{-0.59}$ \\
            $\log_{10}(V(R_{1/2})/\mathrm{km}\,\mathrm{s}^{-1})$             & $1.16^{+0.04}_{-0.04}$ & $1.15^{+0.04}_{-0.05}$ &  $1.05^{+0.06}_{-0.07}$ &  $1.04^{+0.06}_{-0.07}$ &  $1.08^{+0.04}_{-0.03}$ &  $1.07^{+0.04}_{-0.04}$ &  $1.07^{+0.06}_{-0.07}$ &  $1.06^{+0.07}_{-0.08}$ &  $0.94^{+0.10}_{-0.10}$ &  $0.89^{+0.12}_{-0.15}$ \\
            $\log_{10}(M(R_{1/2})/M_\sun)$                                  & $7.13^{+0.08}_{-0.08}$ & $7.11^{+0.08}_{-0.09}$ &  $6.65^{+0.12}_{-0.14}$ &  $6.62^{+0.12}_{-0.15}$ &  $6.92^{+0.07}_{-0.07}$ &  $6.90^{+0.08}_{-0.09}$ &  $6.36^{+0.12}_{-0.14}$ &  $6.34^{+0.13}_{-0.15}$ &  $6.42^{+0.20}_{-0.20}$ &  $6.32^{+0.25}_{-0.30}$ \\
            $\log_{10}(\Upsilon(R_{1/2})/M_\sun\,L_\sun^{-1})$             & $1.63^{+0.08}_{-0.08}$ & $1.61^{+0.08}_{-0.09}$ &  $1.99^{+0.12}_{-0.14}$ &  $1.96^{+0.12}_{-0.15}$ &  $2.46^{+0.07}_{-0.07}$ &  $2.44^{+0.08}_{-0.09}$ &  $2.79^{+0.12}_{-0.14}$ &  $2.78^{+0.13}_{-0.15}$ &  $3.16^{+0.20}_{-0.20}$ &  $3.06^{+0.25}_{-0.30}$ \\
            $\log_{10}(V(r_{1/2})/\mathrm{km}\,\mathrm{s}^{-1})$             & $1.20^{+0.04}_{-0.04}$ & $1.19^{+0.04}_{-0.04}$ &  $1.10^{+0.06}_{-0.07}$ &  $1.09^{+0.06}_{-0.07}$ &  $1.13^{+0.03}_{-0.03}$ &  $1.12^{+0.03}_{-0.04}$ &  $1.12^{+0.06}_{-0.07}$ &  $1.12^{+0.06}_{-0.07}$ &  $0.98^{+0.10}_{-0.11}$ &  $0.89^{+0.12}_{-0.15}$ \\
            $\log_{10}(M(r_{1/2})/M_\sun)$                                  & $7.34^{+0.08}_{-0.08}$ & $7.33^{+0.08}_{-0.09}$ &  $6.88^{+0.13}_{-0.15}$ &  $6.87^{+0.13}_{-0.15}$ &  $7.14^{+0.07}_{-0.06}$ &  $7.13^{+0.07}_{-0.07}$ &  $6.60^{+0.12}_{-0.14}$ &  $6.59^{+0.13}_{-0.15}$ &  $6.64^{+0.21}_{-0.22}$ &  $6.32^{+0.25}_{-0.30}$ \\
            $\log_{10}(J(\alpha_\mathrm{c}^J)/M_\sun^2\,\mathrm{kpc}^{-5})$ & $9.49^{+0.16}_{-0.17}$ & $9.48^{+0.16}_{-0.17}$ & $10.44^{+0.31}_{-0.35}$ & $10.44^{+0.31}_{-0.35}$ & $10.38^{+0.13}_{-0.13}$ & $10.37^{+0.12}_{-0.13}$ & $12.04^{+0.32}_{-0.34}$ & $12.07^{+0.31}_{-0.34}$ & $10.91^{+0.50}_{-0.49}$ & $10.82^{+0.55}_{-0.56}$ \\
            $\log_{10}(D(\alpha_\mathrm{c}^D)/M_\sun\,\mathrm{kpc}^{-2})$   & $1.90^{+0.21}_{-0.19}$ & $1.83^{+0.21}_{-0.17}$ &  $2.71^{+0.39}_{-0.37}$ &  $2.63^{+0.33}_{-0.32}$ &  $2.80^{+0.15}_{-0.13}$ &  $2.76^{+0.15}_{-0.12}$ &  $4.22^{+0.51}_{-0.48}$ &  $3.92^{+0.40}_{-0.35}$ &  $3.28^{+0.48}_{-0.41}$ &  $3.24^{+0.46}_{-0.42}$ \\
            \hline
        \end{tabular}
        \tablefoot{%
            $V_\mathrm{max}$: maximum circular velocity;
            $r_{200}$: virial radius;
            $c_{200}$: concentration parameter;
            $M_{200}$: virial mass;
            $\Upsilon(r)$: mass-to-light ratio integrated within radius~$r$;
            $V(r)$: circular velocity at radius~$r$;
            $R_{1/2}$: projected half-light radius;
            $M(r)$: mass within radius~$r$;
            $r_{1/2}$: three-dimensional half-light radius;
            $J(\alpha_\mathrm{c}^J)$: astrophysical $J$ factor within its critical angle;
            $D(\alpha_\mathrm{c}^D)$: astrophysical $D$ factor within its critical angle.%
        }
    \end{sidewaystable}
    The $M_{200}$ and $c_{200}$ values in this Table are calculated by integrating the profiles, as was done for CJAM, and are not taken from the set of profile parameters.
    For the core+tides model, the $M_{200}$ calculated here are clearly lower than the $M_{200}^\mathrm{prog}$ profile parameters that represent the virial mass of the progenitor profile.
    Hya~II has a significantly higher $M_{200}$ and $V_\mathrm{max}$ than the other galaxies, qualitatively similar to what we found with CJAM.
    The large-scale parameters are all noticeably higher than their CJAM counterparts, except for the lower $c_{200}$, whereas the parameters measured at or within the half-light radii are consistent between the two methods.

    To compare the two GravSphere models, we estimate the Bayesian evidence with MCEvidence~\citep{2017arXiv170403472H}, as we did in \citetalias{Zoutendijk-2021-A&A-651-A80} for the pyGravSphere models.
    We display the estimated evidence in Table~\ref{tab:gsmodcomp}.
    \begin{table*}
        \caption{Bayesian evidence and Bayes factors for GravSphere.}
        \label{tab:gsmodcomp}
        \centering
        \begin{tabular}{lc@{\hspace{3pt}}cc@{\hspace{3pt}}cc@{\hspace{3pt}}cc@{\hspace{3pt}}cc@{\hspace{3pt}}c}
            \hline
            \hline
                                 & \multicolumn{2}{c}{Ant~B} & \multicolumn{2}{c}{Leo~T} & \multicolumn{2}{c}{Eri~2} & \multicolumn{2}{c}{Hya~II} & \multicolumn{2}{c}{Gru~1} \\
                                 & cusp      & core+tides    & cusp      & core+tides    & cusp      & core+tides    & cusp      & core+tides     & cusp      & core+tides    \\
            \hline
            $\ln(Z)$             & $-287.43$ & $-288.88$     & $-291.89$ & $-293.63$     & $-391.59$ & $-392.78$     & $-295.97$ & $-297.48$      & $-269.83$ & $-271.11$     \\
            $\Delta\log_{10}(Z)$ & $0$       & $-0.62$       & $0$       & $-0.76$       & $0$       & $-0.52$       & $0$       & $-0.66$        & $0$       & $-0.56$       \\
            \hline
        \end{tabular}
        \tablefoot{%
            $\ln(Z)$: natural logarithm of the Bayesian evidence~$Z$;
            $\Delta\log_{10}(Z)$: decimal logarithm of the Bayes factor, or difference of the decimal logarithms of Bayesian evidence, relative to the best-fitting model of each galaxy.%
        }
    \end{table*}
    For all galaxies, the cusp model is substantially, but not significantly, preferred over the core+tides model.

\begin{figure*}
        \includegraphics[width=\linewidth]{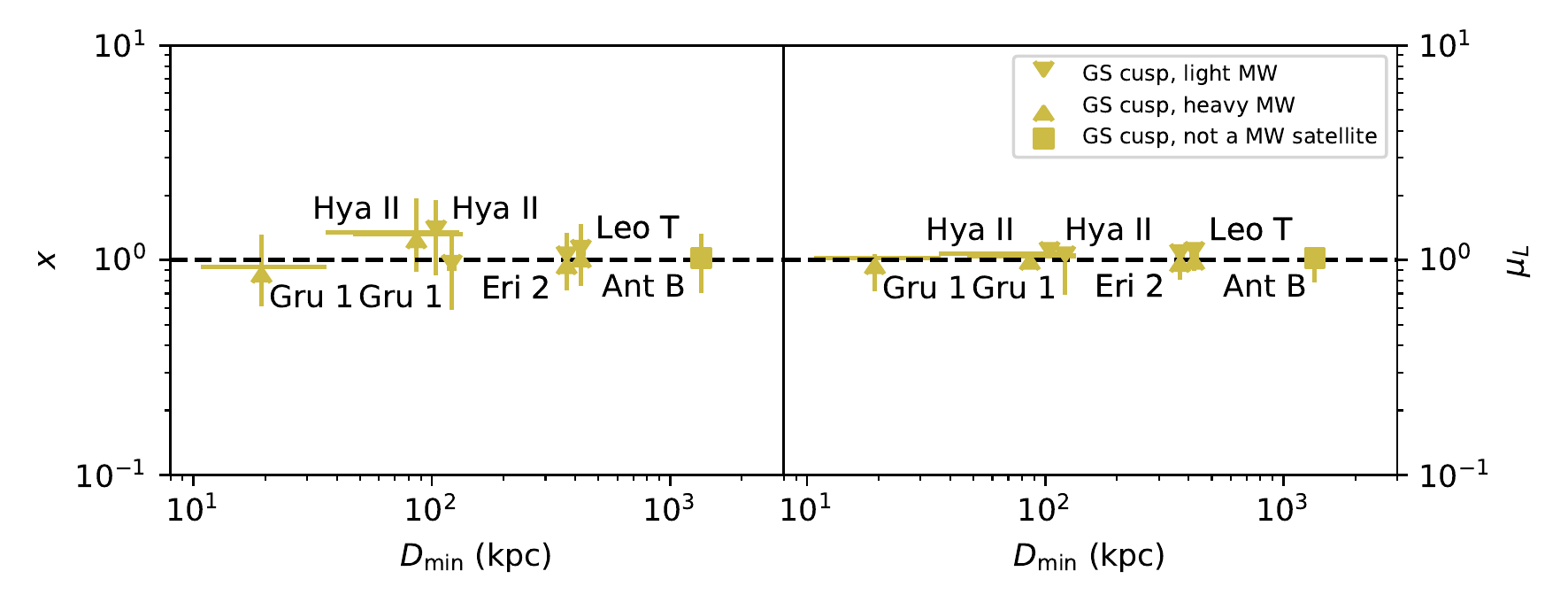}
        \caption{%
            Stripping parameters of the galaxies, as determined with GravSphere, versus their closest distance to the Milky Way ever reached.
            The stripping parameters $x$ and $\mu_L$ are measures of the fraction of dark matter and stars retained after stripping; lower values indicate more stripping has occurred.
            We show results for two models of the Milky Way that differ in the halo mass.
            In most cases, the closest distance is not the pericentre distance but the current distance, as most of these galaxies are still in their first infall.
            We note that Antlia~B is a satellite of NGC~3109, not of the Milky Way, and that its distance to the Milky Way is therefore not relevant for its stripping parameters; its display in this figure is merely illustrative.%
        }
        \label{fig:gsstripdist}
    \end{figure*}

    We perform the same stripping analysis on the cusp model of GravSphere as on that of CJAM.
    We are again successful in finding a progenitor solution for all five galaxies and display the progenitor properties in Table~\ref{tab:gsprog}.
    \begin{table*}
        \caption{Progenitor properties based on GravSphere, according to the method of \citet{Fattahi-2018-MNRAS-476-3816}.}
        \label{tab:gsprog}
        \centering
        \begin{tabular}{lccccc}
            \hline
            \hline
                                                                         & Ant~B                   & Leo~T                   & Eri~2                   & Hya~II                  & Gru~1                   \\
            \hline
            $\log_{10}(M_*^\mathrm{prog}/M_\sun)$                                 &  $5.87^{+0.10}_{-0.00}$ &  $5.14^{+0.07}_{-0.00}$ &  $4.95^{+0.10}_{-0.01}$ &  $4.28^{+0.00}_{-0.00}$ &  $3.80^{+0.16}_{-0.02}$ \\
            $\log_{10}(V_\mathrm{max}^\mathrm{prog}/\mathrm{km}\,\mathrm{s}^{-1})$ &  $1.45^{+0.09}_{-0.07}$ &  $1.40^{+0.07}_{-0.06}$ &  $1.39^{+0.07}_{-0.05}$ &  $1.35^{+0.06}_{-0.04}$ &  $1.33^{+0.05}_{-0.04}$ \\
            $\log_{10}(r_{1/2}^\mathrm{prog}/\mathrm{kpc})$                          & $-0.42^{+0.06}_{-0.06}$ & $-0.67^{+0.06}_{-0.06}$ & $-0.48^{+0.05}_{-0.06}$ & $-0.97^{+0.08}_{-0.07}$ & $-0.70^{+0.06}_{-0.06}$ \\
            $\log_{10}(V_\mathrm{prog}(r_{1/2}^\mathrm{prog})/\mathrm{km}\,\mathrm{s}^{-1})$     &  $1.19^{+0.08}_{-0.07}$ &  $1.07^{+0.10}_{-0.07}$ &  $1.12^{+0.08}_{-0.06}$ &  $1.04^{+0.14}_{-0.12}$ &  $1.00^{+0.09}_{-0.07}$ \\
            $\log_{10}(x)$                                               &  $0.01^{+0.12}_{-0.15}$ &  $0.03^{+0.13}_{-0.15}$ &  $0.00^{+0.12}_{-0.14}$ &  $0.12^{+0.16}_{-0.19}$ & $-0.04^{+0.15}_{-0.19}$ \\
            $\log_{10}(\mu_L)$                                           &  $0.01^{+0.00}_{-0.10}$ &  $0.02^{+0.00}_{-0.07}$ &  $0.01^{+0.01}_{-0.10}$ &  $0.02^{+0.00}_{-0.00}$ &  $0.00^{+0.02}_{-0.16}$ \\
            \hline
        \end{tabular}
        \tablefoot{%
            $M_*^\mathrm{prog}$: progenitor stellar mass;
            $V_\mathrm{max}^\mathrm{prog}$: progenitor maximum circular velocity;
            $r_{1/2}^\mathrm{prog}$: progenitor three-dimensional half-light radius;
            $V_\mathrm{prog}(r_{1/2}^\mathrm{prog})$: progenitor circular velocity within the progenitor three-dimensional half-light radius;
            $x$: fraction of dark-matter mass within the progenitor three-dimensional half-light radius in the present-day galaxy, versus the dark-matter mass within the progenitor three-dimensional half-light radius in the progenitor galaxy;
            $\mu_L$: fraction of the stellar mass of the present-day galaxy, versus the stellar mass of the progenitor galaxy.%
        }
    \end{table*}
    We compare the stripping parameters against the closest distances to the Milky Way ever reached in Fig.~\ref{fig:gsstripdist}.
    This comparison paints a very consistent picture to the CJAM stripping analysis.
    All galaxies are again consistent with having undergone no stripping.
    The uncertainties for GravSphere are smaller than for CJAM, and in the case of GravSphere the Gru~1 results are no longer compatible with a more significant amount of stripping than the other satellites.

    As a result of the systematically different $M_{200}$ and $c_{200}$, the comparisons of the GravSphere results against the galaxy--halo scaling relations look very different than that of the CJAM results.
    These comparisons are show in Figs.~\ref{fig:c200M200} and~\ref{fig:MstarM200} and discussed in Sect.~\ref{ssec:scaling}.

\section{Figures of parameter constraints}
\label{app:corner}
\begin{figure*}
        \includegraphics[width=0.5\linewidth]{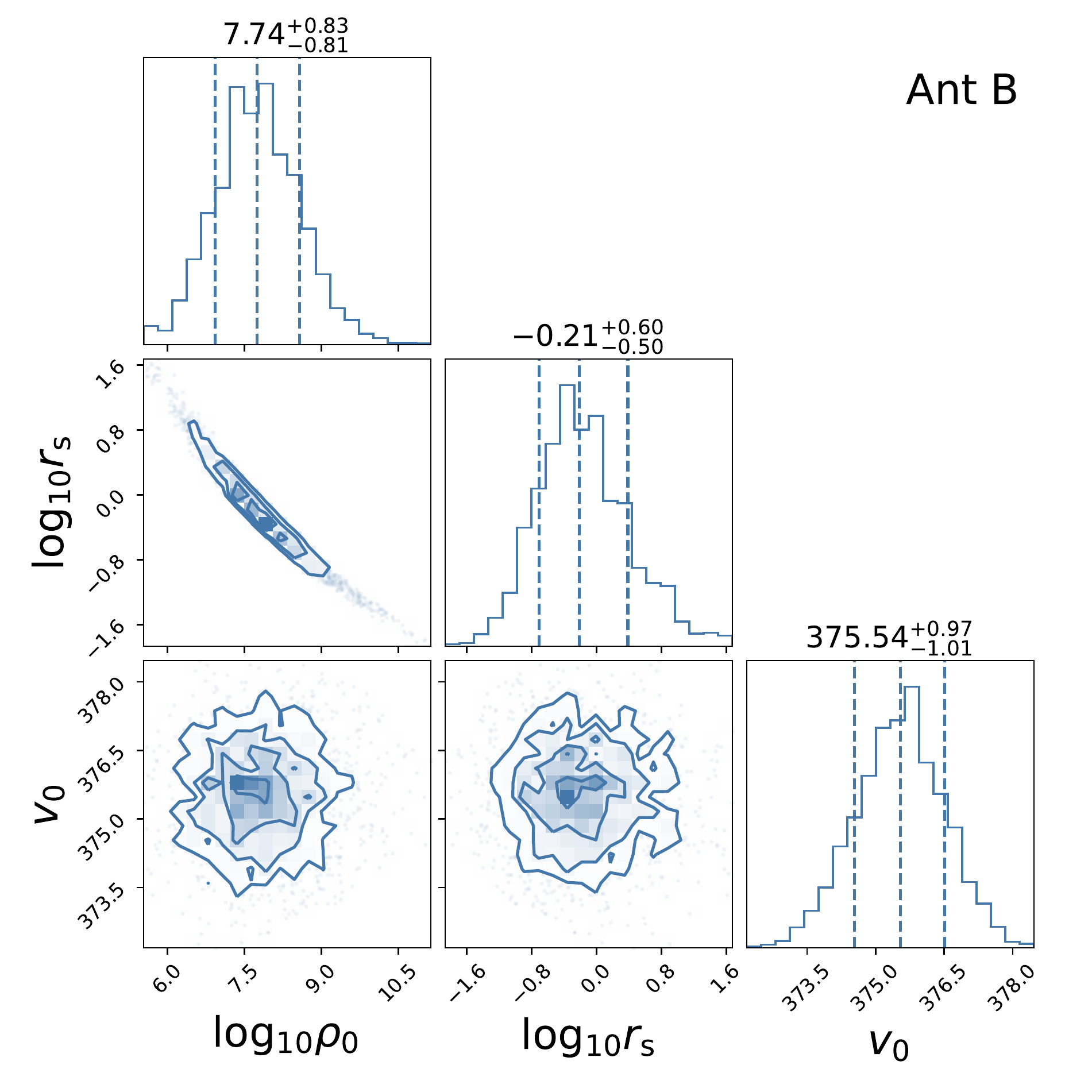}%
        \includegraphics[width=0.5\linewidth]{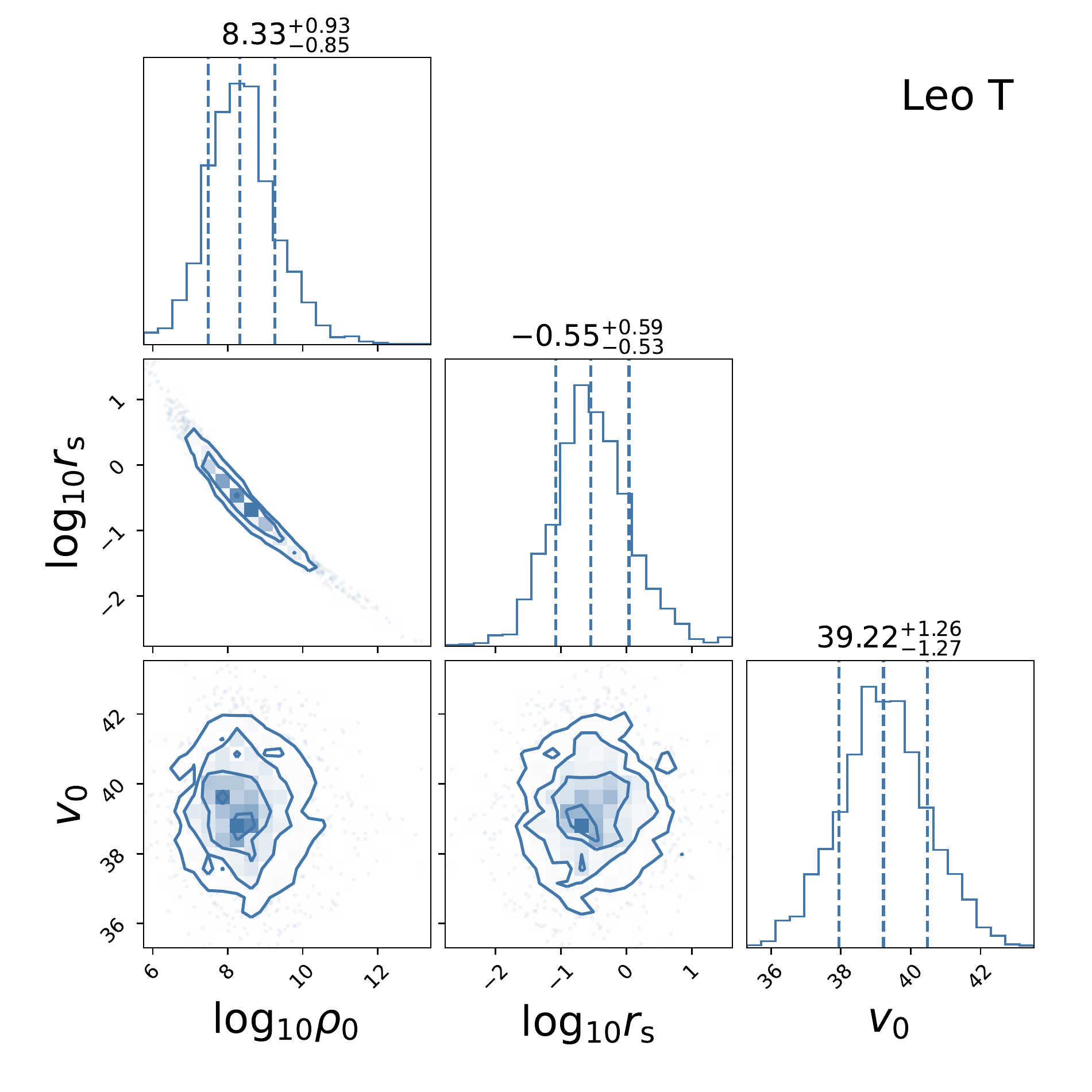}
        \includegraphics[width=0.5\linewidth]{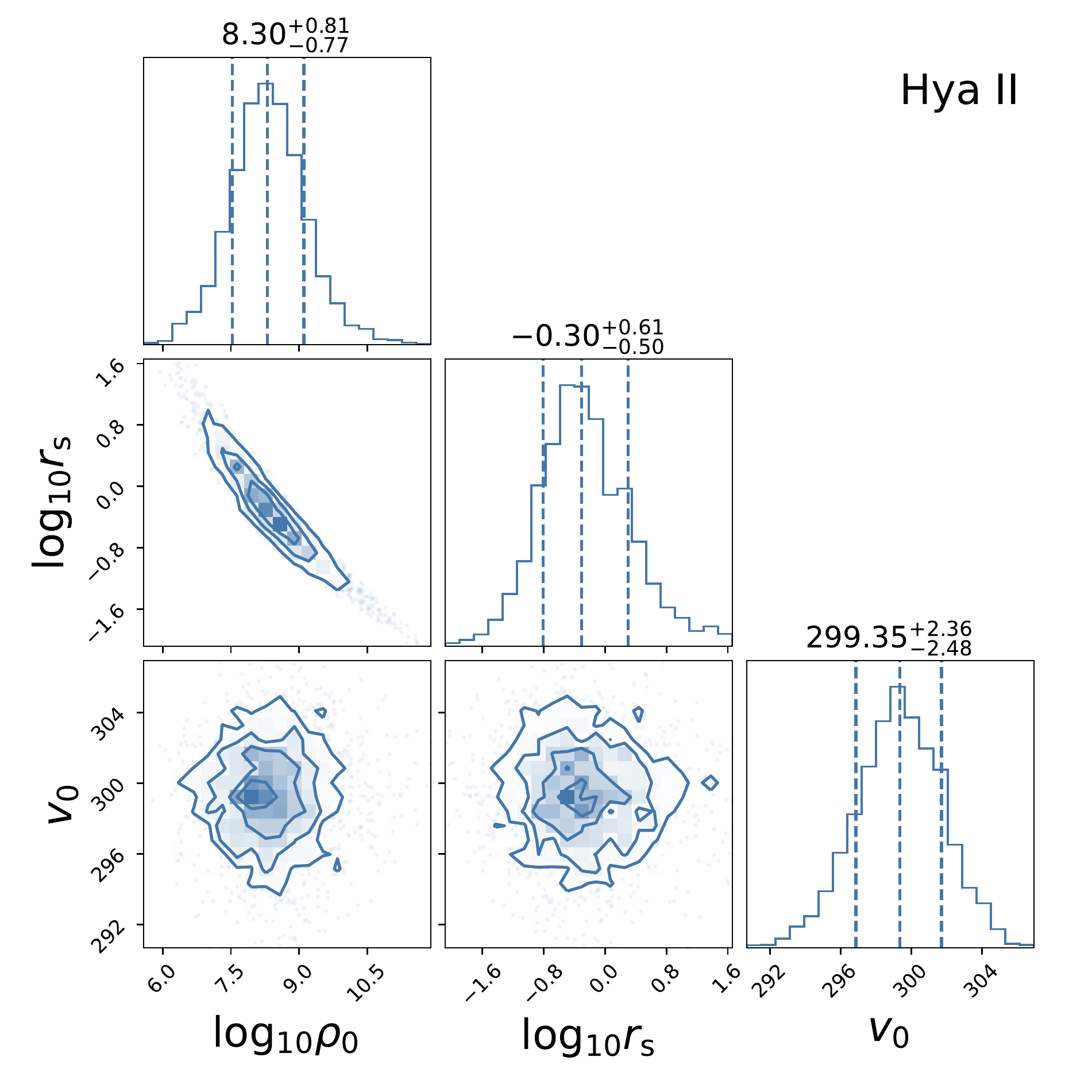}%
        \includegraphics[width=0.5\linewidth]{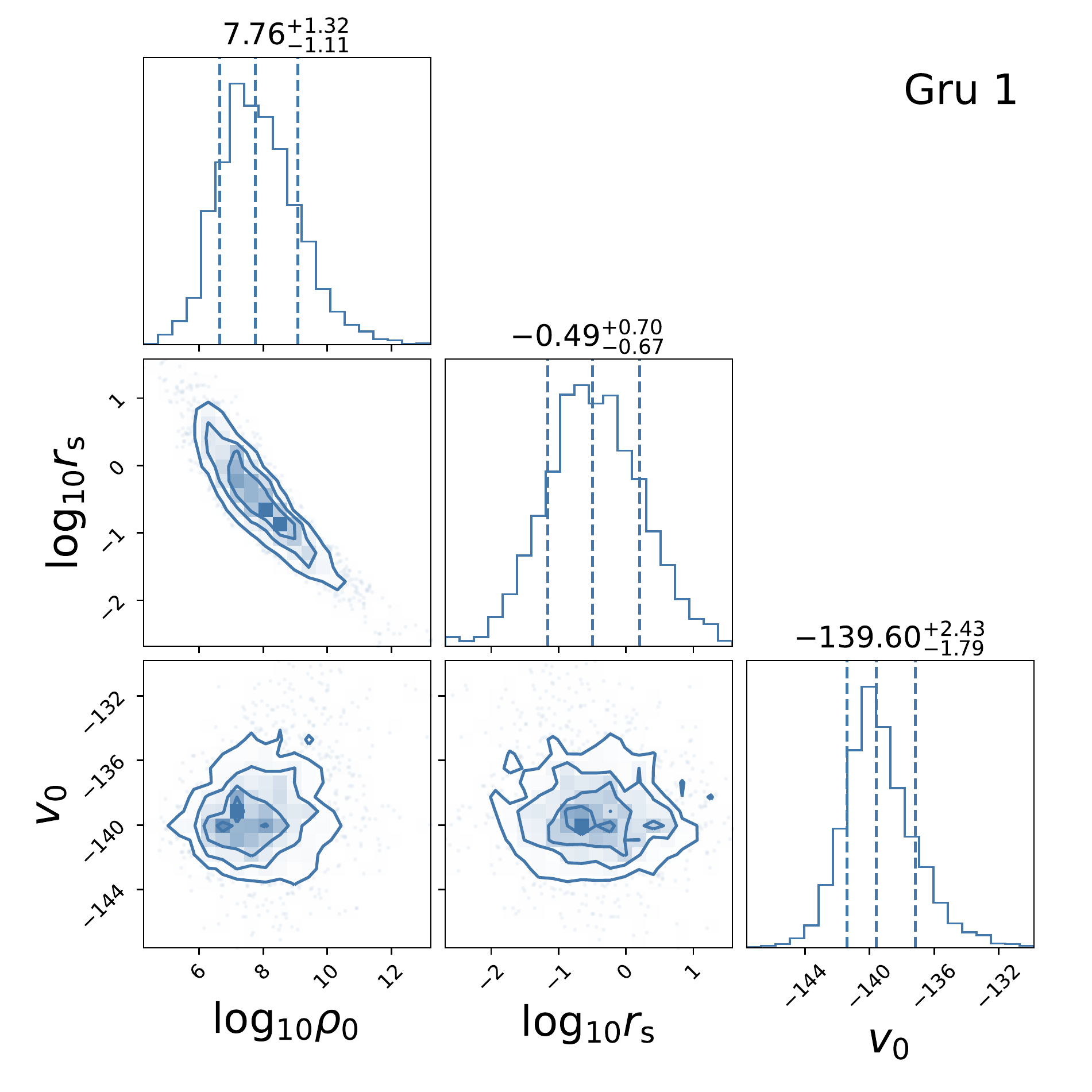}
        \caption{%
            Constraints on the dark-matter density profiles of the four (ultra-)faint dwarf galaxies, for the cusp model, modelled with CJAM.
            Units are omitted from the labels for clarity.
            The parameters are the characteristic dark-matter density~$\rho_0$ in $M_\sun\,\mathrm{kpc}^{-3}$, the scale radius~$r_\mathrm{s}$ in $\mathrm{kpc}$, and the systemic velocity~$v_0$ in $\mathrm{km}\,\mathrm{s}^{-1}$.
            The contours correspond to $0.5\sigma$, $1.0\sigma$, $1.5\sigma$, and $2.0\sigma$ confidence levels, where $\sigma$ is the standard deviation of a two-dimensional normal distribution.
            The vertical dashed lines in the one-dimensional histograms indicate the median and the 68\% confidence interval.%
        }
        \label{fig:cdmaltcorner}
    \end{figure*}

\begin{figure*}
        \includegraphics[width=0.5\linewidth]{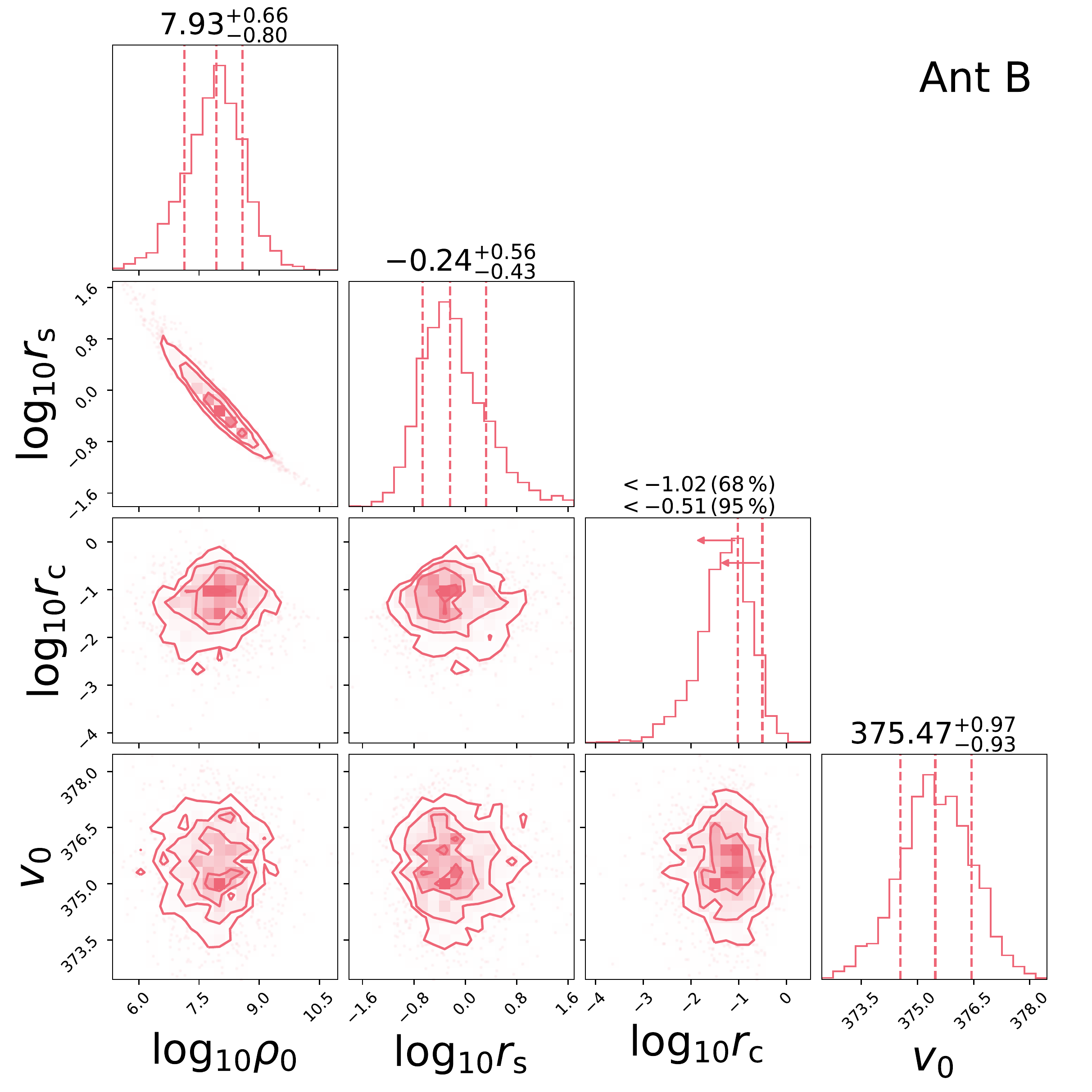}%
        \includegraphics[width=0.5\linewidth]{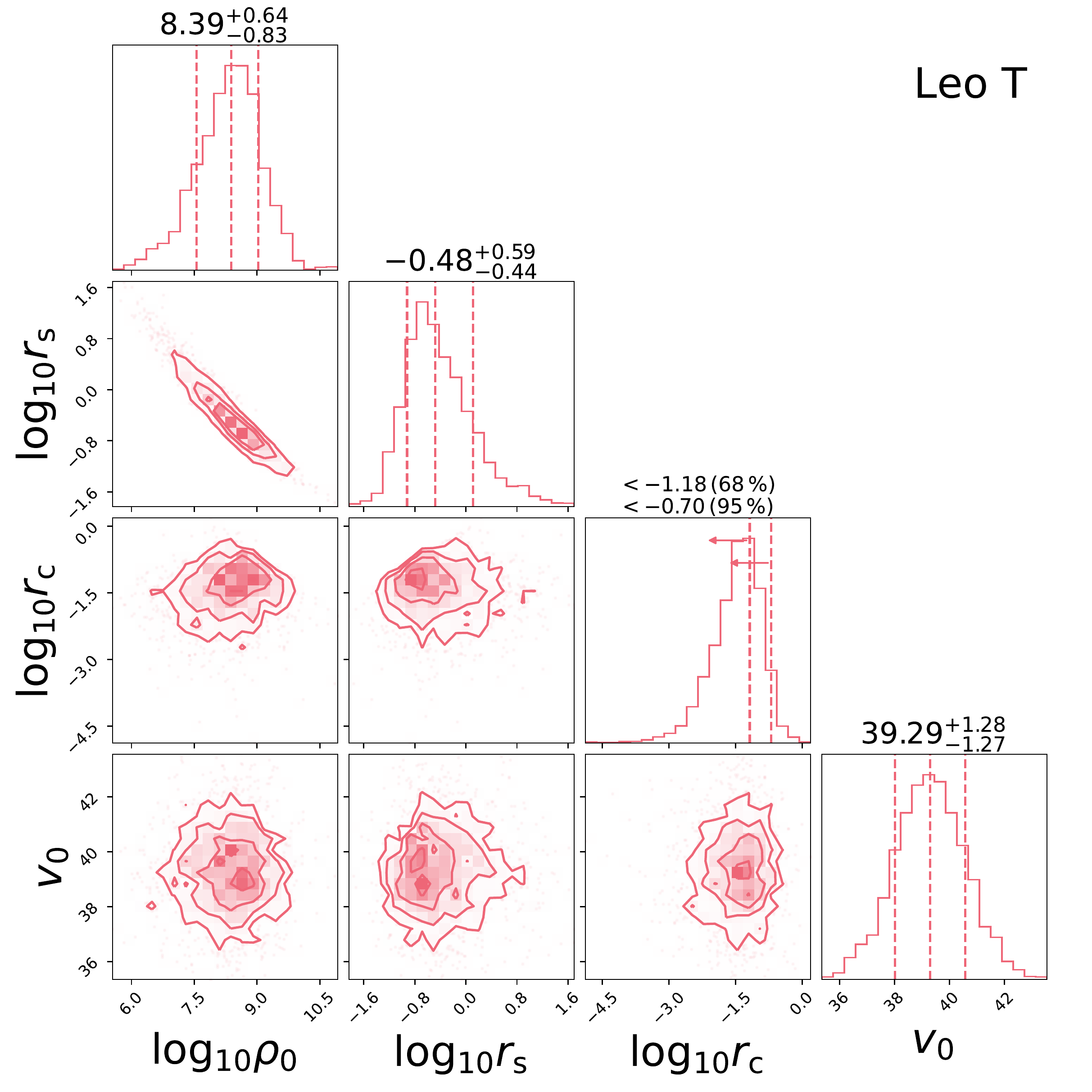}
        \includegraphics[width=0.5\linewidth]{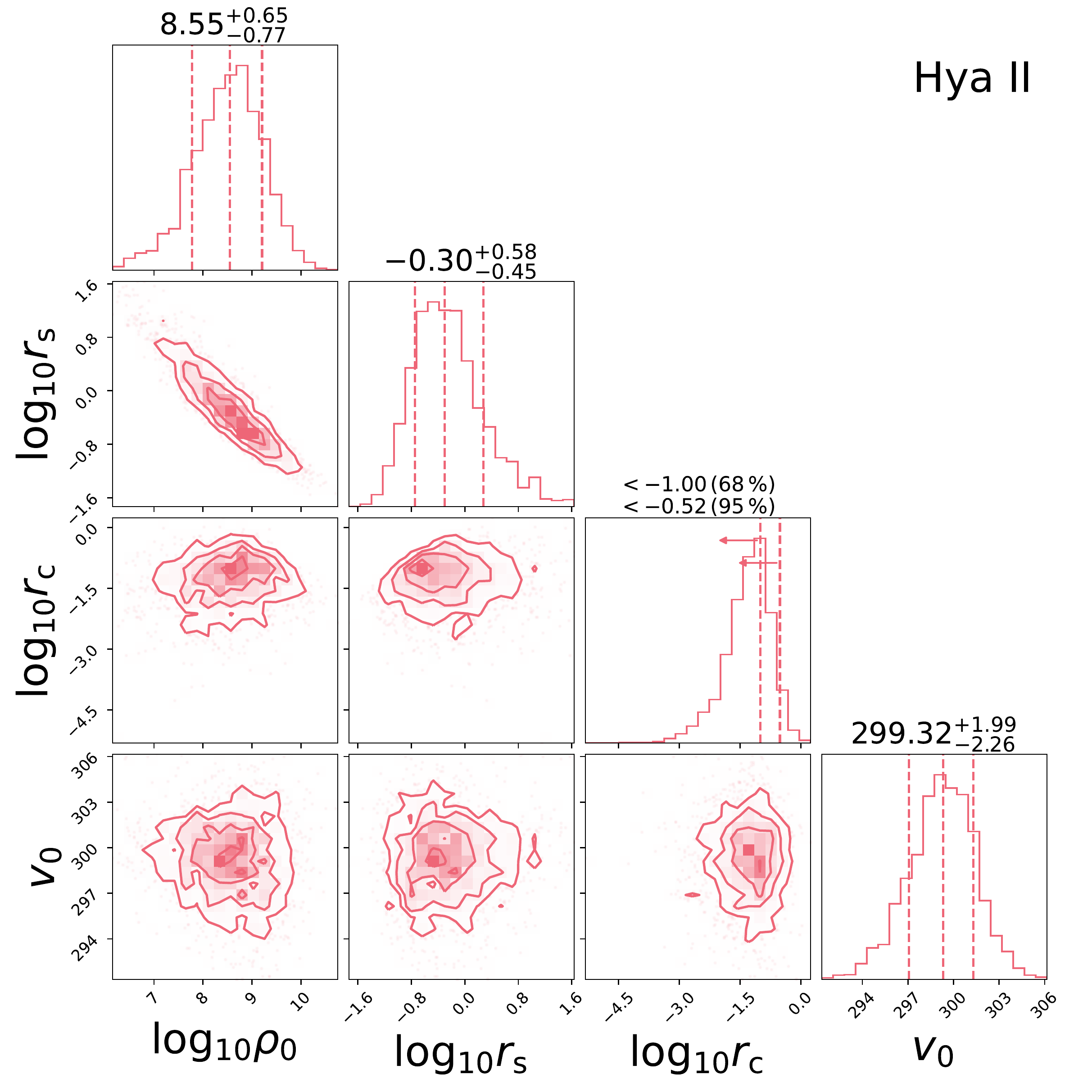}%
        \includegraphics[width=0.5\linewidth]{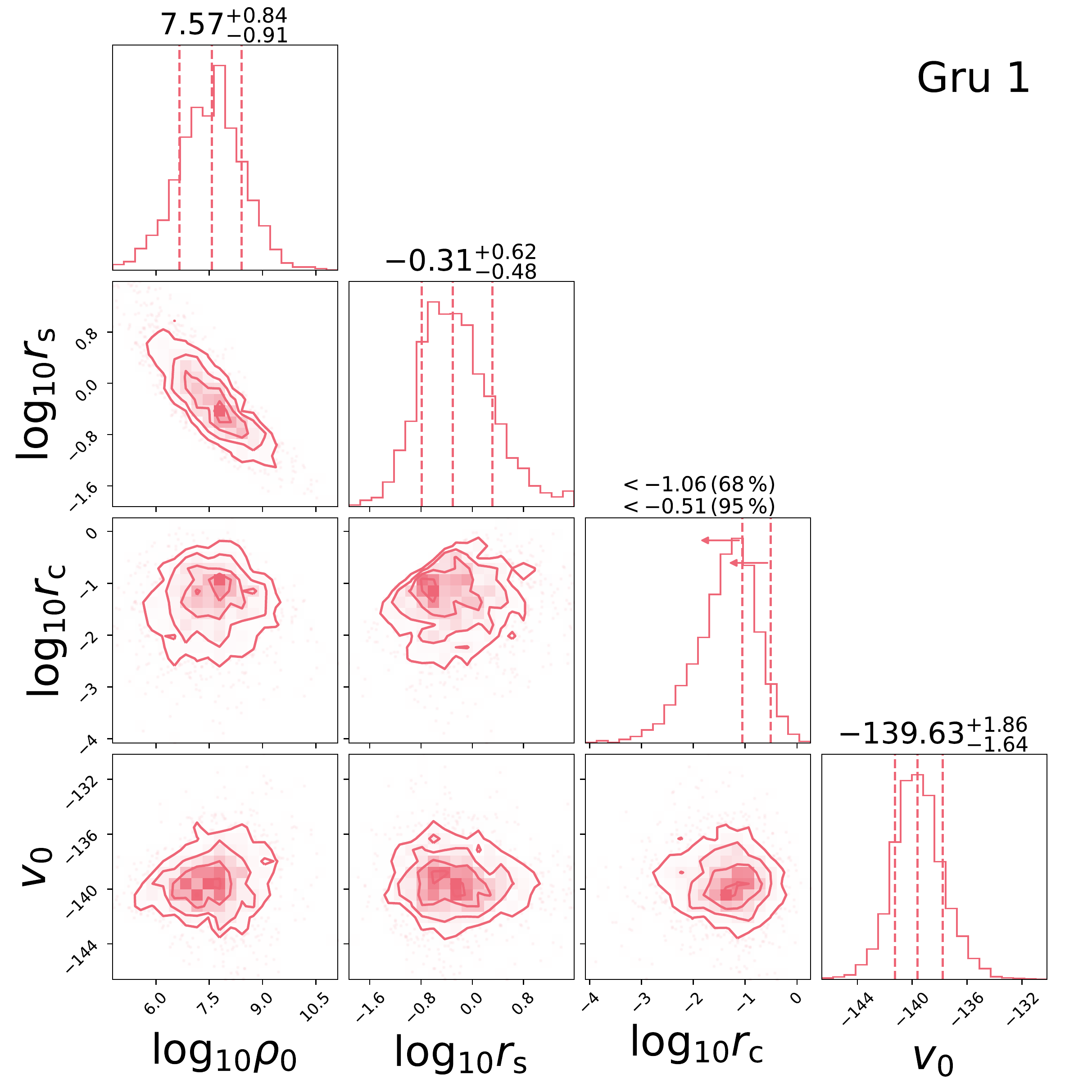}
        \caption{%
            Constraints on the dark-matter density profiles of the four (ultra-)faint dwarf galaxies, for the core model, modelled with CJAM.
            Units are omitted from the labels for clarity.
            The parameters are the characteristic dark-matter density~$\rho_0$ in $M_\sun\,\mathrm{kpc}^{-3}$, the scale radius~$r_\mathrm{s}$ and core radius~$r_\mathrm{c}$ in $\mathrm{kpc}$, and the systemic velocity~$v_0$ in $\mathrm{km}\,\mathrm{s}^{-1}$.
            The contours correspond to $0.5\sigma$, $1.0\sigma$, $1.5\sigma$, and $2.0\sigma$ confidence levels, where $\sigma$ is the standard deviation of a two-dimensional normal distribution.
            The vertical dashed lines in the one-dimensional histograms indicate the median and the 68\% confidence interval (without arrows) or the 68\% and 95\% confidence limits (upper and lower arrows, respectively).%
        }
        \label{fig:sidmaltcorner}
    \end{figure*}

\begin{figure*}
        \includegraphics[width=0.5\linewidth]{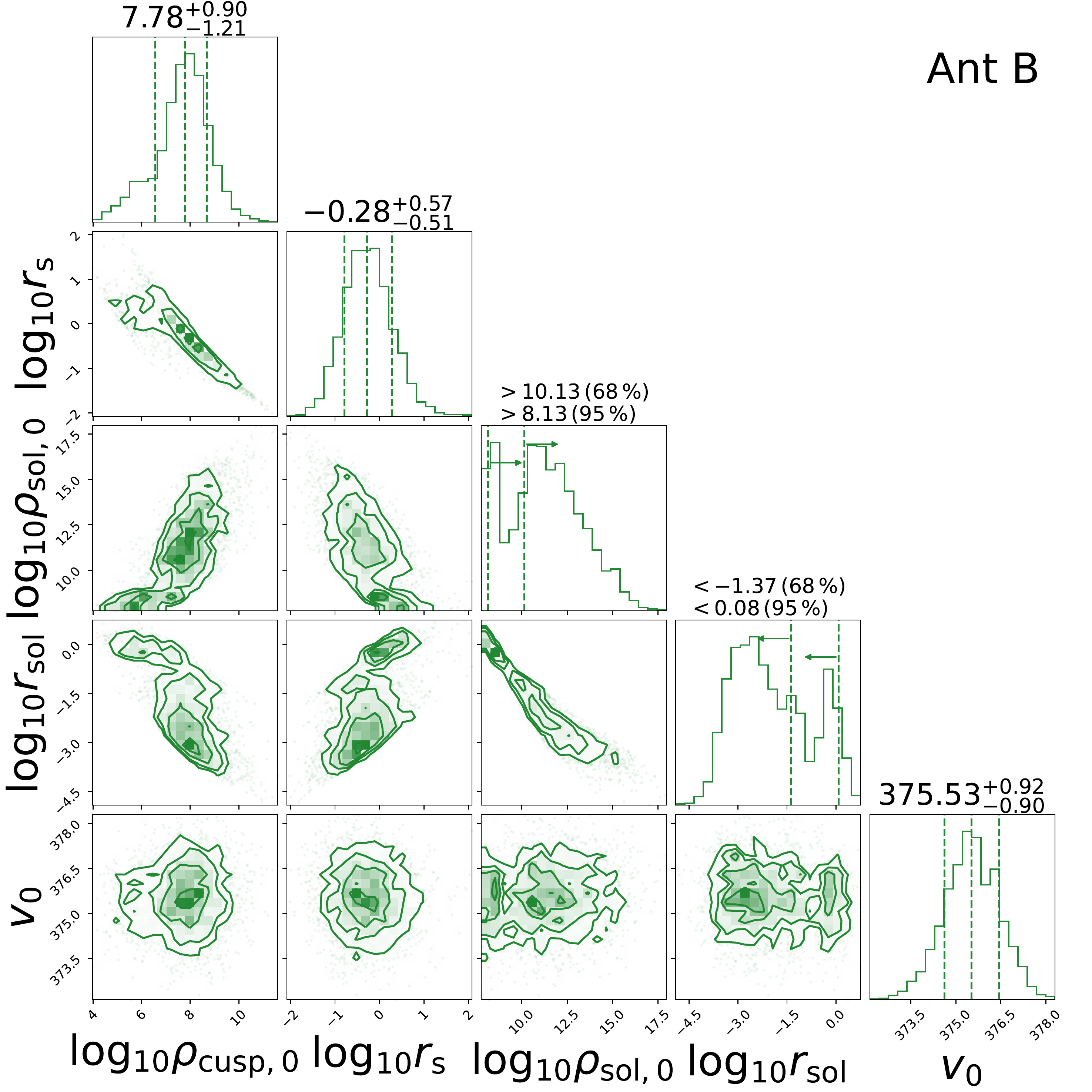}%
        \includegraphics[width=0.5\linewidth]{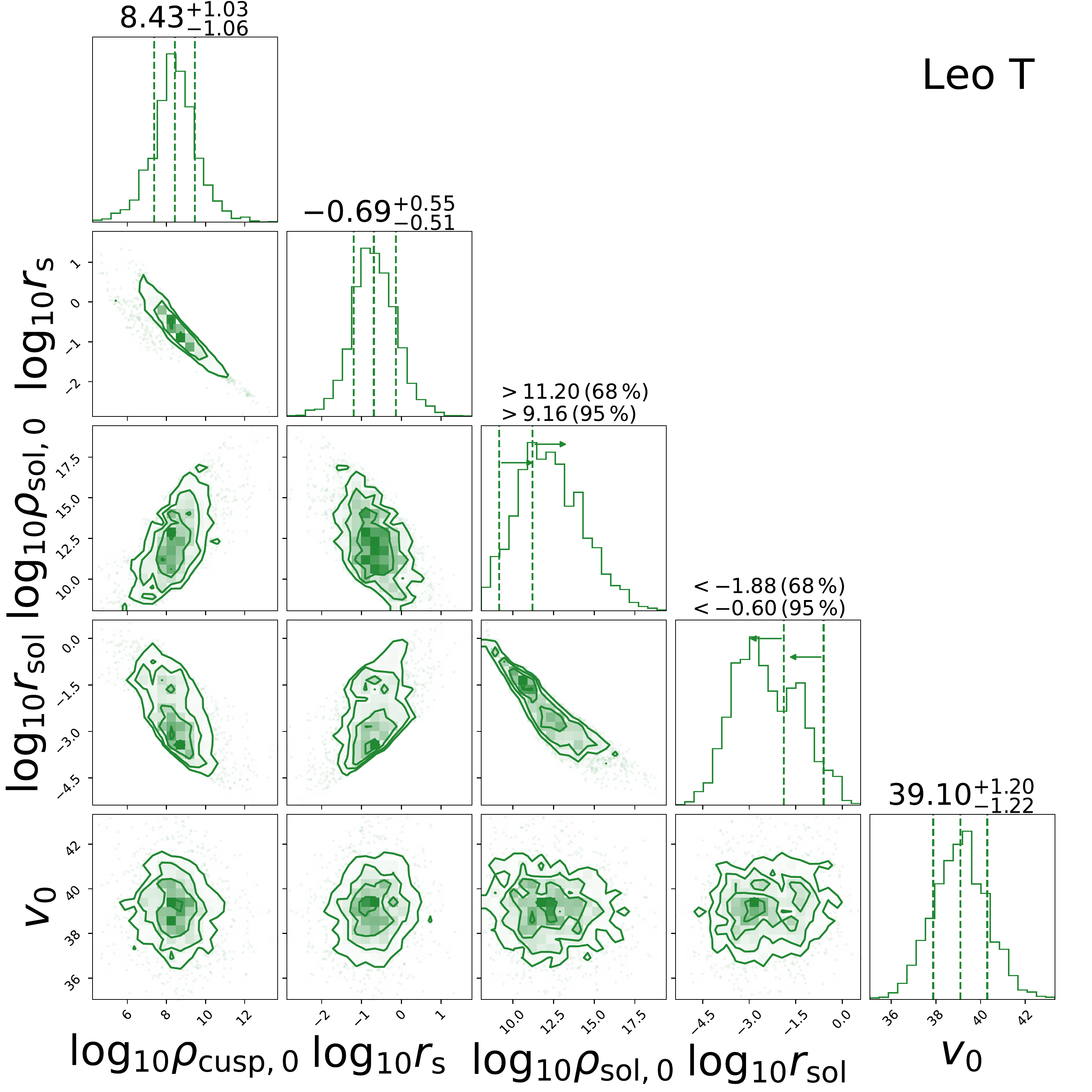}
        \includegraphics[width=0.5\linewidth]{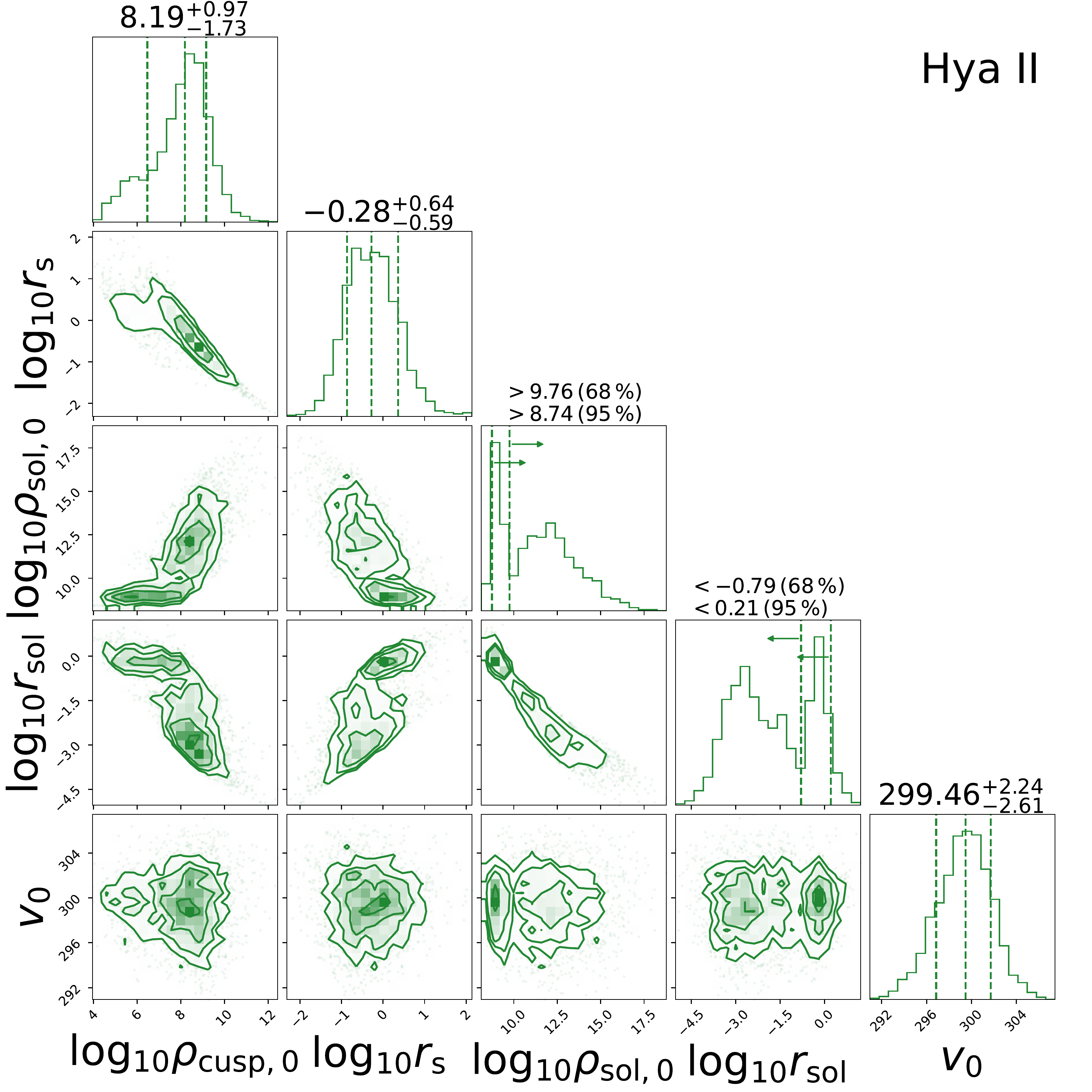}%
        \includegraphics[width=0.5\linewidth]{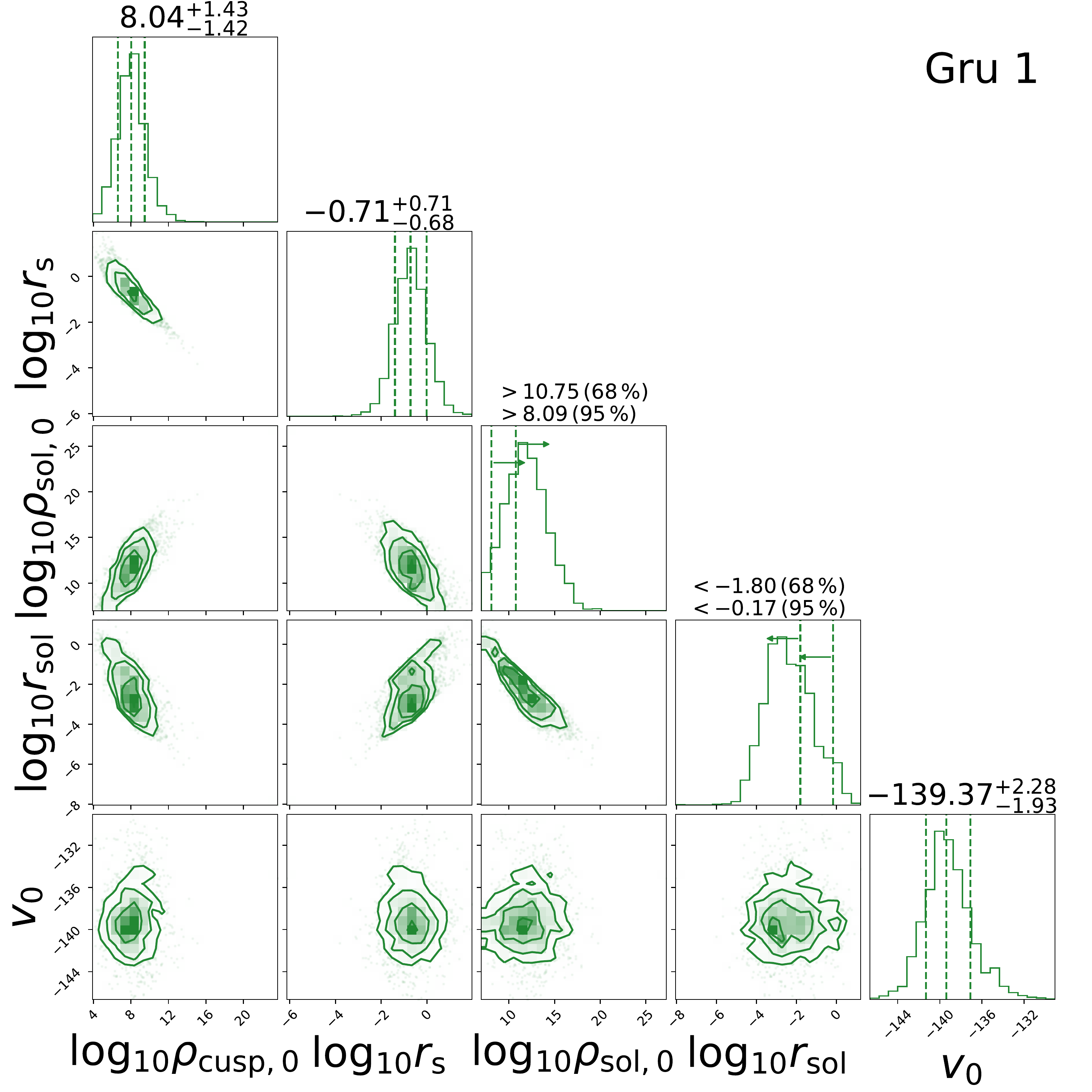}
        \caption{%
            Constraints on the dark-matter density profiles of the four (ultra-)faint dwarf galaxies, for the soliton model, modelled with CJAM.
            Units are omitted from the labels for clarity.
            The parameters are the characteristic dark-matter density~$\rho_{\mathrm{cusp},0}$ in $M_\sun\,\mathrm{kpc}^{-3}$ and the scale radius~$r_\mathrm{s}$ in $\mathrm{kpc}$ of the large-scale, cusp-like component, the central density~$\rho_{\mathrm{sol},0}$ in $M_\sun\,\mathrm{kpc}^{-3}$ of the soliton, the soliton radius~$r_\mathrm{sol}$ in $\mathrm{kpc}$, and the systemic velocity~$v_0$ in $\mathrm{km}\,\mathrm{s}^{-1}$.
            The contours correspond to $0.5\sigma$, $1.0\sigma$, $1.5\sigma$, and $2.0\sigma$ confidence levels, where $\sigma$ is the standard deviation of a two-dimensional normal distribution.
            The vertical dashed lines in the one-dimensional histograms indicate the median and the 68\% confidence interval (without arrows) or the 68\% and 95\% confidence limits (upper and lower arrows, respectively).%
        }
        \label{fig:fdm4altcorner}
    \end{figure*}

\begin{figure*}
        \includegraphics[width=\linewidth]{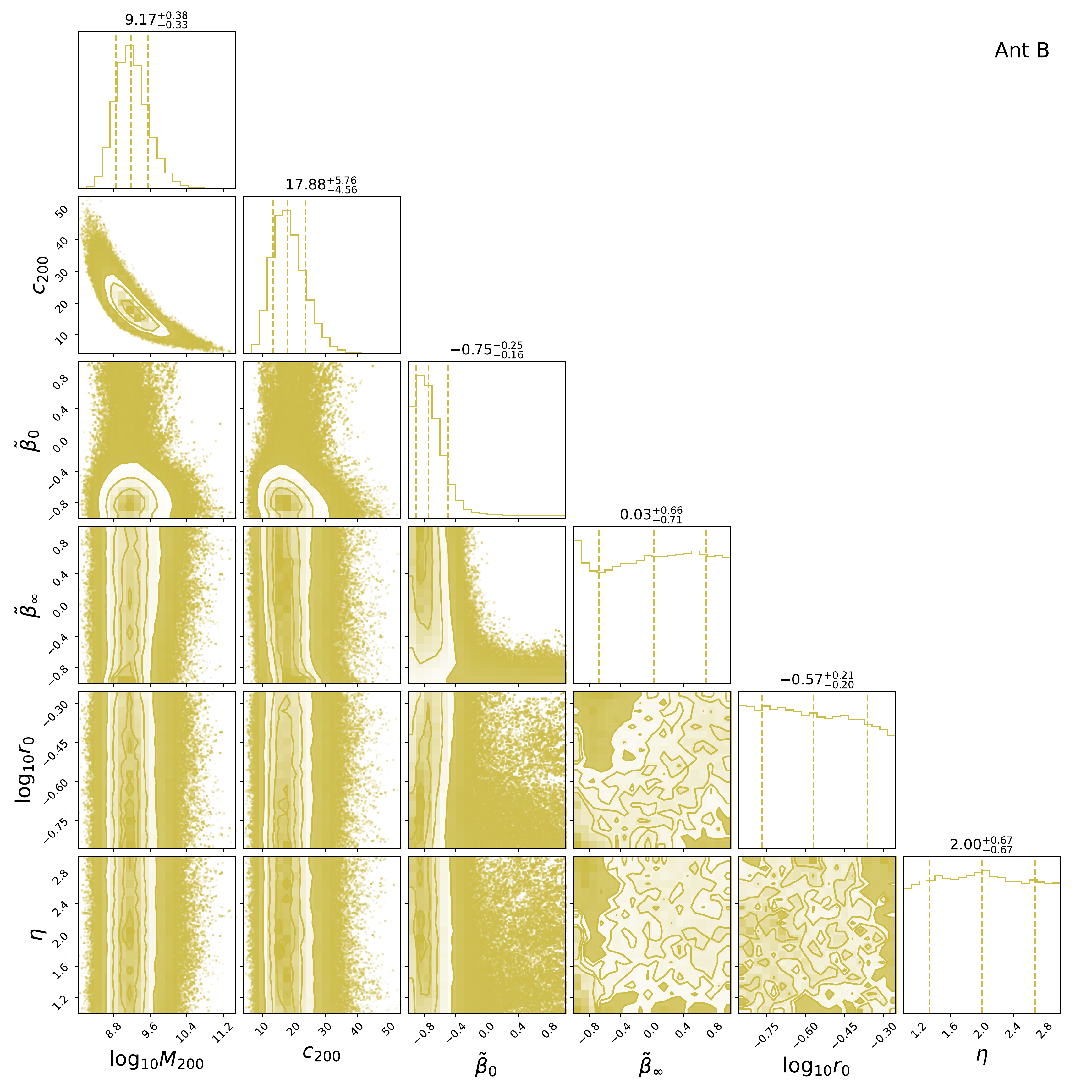}
        \caption{%
            Constraints on the dark-matter density profiles of Antlia~B, for the cusp model, modelled with GravSphere.
            Units are omitted from the labels for clarity.
            The parameters are the virial mass~$M_{200}$ in $M_\sun$, the concentration parameter~$c_{200}$, the inner symmetrized anisotropy~$\tilde\beta_0$ and outer symmetrized anisotropy~$\tilde\beta_\infty$, the anisotropy transition radius~$r_0$ in $\mathrm{kpc}$, and the anisotropy transition rapidity~$\eta$.
            The contours correspond to $0.5\sigma$, $1.0\sigma$, $1.5\sigma$, and $2.0\sigma$ confidence levels, where $\sigma$ is the standard deviation of a two-dimensional normal distribution.
            The vertical dashed lines in the one-dimensional histograms indicate the median and the 68\% confidence interval.%
        }
        \label{fig:gsnfwcornerAntB}
    \end{figure*}

\begin{figure*}
        \includegraphics[width=\linewidth]{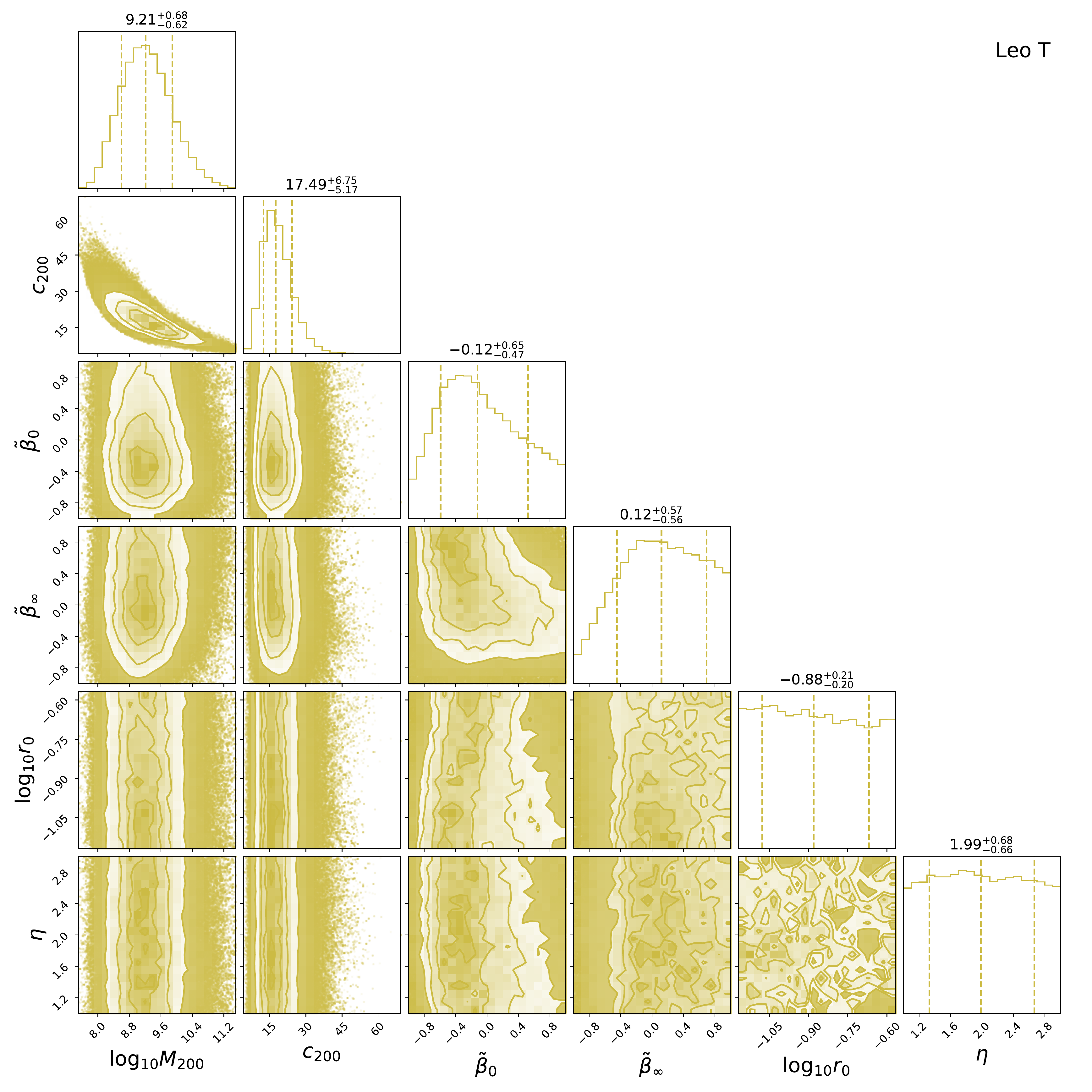}
        \caption{%
            Constraints on the dark-matter density profiles of Leo~T, for the cusp model, modelled with GravSphere.
            Units are omitted from the labels for clarity.
            The parameters are the virial mass~$M_{200}$ in $M_\sun$, the concentration parameter~$c_{200}$, the inner symmetrized anisotropy~$\tilde\beta_0$ and outer symmetrized anisotropy~$\tilde\beta_\infty$, the anisotropy transition radius~$r_0$ in $\mathrm{kpc}$, and the anisotropy transition rapidity~$\eta$.
            The contours correspond to $0.5\sigma$, $1.0\sigma$, $1.5\sigma$, and $2.0\sigma$ confidence levels, where $\sigma$ is the standard deviation of a two-dimensional normal distribution.
            The vertical dashed lines in the one-dimensional histograms indicate the median and the 68\% confidence interval.%
        }
        \label{fig:gsnfwcornerLeoT}
    \end{figure*}

\begin{figure*}
        \includegraphics[width=\linewidth]{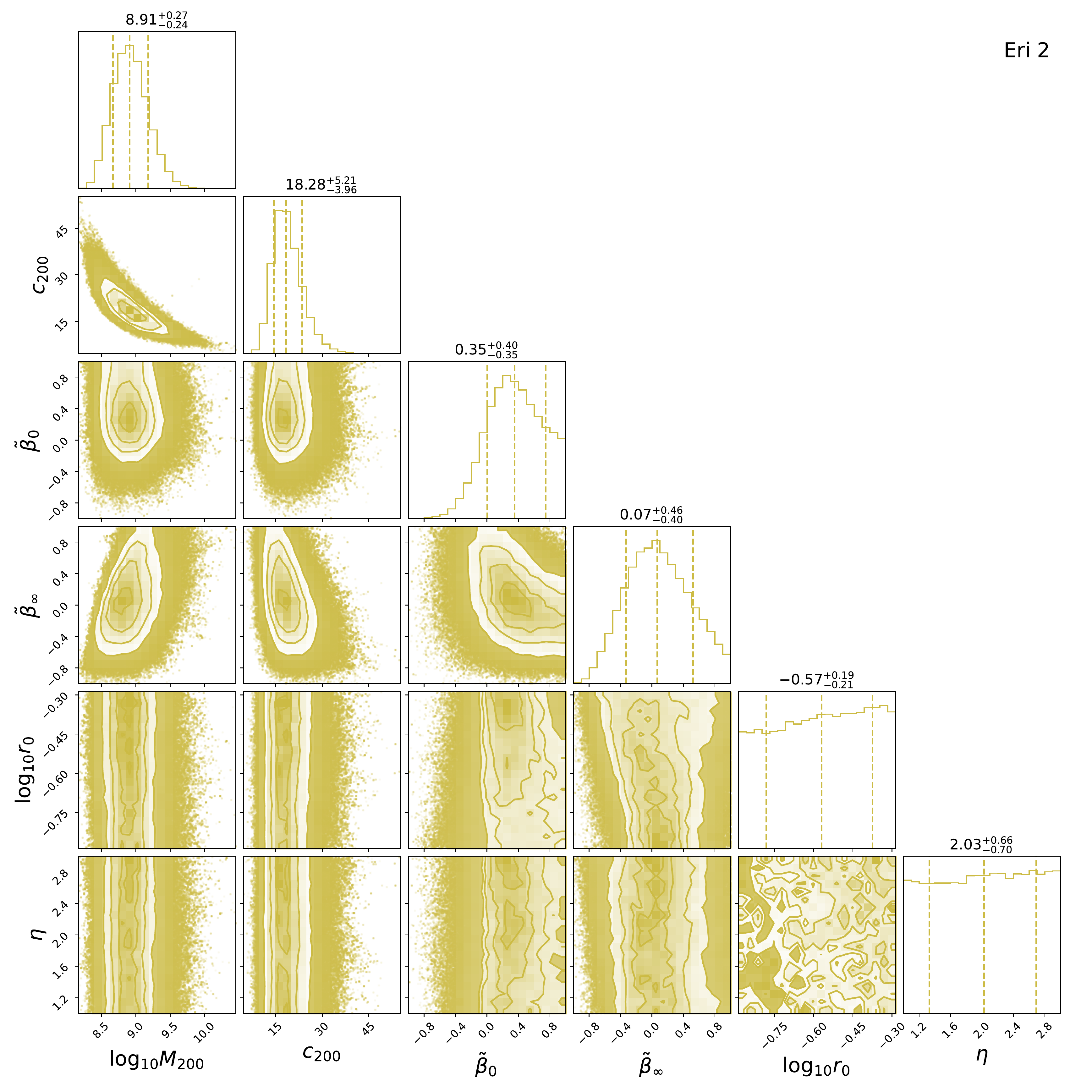}
        \caption{%
            Constraints on the dark-matter density profiles of Eridanus~2, for the cusp model, modelled with GravSphere.
            Units are omitted from the labels for clarity.
            The parameters are the virial mass~$M_{200}$ in $M_\sun$, the concentration parameter~$c_{200}$, the inner symmetrized anisotropy~$\tilde\beta_0$ and outer symmetrized anisotropy~$\tilde\beta_\infty$, the anisotropy transition radius~$r_0$ in $\mathrm{kpc}$, and the anisotropy transition rapidity~$\eta$.
            The contours correspond to $0.5\sigma$, $1.0\sigma$, $1.5\sigma$, and $2.0\sigma$ confidence levels, where $\sigma$ is the standard deviation of a two-dimensional normal distribution.
            The vertical dashed lines in the one-dimensional histograms indicate the median and the 68\% confidence interval.%
        }
        \label{fig:gsnfwcornerEri2}
    \end{figure*}

\begin{figure*}
        \includegraphics[width=\linewidth]{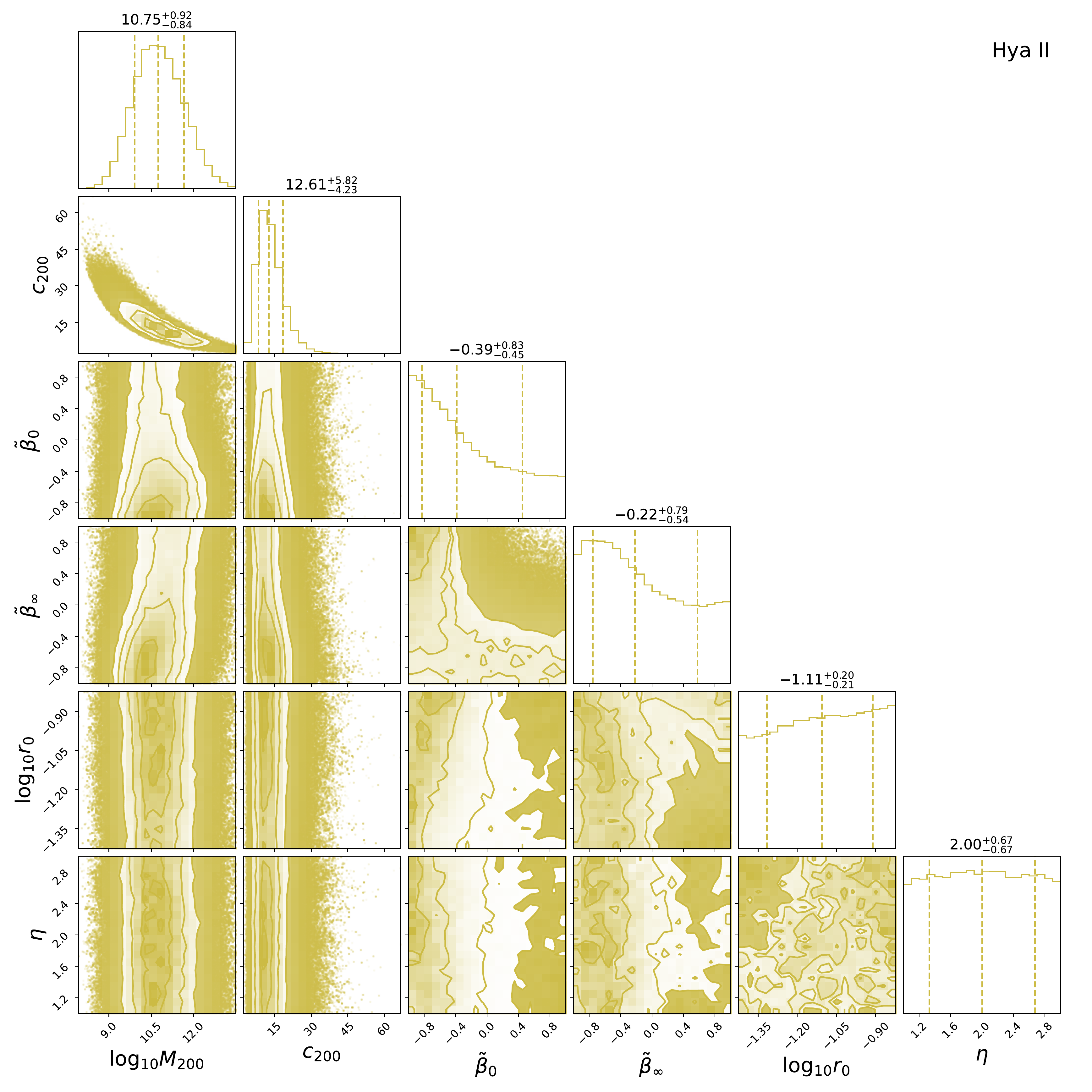}
        \caption{%
            Constraints on the dark-matter density profiles of Hydra~II, for the cusp model, modelled with GravSphere.
            Units are omitted from the labels for clarity.
            The parameters are the virial mass~$M_{200}$ in $M_\sun$, the concentration parameter~$c_{200}$, the inner symmetrized anisotropy~$\tilde\beta_0$ and outer symmetrized anisotropy~$\tilde\beta_\infty$, the anisotropy transition radius~$r_0$ in $\mathrm{kpc}$, and the anisotropy transition rapidity~$\eta$.
            The contours correspond to $0.5\sigma$, $1.0\sigma$, $1.5\sigma$, and $2.0\sigma$ confidence levels, where $\sigma$ is the standard deviation of a two-dimensional normal distribution.
            The vertical dashed lines in the one-dimensional histograms indicate the median and the 68\% confidence interval.%
        }
        \label{fig:gsnfwcornerHyaII}
    \end{figure*}

\begin{figure*}
        \includegraphics[width=\linewidth]{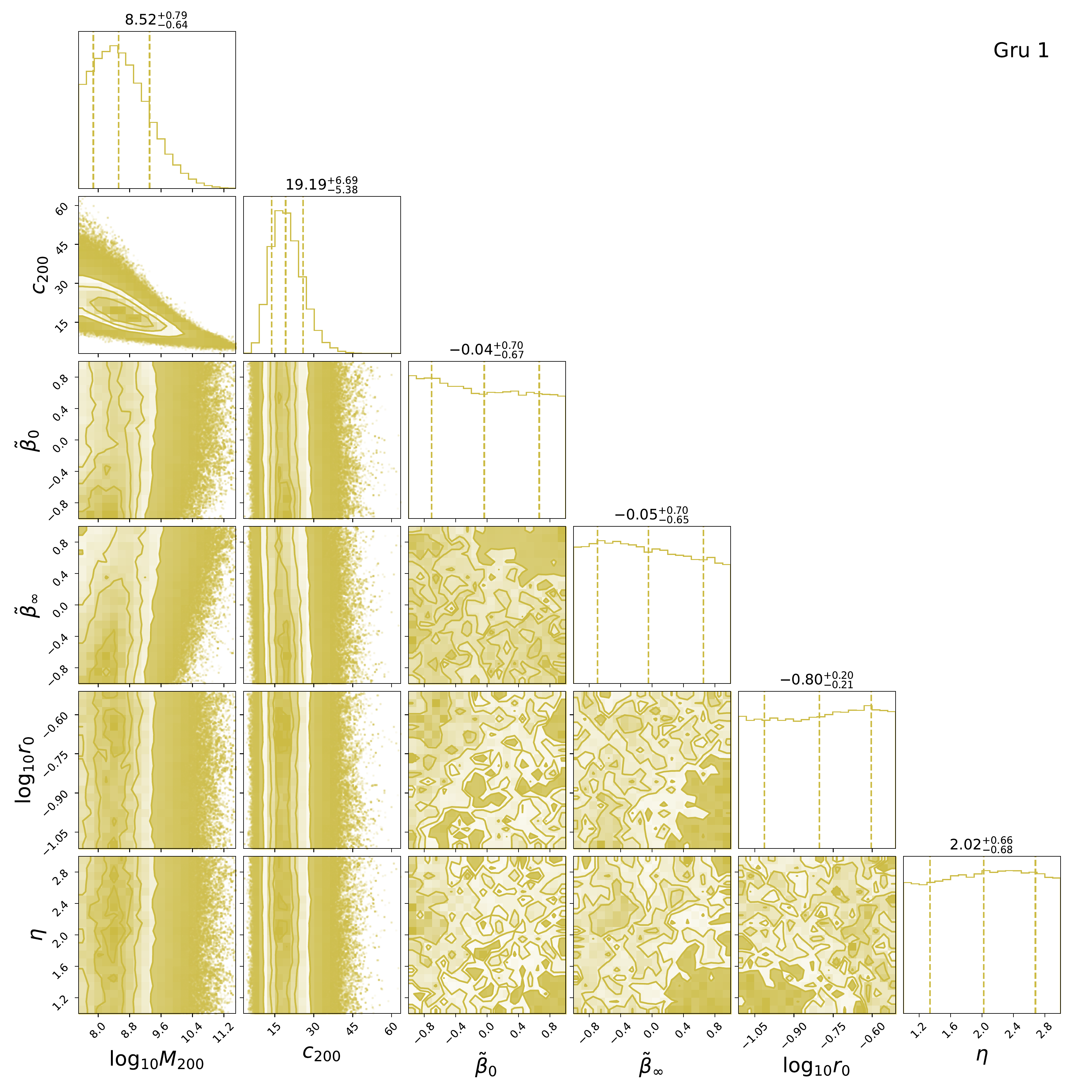}
        \caption{%
            Constraints on the dark-matter density profiles of Grus~1, for the cusp model, modelled with GravSphere.
            Units are omitted from the labels for clarity.
            The parameters are the virial mass~$M_{200}$ in $M_\sun$, the concentration parameter~$c_{200}$, the inner symmetrized anisotropy~$\tilde\beta_0$ and outer symmetrized anisotropy~$\tilde\beta_\infty$, the anisotropy transition radius~$r_0$ in $\mathrm{kpc}$, and the anisotropy transition rapidity~$\eta$.
            The contours correspond to $0.5\sigma$, $1.0\sigma$, $1.5\sigma$, and $2.0\sigma$ confidence levels, where $\sigma$ is the standard deviation of a two-dimensional normal distribution.
            The vertical dashed lines in the one-dimensional histograms indicate the median and the 68\% confidence interval.%
        }
        \label{fig:gsnfwcornerGru1}
    \end{figure*}

\begin{figure*}
        \includegraphics[width=\linewidth]{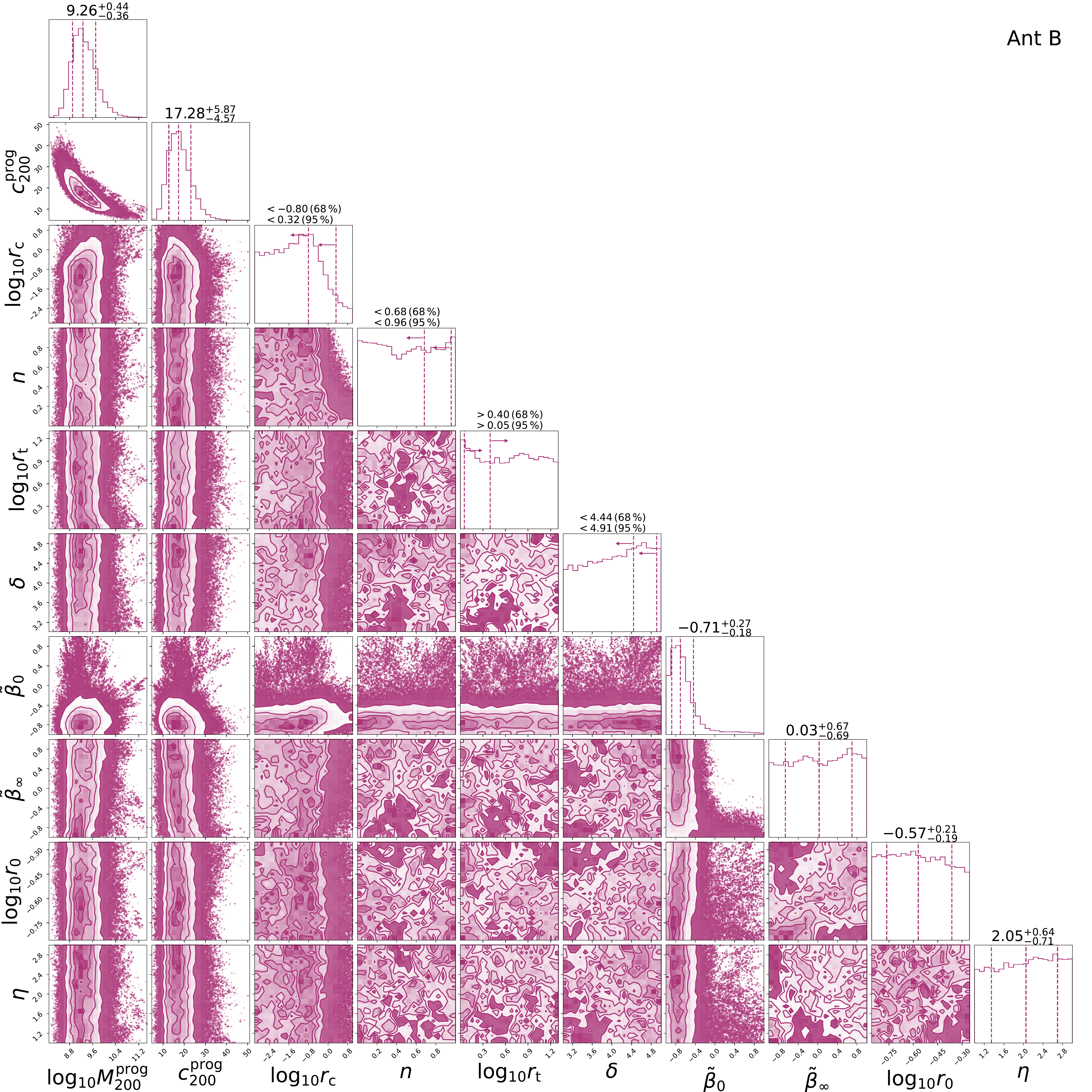}
        \caption{%
            Constraints on the dark-matter density profiles of Antlia~B, for the core+tides model, modelled with GravSphere.
            Units are omitted from the labels for clarity.
            The parameters are the progenitor virial mass~$M_{200}^\mathrm{prog}$ in $M_\sun$, the progenitor concentration parameter~$c_{200}^\mathrm{prog}$, the core radius~$r_\mathrm{c}$ in $\mathrm{kpc}$, the coredness parameter~$n$, the tidal radius~$r_\mathrm{t}$ in $\mathrm{kpc}$, the negative logarithmic density slope~$\delta$ beyond $r_\mathrm{t}$, the inner symmetrized anisotropy~$\tilde\beta_0$ and outer symmetrized anisotropy~$\tilde\beta_\infty$, the anisotropy transition radius~$r_0$ in $\mathrm{kpc}$, and the anisotropy transition rapidity~$\eta$.
            The contours correspond to $0.5\sigma$, $1.0\sigma$, $1.5\sigma$, and $2.0\sigma$ confidence levels, where $\sigma$ is the standard deviation of a two-dimensional normal distribution.
            The vertical dashed lines in the one-dimensional histograms indicate the median and the 68\% confidence interval (without arrows) or the 68\% and 95\% confidence limits (upper and lower arrows, respectively).%
        }
        \label{fig:gscorenfwcornerAntB}
    \end{figure*}

\begin{figure*}
        \includegraphics[width=\linewidth]{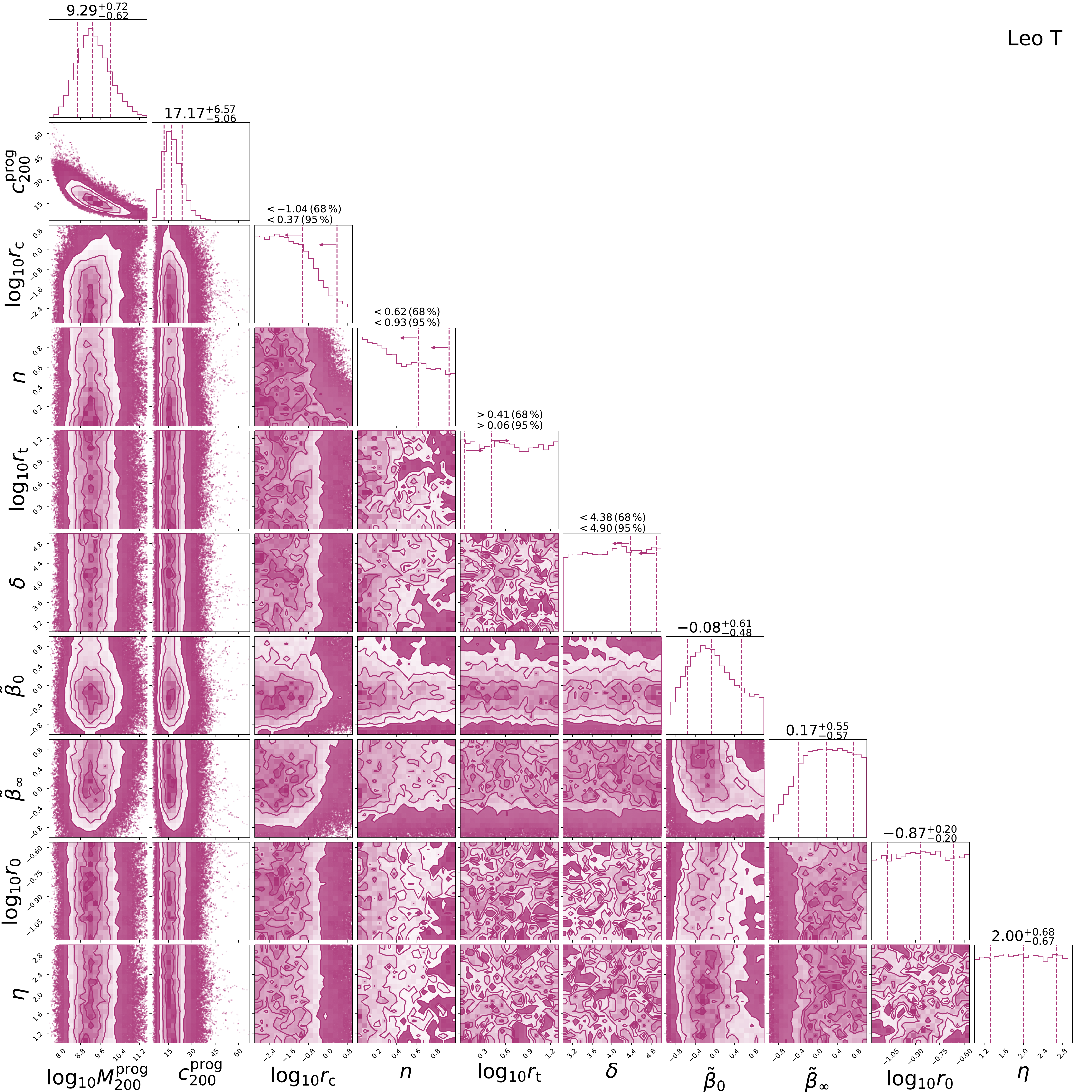}
        \caption{%
            Constraints on the dark-matter density profiles of Leo~T, for the core+tides model, modelled with GravSphere.
            Units are omitted from the labels for clarity.
            The parameters are the progenitor virial mass~$M_{200}^\mathrm{prog}$ in $M_\sun$, the progenitor concentration parameter~$c_{200}^\mathrm{prog}$, the core radius~$r_\mathrm{c}$ in $\mathrm{kpc}$, the coredness parameter~$n$, the tidal radius~$r_\mathrm{t}$ in $\mathrm{kpc}$, the negative logarithmic density slope~$\delta$ beyond $r_\mathrm{t}$, the inner symmetrized anisotropy~$\tilde\beta_0$ and outer symmetrized anisotropy~$\tilde\beta_\infty$, the anisotropy transition radius~$r_0$ in $\mathrm{kpc}$, and the anisotropy transition rapidity~$\eta$.
            The contours correspond to $0.5\sigma$, $1.0\sigma$, $1.5\sigma$, and $2.0\sigma$ confidence levels, where $\sigma$ is the standard deviation of a two-dimensional normal distribution.
            The vertical dashed lines in the one-dimensional histograms indicate the median and the 68\% confidence interval (without arrows) or the 68\% and 95\% confidence limits (upper and lower arrows, respectively).%
        }
        \label{fig:gscorenfwcornerLeoT}
    \end{figure*}

\begin{figure*}
        \includegraphics[width=\linewidth]{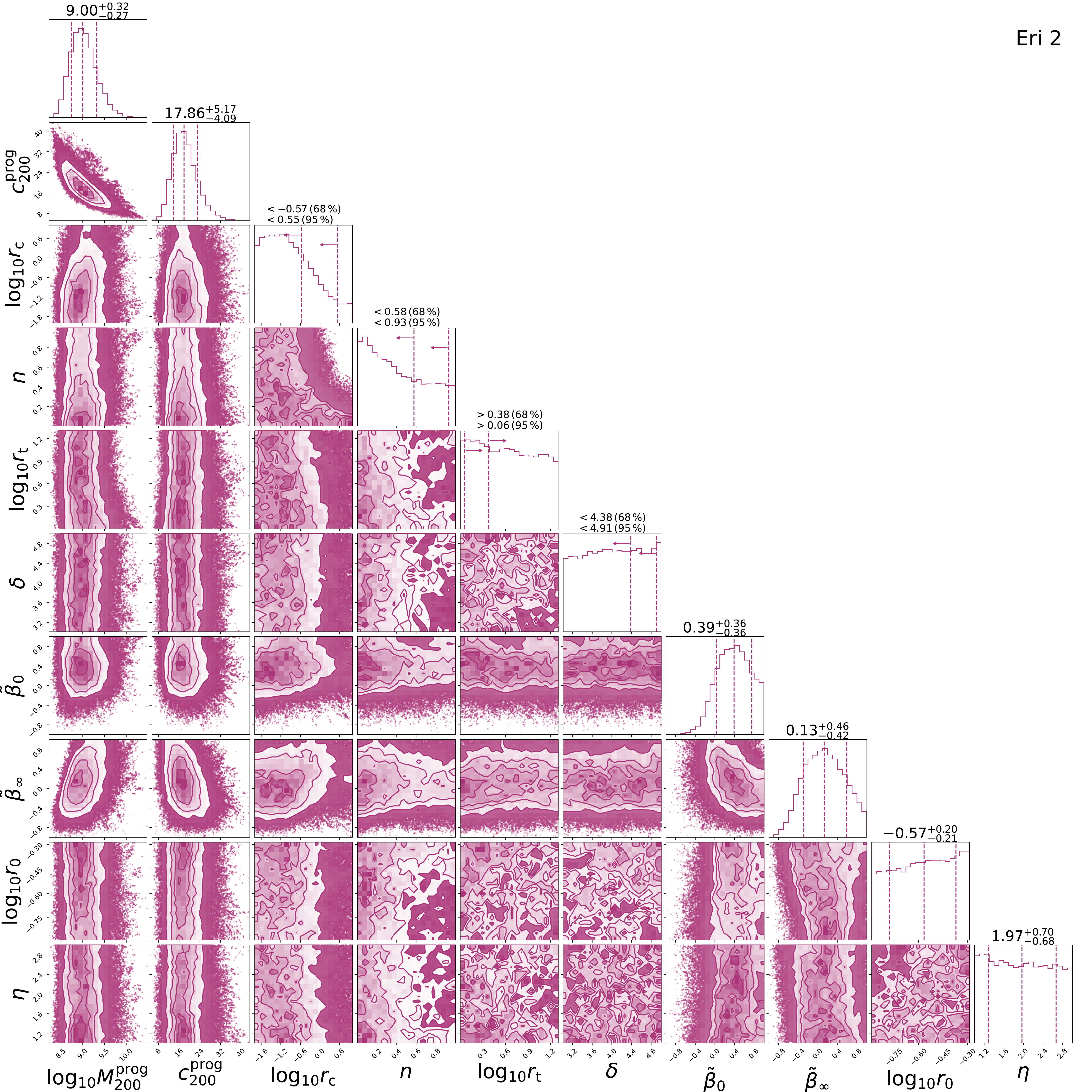}
        \caption{%
            Constraints on the dark-matter density profiles of Eridanus~2, for the core+tides model, modelled with GravSphere.
            Units are omitted from the labels for clarity.
            The parameters are the progenitor virial mass~$M_{200}^\mathrm{prog}$ in $M_\sun$, the progenitor concentration parameter~$c_{200}^\mathrm{prog}$, the core radius~$r_\mathrm{c}$ in $\mathrm{kpc}$, the coredness parameter~$n$, the tidal radius~$r_\mathrm{t}$ in $\mathrm{kpc}$, the negative logarithmic density slope~$\delta$ beyond $r_\mathrm{t}$, the inner symmetrized anisotropy~$\tilde\beta_0$ and outer symmetrized anisotropy~$\tilde\beta_\infty$, the anisotropy transition radius~$r_0$ in $\mathrm{kpc}$, and the anisotropy transition rapidity~$\eta$.
            The contours correspond to $0.5\sigma$, $1.0\sigma$, $1.5\sigma$, and $2.0\sigma$ confidence levels, where $\sigma$ is the standard deviation of a two-dimensional normal distribution.
            The vertical dashed lines in the one-dimensional histograms indicate the median and the 68\% confidence interval (without arrows) or the 68\% and 95\% confidence limits (upper and lower arrows, respectively).%
        }
        \label{fig:gscorenfwcornerEri2}
    \end{figure*}

\begin{figure*}
        \includegraphics[width=\linewidth]{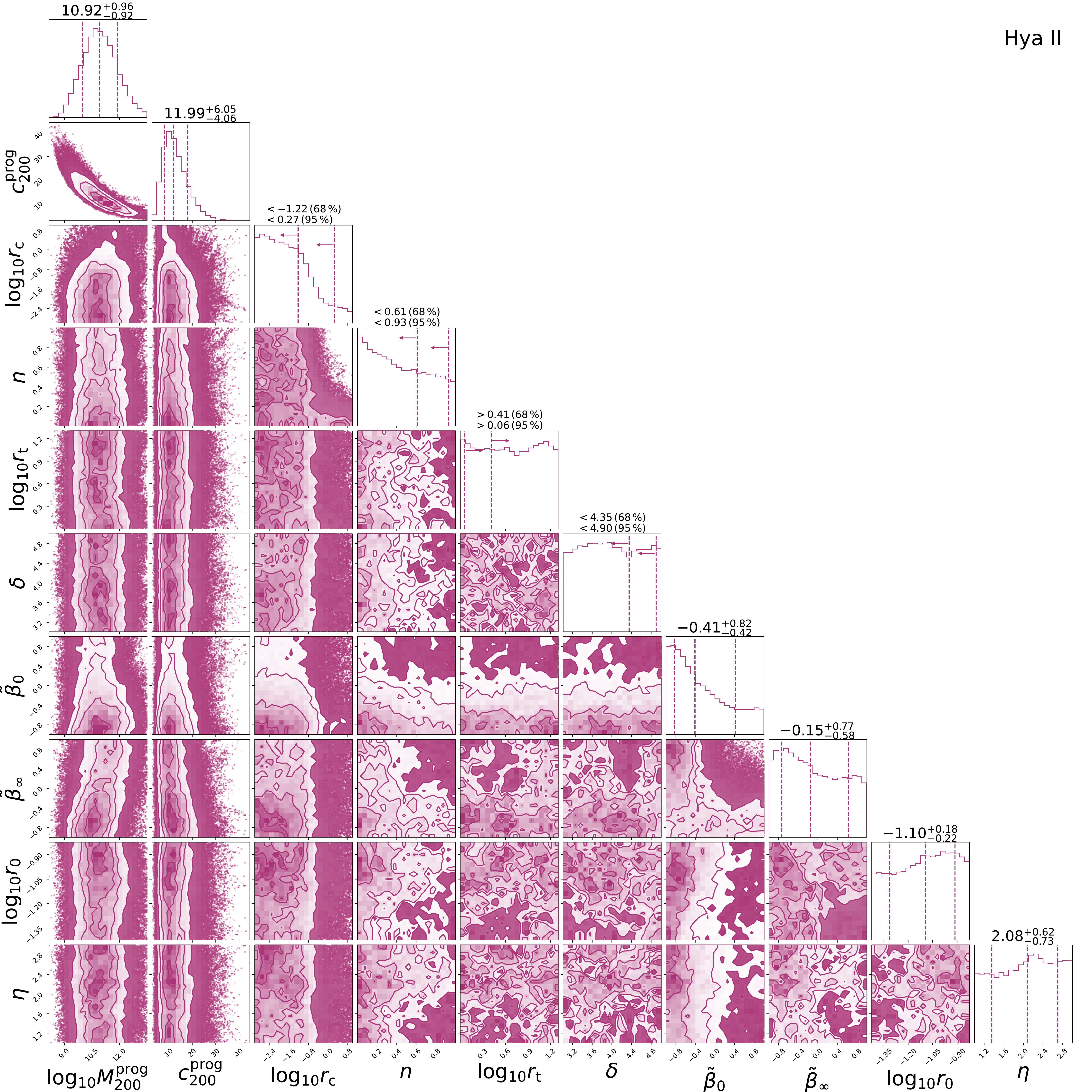}
        \caption{%
            Constraints on the dark-matter density profiles of Hydra~II, for the core+tides model, modelled with GravSphere.
            Units are omitted from the labels for clarity.
            The parameters are the progenitor virial mass~$M_{200}^\mathrm{prog}$ in $M_\sun$, the progenitor concentration parameter~$c_{200}^\mathrm{prog}$, the core radius~$r_\mathrm{c}$ in $\mathrm{kpc}$, the coredness parameter~$n$, the tidal radius~$r_\mathrm{t}$ in $\mathrm{kpc}$, the negative logarithmic density slope~$\delta$ beyond $r_\mathrm{t}$, the inner symmetrized anisotropy~$\tilde\beta_0$ and outer symmetrized anisotropy~$\tilde\beta_\infty$, the anisotropy transition radius~$r_0$ in $\mathrm{kpc}$, and the anisotropy transition rapidity~$\eta$.
            The contours correspond to $0.5\sigma$, $1.0\sigma$, $1.5\sigma$, and $2.0\sigma$ confidence levels, where $\sigma$ is the standard deviation of a two-dimensional normal distribution.
            The vertical dashed lines in the one-dimensional histograms indicate the median and the 68\% confidence interval (without arrows) or the 68\% and 95\% confidence limits (upper and lower arrows, respectively).%
        }
        \label{fig:gscorenfwcornerHyaII}
    \end{figure*}

\begin{figure*}
        \includegraphics[width=\linewidth]{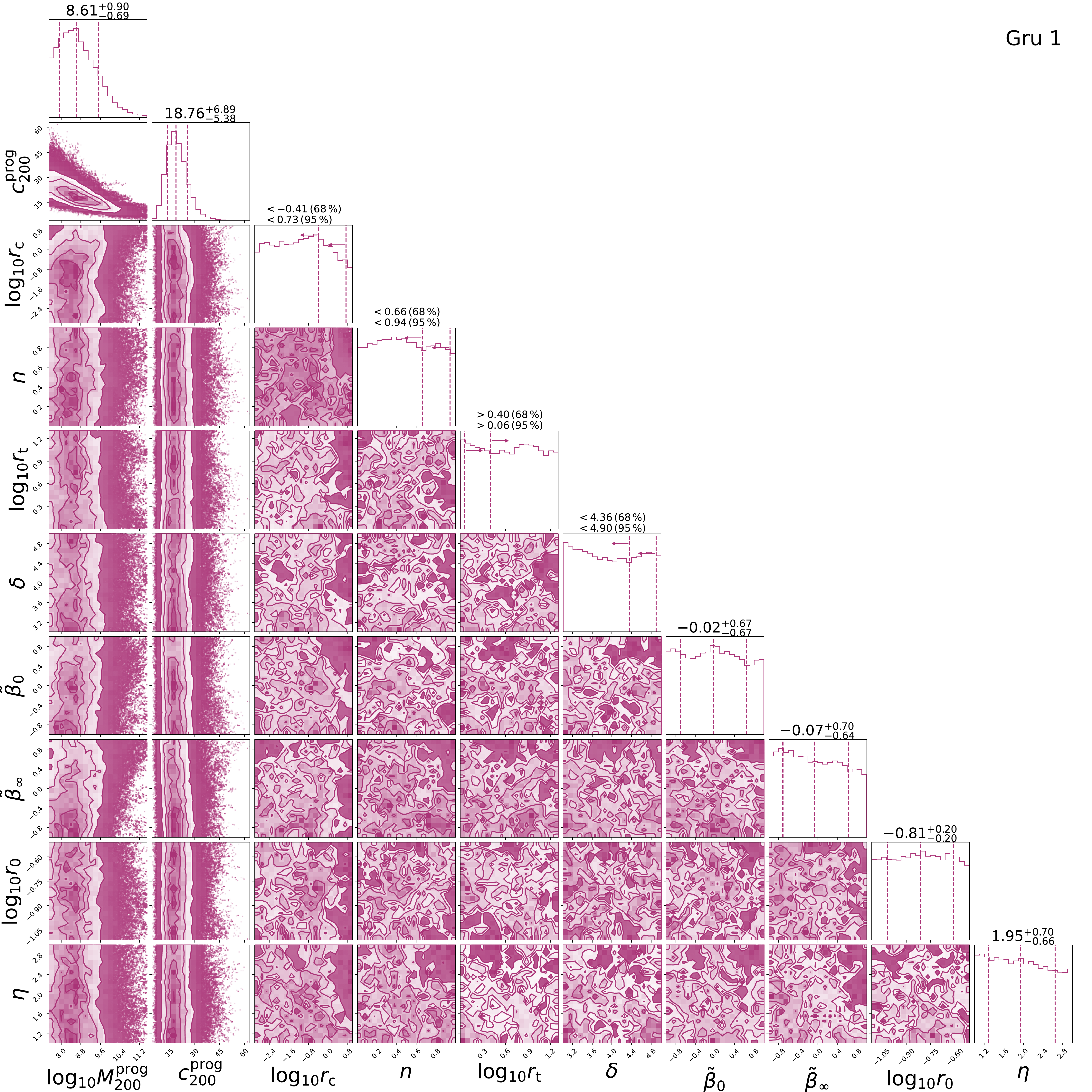}
        \caption{%
            Constraints on the dark-matter density profiles of Grus~1, for the core+tides model, modelled with GravSphere.
            Units are omitted from the labels for clarity.
            The parameters are the progenitor virial mass~$M_{200}^\mathrm{prog}$ in $M_\sun$, the progenitor concentration parameter~$c_{200}^\mathrm{prog}$, the core radius~$r_\mathrm{c}$ in $\mathrm{kpc}$, the coredness parameter~$n$, the tidal radius~$r_\mathrm{t}$ in $\mathrm{kpc}$, the negative logarithmic density slope~$\delta$ beyond $r_\mathrm{t}$, the inner symmetrized anisotropy~$\tilde\beta_0$ and outer symmetrized anisotropy~$\tilde\beta_\infty$, the anisotropy transition radius~$r_0$ in $\mathrm{kpc}$, and the anisotropy transition rapidity~$\eta$.
            The contours correspond to $0.5\sigma$, $1.0\sigma$, $1.5\sigma$, and $2.0\sigma$ confidence levels, where $\sigma$ is the standard deviation of a two-dimensional normal distribution.
            The vertical dashed lines in the one-dimensional histograms indicate the median and the 68\% confidence interval (without arrows) or the 68\% and 95\% confidence limits (upper and lower arrows, respectively).%
        }
        \label{fig:gscorenfwcornerGru1}
    \end{figure*}

    In this appendix we display the constraints on the parameters of the dark-matter density profiles.
    Figure~\ref{fig:cdmaltcorner} shows the constraints on the CJAM cusp model for each galaxy, Fig.~\ref{fig:sidmaltcorner} on the CJAM core model, and Fig.~\ref{fig:fdm4altcorner} on the CJAM soliton model.
    We show the constraints on the GravSphere cusp model in Figs.~\ref{fig:gsnfwcornerAntB}--\ref{fig:gsnfwcornerGru1} and on the GravSphere core+tides model in Figs.~\ref{fig:gscorenfwcornerAntB}--\ref{fig:gscorenfwcornerGru1}, also for Eri~2.
\end{document}